\documentclass[a4,letterpaper,11pt]{article}
    
    \pdfoutput = 1 

    \usepackage{jheppub} 
    \usepackage{xcolor}
    \usepackage[T1]{fontenc}
    \usepackage{lmodern}
    \usepackage{graphicx}
    \usepackage{xspace}
    \usepackage{slashed}
    \usepackage{multirow}
    \usepackage{epstopdf}
    \usepackage{amsmath}
    \usepackage{slashed}
    \usepackage{amssymb}
    \usepackage{mathtools}
    \usepackage{graphicx}
    \usepackage{subfigure}
    \usepackage{setspace}
    \usepackage{sidecap}
    \usepackage{color}
    \usepackage{hyperref}
    \usepackage{tabu} 
    \usepackage{hyperref}
    \usepackage[export]{adjustbox}
    \usepackage{tensor}
    \usepackage{empheq}
    \usepackage{verbatim}
    
    \newcommand{\lb}{\Big{\lbrack}}
    \newcommand{\rb}{\Big{\rbrack}}
    \newcommand{\lp}{\Big{(}}
    \newcommand{\rp}{\Big{)}}
    \newcommand{\lbc}{\Big{\lbrace}}
    \newcommand{\rbc}{\Big{\rbrace}}
    \newcommand{\nn}{\nonumber}
    \newcommand{\bn}{\bar{n}}
    \newcommand{\Bvert}{\Big{\vert}}
    \newcommand{\Rangle}{\Big{\rangle}}
    \newcommand{\Langle}{\Big{\langle}}
    \newcommand{\bmat}[1]{{\boldsymbol{\mathrm{#1}}}}
    
    \newcommand{\Vq}[1]{ \lp i g \gamma^{\mu_{#1}} T^{a_{#1}} \rp }
    \newcommand{\Vbq}[1]{\lp i g \gamma^{\nu_{#1}} T^{b_{#1}} \rp }
    \newcommand{\Do}[1]{\frac{i(1+\slashed{v})}{2p_{t}^{0}(#1)} }
    \newcommand{\uo}{\lp u^{(0)} \rp^{\dag}}
    \newcommand{\Dbo}[1]{\frac{i(1-\slashed{v})}{2p_{t}^{'0}(#1)} }
    \newcommand{\vo}{ v^{(0)} }

    \newcommand{\lrP}{\raise 0.9ex\hbox{$^\leftrightarrow$} \hspace{-10.5pt} \mathcal{P}}
    \newcommand{\Q}[4]{ {}^{#1} #2 ^{[#4]}_{#3} }
    \newcommand{\bmatn}[1]{\vert \bmat{#1} \vert}
    \newcommand{\UV}{\text{UV}}
    \newcommand{\IR}{\text{IR}}
    \newcommand{\ve}{\text{v}}
    \newcommand{\wave}[1]{\text{#1}}
    
\begin{document}
    \pagestyle{myplain}
    
    \title{An effective field theory approach to quarkonium at small transverse momentum}
    \author[a]{Sean Fleming,}
    \author[b]{Yiannis Makris}
    \author[c]{and Thomas Mehen}

    \affiliation[a]{Department of Physics, University of Arizona, Tucson, AZ 85721, USA}
    \affiliation[b]{INFN Sezione di Pavia, via Bassi 6, I-27100 Pavia, Italy}
    \affiliation[c]{Department of Physics, Duke University, Durham, NC 27708}
    
    \emailAdd{fleming@physics.arizona.edu}
    \emailAdd{yiannis.makris@pv.infn.it}
    \emailAdd{mehen@phy.duke.edu}
    
    \preprint{LA-UR-19-31110}
    \abstract{In this work we apply effective field theory (EFT) to observables in quarkonium production and decay that are sensitive to soft gluon radiation, in particular measurements that are sensitive to small transverse momentum. Within the EFT framework we study $\chi_Q$ decay to light quarks followed by the fragmentation of those quarks to light hadrons. We derive a factorization theorem that involves transverse momentum distribution (TMD) fragmentation functions and new quarkonium TMD shape functions. We derive renormalization group equations, both in rapidity and virtuality, which are used to evolve the different terms in the factorization theorem to  resum large logarithms. This theoretical framework will provide a systematic treatment of quarkonium production and decay processes in TMD sensitive measurements.}

    \maketitle
    
    \section{Introduction}
    \label{sec:intro}
    
    Transverse momentum distributions (TMDs) have been the subject of many theoretical and phenomenological studies in recent years. TMD parton distribution functions (TMDPDFs), which encode information about the three-dimensional distribution of partons in the nucleon, are usually studied through the $k_T$-factorization theorems~\cite{GarciaEchevarria:2011rb, Echevarria:2014rua, Vladimirov:2017ksc, Gaunt:2014ska, Ji:2004wu}. Processes for which such factorization theorems exist are Drell-Yan and Higgs production~\cite{Adamczyk:2015gyk, Aghasyan:2017jop}, semi-inclusive DIS (SIDIS)~\cite{Airapetian:2012ki, Aghasyan:2017ctw}, and electron-positron annihilation to di-hadrons~\cite{Leitgab:2013qh, Lees:2013rqd} (see also Refs.~\cite{Bain:2016rrv, Neill:2016vbi, Makris:2017hjk, Kang:2017glf, Makris:2017arq, Gutierrez-Reyes:2018qez, Makris:2018npl, Gutierrez-Reyes:2019vbx, Gutierrez-Reyes:2019msa} for processes involving jets). In these theorems the differential cross section is written in terms of various perturbative and non-perturbative functions up to power corrections. While the perturbative elements can be calculated in a systematic expansion in the strong coupling, the non-perturbative elements need to be fit from experiments. The true power of factorization is that non-perturbative matrix elements are universal and thus can be extracted from one process and used to make predictions for other experiments. Furthermore, factorization separates different scales appearing in a  cross section so that renormalization group (RG) evolution can be used to resum logarithms of ratios of disparate scales which is often needed for convergence of the perturbative expansion. RG evolution is a key procedure for the consistent extraction and use of the universal non-perturbative matrix elements since different experiments can operate at widely separated scales. 
    
    The extraction of non-perturbative TMDs is achieved by parametrizing the form of these distributions and fitting the transverse momentum differential cross section against experimental data from various processes that are sensitive to the TMDs of interest. For example, the Drell-Yan process gives access to the quark TMDPDFs, but is relatively insensitive to the gluon TMDs. In fact, while unpolarized quark TMDPDFs are extensively studied both experimentally and theoretically, the gluon TMDPDF is not well constrained. This is because quarks can interact via electroweak processes which lead to final states that do not intereact strongly, hence only probing the TMDPDF. Processes in which the final state contains hadrons (which can absorb or emit collinear and soft radiation) are sensitive to other TMD distributions, such as TMD fragmentation functions (TMDFFs) and TMD soft functions. Gluon fusion processes, which potentially can be used to extract the gluon TMDPDF are dominated by hadronic final states, so that understanding soft final state radiation is vital. Processes proposed to study gluon TMDs in collider experiments include inclusive, exclusive, and associated quarkonium production processes~\cite{Accardi:2012qut}.
    
    In collider physics, quarkonia are studied within the framework of NRQCD factorization~\cite{Caswell:1985ui, Bodwin:1994jh,Kang:2014tta, Nayak:2005rw, Nayak:2005rt, Kang:2011zza, Fleming:2012wy} where the cross section is written as a sum of products of short distance matching coefficients and the corresponding long distance matrix elements (LDMEs). The short distance coefficients (SDCs) describe the production of the $Q\bar{Q}$ pair in a particular angular momentum and color configuration. In the case of hadronic initial states, SDCs are expressed as a convolution of the partonic cross section and the collinear PDFs. The partonic cross section is calculated  as an expansion in the strong coupling constant~\cite{Cho:1995vh, Cho:1995ce, Petrelli:1997ge,Braaten:1994kd, Ma:1995vi, Braaten:1996rp, Braaten:1994vv, Bodwin:2015iua}. In contrast, LDMEs describe the decay of the $Q\bar{Q}$ pair into the final color-singlet quarkonium state, $\mathcal{Q}$, through soft and ultra-soft gluon emissions. LDMEs are fundamentally non-perturbative objects, they need to be extracted from experiment~\cite{Butenschoen:2011yh, Butenschoen:2012qr, Chao:2012iv, Bodwin:2014gia,  Bodwin:2015iua}, and are thought to be universal. Although in principle all possible intermediate $Q\bar{Q}$ configurations contribute to the final quarkonium state, LDMEs scale with powers of the heavy quarks relative velocity in the quarkonium rest frame, $\ve$, and thus at a fixed order in the $\ve$ expansion only a finite number of LDME contribute. 
    
    Despite many successes, the NRQCD factorization approach is only effective when the quarkonium is produced with relatively large transverse momenta. Intuitively, in this kinematic regime, emissions of soft and ultra-soft gluons from the heavy quark pair cannot alter the large transverse momentum of the quarkonium state. Ignoring these soft emissions, the quarkonium transverse momentum is then determined from the hard process and infrared (IR) divergences that are present in perturbative calculations of the partonic process are absorbed into the non-perturbative LDMEs and collinear PDFs. However, when quarkonia are produced with small transverse momenta, this soft gluon factorization assumption must be relaxed. In Refs.~\cite{Beneke:1997qw, Fleming:2003gt, Fleming:2006cd} the quarkonium photo/lepto-production was studied in the endpoint region which is sensitive to soft radiation where NRQCD factorization breaks down. It was found that promoting the LDMEs into quarkonium shape functions is necessary for accurately accounting for soft radiation from the heavy quark pair. 
    
    A central point of this paper is to introduce TMD quarkonium shape functions which will appear generically for TMD observables involving quarkonia.\footnote{While this paper was being finished Ref.~\cite{Echevarria:2019ynx} appeared which introduces the color singlet TMD shape function for $\eta_c$ production.} For previous studies of quarkonium production at small transverse momentum and/or small-$x$, see Refs. \cite{Ma:2014mri, Ma:2015sia,Sun:2012vc,Dunnen:2014eta,Bacchetta:2018ivt, Berger:2004cc, Berger:2004ct, Qiu:2017xbx}.   Although NRQCD already contains many of the necessary elements for deriving factorization involving an heavy quark-antiquark pair, it must be extended to include other effects. For example, initial or final state collinear modes are not included in NRQCD but are crucial to describe recoil from collinear radiation. Such collinear modes are extensively studied in the framework of soft collinear effective theory (SCET)~\cite{Bauer:2000ew, Bauer:2000yr, Bauer:2001ct, Bauer:2001yt}, and were combined with NRQCD in Ref.~\cite{Fleming:2006cd}. As a matter of fact, a formulation of NRQCD using the label-momentum formalism~\cite{Luke:1999kz} (known as vNRQCD) can be easily expanded to incorporate collinear degrees of freedom.
    
    Recently~\cite{Rothstein:2018dzq}, vNRQCD was formulated in a manifestly soft gauge invariant form by introducing the soft Wilson-lines in the vNRQCD Lagrangian. These are Wilson-lines along the time-like direction (in the quarkonium rest frame) constructed from soft gluons. In this paper we show that the insertion of these Wilson-lines in the quarkonium production/decay operators---whose vacuum expectation values are the LDMEs---is required for properly matching NRQCD onto QCD and gives the operators the correct gauge transformation properties. This is another confirmation that vNRQCD is the correct starting point.
    
    This paper is organized as follows: In Sect.~\ref{sec:efts} we introduce some of the important ingredients of SCET and NRQCD. Next, in Sect.~\ref{sec:ops} we perform a tree level matching with an arbitrary number of soft gluon emissions from the heavy quark-antiquark lines. We do this for the process $q\bar{q} \to Q\bar{Q}$ as a simple example. This approach is important for identifying the various operators relevant at each order of the matching and seeing how they might be related. In the same section we demonstrate  that the resulting operators are invariant under collinear, soft, and ultra-soft gauge transformations. In Sect.~\ref{sec:NLL} we give the factorization of the process $\chi_{J} \to q\bar{q}$, followed by the fragmentation of the light quarks to hadrons. We study a TMD dependent observable so that the differential rate involves the quarkonium TMD shape functions as well as TMDFFs for the final state. Using the factorization theorem we demonstrate how our approach can be used to obtain the relevant renormalization group equations and perform a NLL resummation.  We conclude in Sect.~\ref{sec:conclusions}.

    \section{SCET and vNRQCD}
    \label{sec:efts}
    
    SCET and vNRQCD are effective field theories that describe the degrees of freedom of QCD in the soft and collinear, and  non-relativistic limits, respectively. Both theories are well established and address a wide spectrum of physical phenomena. We aim to study a TMD dependent observable in $\chi_J$ decaying to light quarks followed by their fragmentation to light hadrons. vNRQCD is needed for the heavy quarkonium in the initial state and SCET is needed for the energetic hadrons as well as soft radiation in the final state.  Our aim in this section is not to give a comprehensive review of those EFTs but rather to introduce the elements and notation necessary for the analysis we pursue. 
    
    \subsection{Non-relativistic QCD}
        There are three important scales that appear when studying the dynamics of non-relativistic heavy quarks: the mass of the heavy quark, $m$, the size of their momentum in the quarkonium rest frame, $m \ve$, and their kinetic energy, $m \ve^2$. The distance $r\sim 1/(m\ve)$ gives an estimate on the size of the quarkonium state and the separation between the heavy quark-antiquark pair. The non-relativistic kinetic energy $\Delta E \sim m \ve^2$ is of the same order as the energy splittings of radial excitations. We refer to $m \ve$ and $m \ve^2$ as the soft and ultra-soft scales respectively. Correspondingly, gluons that have all of their four-momentum components scaling as $m \ve$ and $m \ve^2$ are called soft and ultra-soft gluons respectively. While the ultra-soft scale is well within the non-perturbative regime the soft scale is about 1.4 GeV for bottomonium and and about 800 MeV for charmonium. 
    
    The effective theory of vNRQCD is a version of non-relativistic QCD introduced in Ref.~\cite{Luke:1999kz} and recently formulated in a manifestly gauge invariant form in Ref.~\cite{Rothstein:2018dzq}. What we find appealing about this version of NRQCD is the clear distinction of soft and ultra-soft degrees of freedom and the use of  label-momentum notation. Both  aspects are crucial for this work. We work in the limit where the measurement is sensitive to the kinematics of the heavy quark-antiquark pair (in the quarkonium rest frame) and therefore it is critical to  separate the various infrared degrees of freedom. Using the four-vector $v^{\mu} = (1,\bmat{0})$, the four-momenta of the heavy quark, $p_{Q}^\mu$, can be decomposed as follows, 
    \begin{equation}
      \label{eq:momenta-deco}
      p^{\mu}_{Q} = m v^{\mu} + k^{\mu}\, ,
    \end{equation}
    where $k^0$ is the kinetic energy and $\bmat{k}$ is the three momenta of the heavy quark. Since the heavy quarks we  consider are on-shell i.e., $p_{Q}^2=m^2$, then in the non-relativistic limit, where the three momenta are small compared to the mass,  $ \vert \bmat{k} \vert \sim m \ve $, with $\ve \ll 1$  we have
    \begin{align}
      p^2_{Q} = m^2 + m k^0 + (k^0)^2 - \bmat{k}^2 = m^2
      \, ,
    \end{align}
    which has a solution only if $k^0 \sim m \ve^2$. Thus, the components of $k^\mu$ scale as
    \begin{equation}
      k^{\mu} \sim m (\ve^2, \ve, \ve,\ve)\, .
    \end{equation}
    In the presence of both soft and ultra-soft modes it is important to separate the small components of the  four-momenta, $k^\mu$, into soft (label, denoted $\ell^\mu$) and ultra-soft (residual, denoted $r^\mu$) parts:
    \begin{equation}
      \label{eq:deco2}
      p^{\mu}_{Q} = m v^{\mu} + \ell^{\mu}  + r^{\mu} \, ,
    \end{equation}
    where
    \begin{align}
      r^{\mu} &\sim m(\ve^2,\ve^2,\ve^2,\ve^2)\;, &  \ell^{\mu} &\sim  m(0,\ve,\ve,\ve) \,.
    \end{align}
    
    The QCD heavy quark field, $\Psi$, can then be decomposed into the vNRQCD heavy quark field, $\psi_{\bmat{\ell}}(x)$, as follows,
    \begin{equation}
      \label{eq:fieldredef}
      \Psi(x) = \sum_\bmat{\ell} e^{-i m t -i  \bmat{\ell} \cdot \bmat{x}}\, \frac{1+v\!\!\!\slash}{2}\psi_{\bmat{\ell}}(x)\,.
    \end{equation}
    The soft, $A_{\ell}^{\mu}$, and ultra-soft, $A_{us}^{\mu}$, gluon fields have momenta which scale (all four components) as soft ($\sim m\ve$) or ultra-soft ($\sim m\ve^2$), respectively.
    
    The Lagrangian of the EFT can then be written in terms of those fields in the following form~\cite{Luke:1999kz}
    \begin{equation}
      \mathcal{L}_{\text{vNRQCD}} = \sum_{\bmat{p}} \psi^{\dag}_{\bmat{p}} \lp iD^0 - \frac{(\bmat{\mathcal{P}}-i\bmat{D})^2}{2m} \rp \psi_{\bmat{p}} +\mathcal{L}^{(n>2)} + (\psi \to \chi, T \to \bar{T})  +  \mathcal{L}_s(A_q^{\mu})  +  \mathcal{L}^{V}(\psi,\chi,A_q^{\mu}) \;,
      \label{NRQCDLag}
    \end{equation}
    where $\chi$ is the heavy antiquark field and $\mathcal{L}^{(n)}$ are higher order terms and the superscript $n$ denotes the suppression in powers of $\ve$ of  $\mathcal{L}^{(n)}$ relative to the leading order Lagrangian. In Eq.~(\ref{NRQCDLag}),\
    $\mathcal{L}_s$ is the soft gluon and ghost part of the Lagrangian, and $ \mathcal{L}^{V}$ contains the potential terms which have the following generic structure,
    \begin{align}
      \text{Double soft gluon emissions:}&\;\; \sum_{\bmat{p},\bmat{p}',\ell,\ell'}\psi^{\dag}_{\bmat{p}} \lp A^{\mu}_{\ell} \; A^{\nu}_{\ell'} \rp   \psi_{\bmat{p}'} U_{\mu \nu} (\bmat{p},\bmat{p}',\ell,\ell') \;, \nn\\ 
      \text{Heavy quark-antiquark  potential:}&\;\; \sum_{\bmat{p},\bmat{p}'}\lp \psi^{\dag}_{\bmat{p}} T^A   \psi_{\bmat{p}'} \rp  \lp \chi^{\dag}_{-\bmat{p}} \bar{T}^A \chi_{-\bmat{p}'} \rp V (\bmat{p},\bmat{p}') \nn\;.
    \end{align}
    Recent developments in ref.~\cite{Makris:2019ttx}   include Glauber and Coulomb interactions of the heavy quark fields with the collinear and soft partons correspondingly. The label momentum operator~\cite{Bauer:2001ct}, $\mathcal{P}^{\mu} = (\mathcal{P}^0, \bmat{\mathcal{P}})$, is defined such that it projects only onto the label momentum space,
    \begin{align}
      \mathcal{P}^{\mu}  \psi_{\bmat{\ell}}(x) &= \ell^{\mu}  \psi_{\bmat{\ell}}(x)\;, &  \mathcal{P}^{\mu} A_{\ell}^{\nu} &= \ell^{\mu} A_{\ell}^{\nu} \;.
    \end{align}
    and the covariant derivative is: $iD^{\mu} \equiv i\partial^{\mu} - g A_{us}^{\mu}(x)$.
    
    It is important to note that in the Lagrangian there is no interaction with a heavy quark/anti-quark and a single soft gluon, since such terms will put the heavy quarks too far off-shell. At order $\alpha_s$ interactions with two soft gluons are allowed and are contained in the potential part of the Lagrangian. These interaction make it impossible to decouple the soft gluons form the heavy quarks, which is why they must appear in the same TMD matrix element. Another important observation is that the first term in the Lagrangian of Eq.~(\ref{NRQCDLag}) is the only term in which the ultrasoft gluon couples to the heavy quarks at leading order. As a result 
    the ultra-soft gluon can be decoupled from the heavy quarks using a BPS~\cite{Bauer:2001yt, Rothstein:2018dzq} field redefinition,
    \begin{equation}
      \psi_{\bmat{\ell}}^{(0)} (x) \equiv Y_{v}(x)  \psi_{\bmat{\ell}}(x)\;,
    \end{equation}
    where $ Y_{v}(x)$ is the ultra-soft gluon time-like Wilson line,
    \begin{equation}
       Y_{v}(x) = \exp \lp - ig \int_{-\infty}^{0} dt' A_{us}^{0}(t+t', \bmat{x}) \rp\;. 
    \end{equation}
    Of course, this field redefinition will simplify the leading Lagrangian but it will introduce  ultra-soft Wilson lines into $\mathcal{L}_{s}$ and the sub-leading terms $\mathcal{L}^{(n>0)}$. We will not be working with most of those terms but the next-to-leading contribution,
    \begin{equation}
      \mathcal{L}^{(1)} = - \sum_{\bmat{\ell}} \psi_{\bmat{\ell}}^{\dag}(x) \frac{-i\bmat{D} \cdot \bmat{\mathcal{P}}}{m}  \psi_{\bmat{\ell}}(x)\;,
      \end{equation}
    is of interest.  After the BPS field redefinition this term becomes,
    \begin{equation}
      \mathcal{L}^{(1)} = - g \sum_{\bmat{\ell}} {\psi_{\bmat{\ell}}^{(0)}}^{\dag}(x) \frac{\bmat{B}_{us} \cdot \bmat{\mathcal{P}}}{m}  \psi_{\bmat{\ell}}^{(0)}(x)\;.
      \end{equation}
    where $B^{\mu}_{us}(x) = - g^{-1} Y_{v}^{\dag}(x)(i D^{\mu}) Y_{v}(x) $, is the ultra-soft gluon building block.  
    
    
    \subsection{Soft collinear effective theory}
    The soft collinear effective theory has been used successfully in a wide variety of topics including TMD phenomenology. Particularly, the version of SCET that we are considering is SCET$_\text{II}$ which involves the collinear and soft degrees of freedom (in contrast to SCET$_\text{I}$ which involves the ultra-soft). The power-counting parameter of SCET is usually denoted by $\lambda$. The scaling of the relevant degrees of freedom, in light-cone coordinates $(p^+,p^-,p^{\perp})$, can be written in terms of $\lambda$ as
    \begin{align}
      \text{collinear:}&\;\;\;p_c^{\mu} \sim Q(\lambda^2,1,\lambda)_n\nn \\
      \text{soft:}&\;\;\;p_s^{\mu} \sim Q(\lambda,\lambda,\lambda)_n \,.
    \end{align}
    $Q$ is the typical hard scale and $n^{\mu} = (1,0,0,1)$ is the four-vector along which we expand in light-cone coordinates,
    \begin{equation}
      p^{\mu} =  \frac{\bar{n}^{\mu}}{2} p^{-} + \frac{n^{\mu}}{2} p^{+} + p_{\perp}^{\mu}\,,
    \end{equation}
    where $p^{-} = n \cdot p$, $p^{+} = \bar{n} \cdot p$, and $\bar{n}^{\mu} = (1,0,0,-1)$ such that $n^2 = \bar{n}^2 =0$ and $n \cdot \bar{n} = 2$. We separate the label and residual components of momenta as follows:
    \begin{align}
      p^{\mu}_c& = \ell^{\mu} + r^{\mu}\;& \text{where} & & \ell^{\mu}&\sim Q(0, 1, \lambda)\;,& \text{and}& & r^{\mu}&\sim Q(\lambda^2, \lambda^2, \lambda^2)\,.
    \end{align}
    The EFT collinear quark fields $\xi_{n,\ell}$ are then defined as 
    \begin{equation}
      \xi_{n,\ell}(x) = P_n \;q_{n,\ell}(x)\;,
      \end{equation}
    where $P_n = \slashed{n} \slashed{\bar{n}}/4$ and $q_{n,\ell}$ is the QCD label momentum partitioned field
    \begin{equation}
      \Psi(x) = \sum_{\ell \neq 0} e^{-i \ell \cdot x} q_{n,\ell}(x)\;.
    \end{equation}
    The collinear ($A_{n,\ell}^{\mu}$) and soft ($A_{s,\ell}^{\mu}$) gluons are defined similarly:
    \begin{equation}
      A^{\mu}(x) = \sum_{\ell \neq 0} e^{-i \ell \cdot x} A_{\ell}^{\mu}(x)\;,
    \end{equation}
    where $ A^{\mu}(x)$ are the full theory bosons and $ A^{\mu}_{\ell} = A^{\mu}_{n,\ell}  + A^{\mu}_{s,\ell}$. The SCET Lagrangian can be found in Refs.~\cite{Bauer:2000yr,Bauer:2001ct,Bauer:2001yt}.
    
    Invariance under collinear, soft, and ultra-soft gauge transformations requires that particular combinations of fields appear (rather than $\xi_n$, $A_n$ , and $A_{s}$):
    \begin{align}
      \xi_{n,\ell}(x)  &\to   S_{n}(x) \chi_{n,\ell}(x) \equiv S_{n}(x) W_{n}^{\dag}(x) \xi_{n,\ell}(x) \;,\nn\\
      A_{n,\ell}^{\mu}(x) &\to \mathcal{S}^{ab}_n B_{n,\perp}^{b \mu}(x) \equiv -\frac{1}{g}  \mathcal{S}^{ab}_n \text{Tr} \lp T^{b} W_n^{\dag}(x)(\mathcal{P}_{\perp}^{\mu} - g A_{n,\perp}^{\mu}) W_{n}(x) \rp  \;,
    \end{align}
    where $W_n$ and $S_n$ are the collinear and soft  Wilson lines respectively,  defined as:
    \begin{align}
      W_n(x) &= \text P \exp \lp -i g \int_{-\infty}^{0} ds \bar{n} \cdot A_n (x + \bar{n} s)  \rp  \;, \nn \\
      S_n(x) &= \text P \exp \lp -i g \int_{-\infty}^{0} ds n \cdot A_s (x + n s)  \rp  \;.
      \end{align}
    Note that collinear fields cannot directly interact with soft gluons. As a result the leading SCET Lagrangian does not include interactions of soft and collinear gluons or quarks. On the other hand in full QCD it is possible to have such interactions through Glauber gluon exchanges\footnote{For Glauber gluon exchanges in SCET see Refs.~\cite{Idilbi:2008vm,DEramo:2010wup,Ovanesyan:2011xy,Ovanesyan:2012fr,Benzke:2012sz,Rothstein:2016bsq}.}. These interactions can be included in the effective theory systematically. This part of the Lagrangian was extensively studied recently in Ref.~\cite{Rothstein:2016bsq}. In this work we focus on processes that are not sensitive to Glauber gluon exchanges such us lepton annihilation to di-hadrons or semi inclusive DIS.
    
    In this section we have introduced two power counting parameters, one for vNRQCD ($\ve$), and one for SCET ($\lambda$). Its is natural to ask about the hierarchy between these small parameters. We are interested in two distinct cases: 
    \begin{itemize}
    \item $\ve \sim \lambda$: The two EFTs overlap in the region of soft dynamics. Particularly the soft gluon of one theory is the same as in the other. In this case the two EFTs merge into a single one containing all degrees of freedom. We will refer to this EFT as SCET$_{\text{Q}}$.
    \item$\ve \ll \lambda$: There is no overlap and thus we need to work with both EFTs. Factorization theorems for this region can be achieved by matching from SCET$_{\text{Q}}$ onto vNRQCD.
    \end{itemize}
    For the rest of this paper we will be assuming $\ve \sim \lambda$ and use SCET$_{\text Q}$.

    
    \section{Operators, Matching, Reparametrization Invariance, and Gauge Invariance}
    \label{sec:ops}
    
    In this section we consider the operators that appear in the hard sector of the Lagrangian by performing tree level matching of the QCD diagrams with an arbitrary number of soft gluon emissions from the heavy quark and antiquark lines.
    We then study the simple but non-trivial example of $q\bar{q} \to Q \bar{Q}$ $+$ gluons. We match these QCD diagrams onto NRQCD operators with soft Wilson lines and the $Q\bar{Q}$ in relative S- and P-waves. We apply reparameterization invariance (RPI) of the EFT to show that two of the operators appearing in the P-wave channel are related to the S-wave operator by RPI transformations. Finally, we discuss gauge invariance in the effective theory and show that our results respect all gauge symmetries of the EFT.

    
    \subsection{Diagrammatic analysis}
\begin{figure}[!ht]
      \centerline{\includegraphics[width = 0.92 \textwidth]{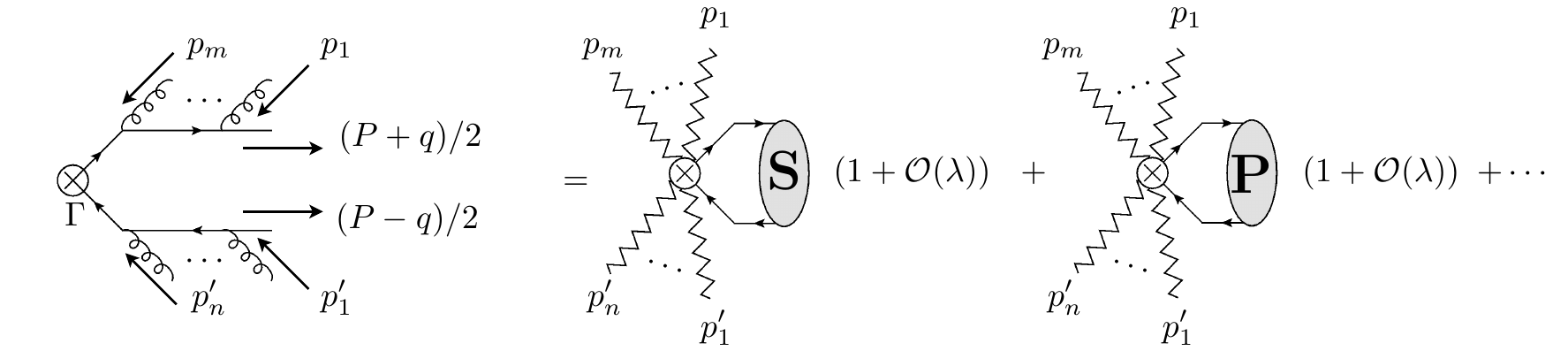}}
      \caption{A diagrammatic illustration of the matching procedure from full QCD onto EFT operators.}
      \label{fig:matching-tree}
    \end{figure}

     To keep the discussion  as generic as possible, we consider the QCD diagram on the left-hand side of Fig.~\ref{fig:matching-tree}, where the vertex with the $\otimes$ symbol corresponds to an arbitrary color and spin structure, denoted in the equations below by $\Gamma$. We will evaluate a diagram with $n$ gluons attached to the antiquark line and $m$ to the quark line, then sum over $n$ and $m$. The diagram in the full theory is given by
    \begin{multline}
      d_{\Gamma}(m,n) =  \bar{u}(p_Q) \Vq{1} D(p_Q+p_1) \Vq{2} \cdots \Vq{m} D(p_Q+p_t(m)) \\
      \times \Gamma(p_t(m),p_Q,p'_t(n),p_{\bar{Q}})   D(-p_{\bar{Q}}-p'_t(n)) \Vbq{n} \cdots \Vbq{2} \\
      \times D(-p_{\bar{Q}}-p'_1) \Vbq{1} v(p_{\bar{Q}})  \times A_{\nu_1}^{b_1}(p'_1) A_{\nu_2}^{b_2}(p'_2)\cdots A_{\mu_1}^{a_1}(p_1)  \;,
    \end{multline}
    where
    \begin{equation}
      p^{\mu}_t(k) \equiv \sum_{\ell = 1}^{k} p_\ell^{\mu} \;,
    \end{equation}
    and similarly for the primed gluon momenta $p^{(\prime)}_i$ labelled in Fig.~\ref{fig:matching-tree}.  The heavy quark and anti-quark momenta are
    \begin{equation}
    p_Q = (P +q)/2\,, \qquad p_{\bar{Q}}=(P -q)/2 \,,
    \end{equation}
    where $P$
    is the total momentum of the heavy quark and antiquark and $q$ is their relative momentum. 
    We will calculate these diagrams for S-wave and P-wave $Q\bar{Q}$ states. For the S-waves we can set $q$ to zero and P-waves are extracted by expanding to linear order in $q$. Each coefficient of the $q$-expansion is further expanded in the small parameter $\lambda$ and here we consider only the leading non-trivial terms in the $\lambda$-expansion. The leading term in the  $q$-expansion, $\mathcal{O}(q^{0})$, gives the overlap with the S-wave states where the  $\mathcal{O}(q^{1})$ terms give the overlap with the P-wave states
    \begin{equation}
      d_{\Gamma}(m,n) = d^{(0)}_{\Gamma}(m,n)(1+\mathcal{O}(\lambda))+ d^{(1)}_{\Gamma}(m,n)(1+\mathcal{O}(\lambda))+\cdots
    \end{equation}
    with $d_{\Gamma}(m,n)$ the result of evaluating the diagram. Supercripts denote the order in $q$ of the expansion, and the ellipsis corresponds to terms of order $\mathcal{O}(q^2)$ or higher. In the $q$-expansion we need to consider expanding all elements that could depend on the momenta of the heavy quark and antiquark, $p_Q$ and $p_{\bar{Q}}$ respectively. Those are the spinors, $u$ and $v$, the propagators, $D$, and the vertex, $\Gamma$.  The spinor and propagator expansions in $q$ and $\lambda$ are given in Appendix~\ref{app:defs}. We use the following notation for the vertex expansion,
    \begin{equation}
      \Gamma(p_t(m),p_Q,p'_i(n),p_{\bar{Q}}) = \Gamma^{(0)}(p_t(m),p'_t(n))(1+\mathcal{O}(\lambda))+\bmat{q} \cdot \bmat{ \Gamma}^{(1)}(p_t(m),p'_t(n))(1+\mathcal{O}(\lambda)) + \cdots 
    \end{equation}
    
    First we consider the projection onto S-wave states
    \begin{eqnarray}
      \label{eq:d01}
      d^{(0)}_{\Gamma}(m,n) &=&  \uo \Vq{1} \Do{1}\Vq{2} \Do{2} \cdots \Vq{m} \Do{m} \nonumber \\
    &\times&  \Gamma^{(0)} (p_t(m),p'_t(n))   \Dbo{n}\Vbq{n} \cdots \Vbq{2}\Dbo{1} \Vbq{1} \vo
    \nonumber  \\ 
    &\times& A_{1', \nu_1}^{b_1} A_{2',\nu_2}^{b_2} \cdots A_{1,\mu_1}^{a_1} \;,
    \end{eqnarray}
    where we use the label-momentum notation also used in the context of vNRQCD:
    \begin{align}
      A_{i,\mu}^{a} & \equiv A_{\mu}^{a} (p_i)\;,& A_{i',\nu}^{b} & \equiv A_{\nu}^{b} (p'_i) \;.
    \end{align}
    Further we use 
    \begin{align}
      \label{eq:analysis-1}
      \gamma^{\mu_i}(1+\slashed{v}) &= (1-\slashed{v})\gamma^{\mu_i} + 2 v^{\mu_i} \;,&
      (1-\slashed{v}) \gamma^{\nu_i} &= \gamma^{\nu_i} (1+\slashed{v}) - 2 v^{\nu_i} \;,
    \end{align}
    along with Eq.~(\ref{eq:EoM-RF}) to rewrite $d^{(0)}_{\Gamma}$ as follows
    \begin{equation}
      d^{(0)}_{\Gamma}(m,n) =  \uo \lb (-g)^{m}\prod_{s=1}^{m} \frac{A^{0}_s}{p_{t}^{0}(s)} \rb
      \Gamma^{(0)} (p_t(m),p'_t(n)) \lb g^{n} \prod_{s=1}^{n} \frac{A^{0}_{(n+1-s)'}}{p_{t}^{'0}(s)} \rb \vo \;.
    \end{equation}
    Since gluons are bosons we can sum over all possible permutations within each product and normalize with the number of permutations. Furthermore since we are considering an arbitrary number of soft gluons we need to sum over all possible $n$ and $m$ which gives
    \begin{equation}
      \label{eq:dG0}
      d^{(0)}_\Gamma  = \uo S_{v}^{\dag} \;  \Gamma^{(0)} \; S_v \vo \;.
    \end{equation}
    We have omitted the arguments of $\Gamma^{(0)}$ which could depend on the total momenta of soft gluons attached to the heavy quark and antiquark lines. Here  $S_v$ is given by
    \begin{equation}
      S_v =\sum_{n}  \sum_{\text{perms}} \frac{g^n}{n!}\prod_{s=1}^{n} \lb \frac{A^{0}_{n+1-s}}{p_{t}^{0}(s)} \rb \;,
    \end{equation}
    which is the momentum space expansion of the time-like Wilson line of soft gluons, $S_v(x,-\infty)$, defined as,
    \begin{equation}
      S_v(x,-\infty) = \text{P} \lb \exp \lp -i g \int_{-\infty}^{0} d\tau \; v\cdot A_{soft} (x^{\mu} + v^{\mu} \tau )\rp  \rb\;.
    \end{equation}
    
    Eq.~(\ref{eq:dG0}) is our final result for the generic treatment of soft gluon emissions from $S$-wave quarkonia. For the specific mechanisms of interest, e.g., $\tensor*[^3]{S}{^{(8)}_1}, \; \tensor*[^1]{S}{^{(1)}_0},\ldots $, to obtain the exact operator form we need to specify the hard process and then evaluate the vertex $\Gamma$ and its expansion in $q$ and $\lambda$. We will do a particular example in the next subsection.

    Next we perform a tree-level, diagrammatic analysis for the case where the $Q\bar{Q}$ are in a P-wave.  To achieve this, we need to expand in the relative momenta of the heavy quark-antiquark pair  and keep only the linear terms in $q$. As mentioned earlier there are three sources that will contribute: the spinors $u$ and $v$, the propagators $D$, and the vertex $\Gamma$. The procedure for the expansion of the vertex is identical to the analysis following Eq.~(\ref{eq:d01})
    \begin{equation}
      d^{(1)}_{V}  = \uo S_{v}^{\dag} \;  \bmat{q} \cdot \bmat{\Gamma}^{(1)} \; S_v \vo \;.
    \end{equation}
    From the expansion of the propagators, $D(p_Q+p)$, we have two contributions, the one proportional to $p^{i}$ and the other to $\gamma^{i}$, see Eqs.~(\ref{eq:prob}) and (\ref{eq:anti-prob}). We consider the former independently and the latter along with the contributions from the spinor expansions; we refer to these terms as $d^{(1)}_{D}$ and $d^{(1)}_{\gamma}$ respectively. The term $d_{D}^{(1)}$ can be written as the sum of contributions where the expanded propagators lie along in the quark line $d_{D}^{(1)}(\text{quark})$ or antiquark line $d_{D}^{(1)}(\text{antiquark})$
    \begin{equation}
      \label{eq:sumD}
      d_{D}^{(1)} = d_{D}^{(1)}(\text{quark}) + d_{D}^{(1)}(\text{antiquark}) \;.
    \end{equation}
    Here we demonstrate the calculation for $d_{D}^{(1)}(\text{quarks})$ only. For all other terms refer to Appendix~\ref{app:details}. We begin by expanding the $i$-th propagator along the quark line at leading order in $\bmat{q}$. Then using Eq.~(\ref{eq:analysis-1}) and summing over $i$ we have 
    \begin{multline}
      d_{D}^{(1)}(m,n;\text{quark}) = \uo  \sum_{i=1}^{m}  \lb (-g)^{i} \bmat{q}\cdot \bmat{p}_t(i) \prod_{s=1}^{i} \frac{A^{0}_s}{ p_{t}^{0}(s)} \rb \frac{1}{2 m p_{t}^{0}(i)} \lb (-g)^{m-i} \prod_{s=i+1}^{m} \frac{A^{0}_s}{p_t^0(s)} \rb  \\
      \Gamma^{(0)} (p_t(m),p'_t(n)) \lb g^{n} \prod_{s=1}^{n} \frac{A^{0}_{(n+1-s)'}}{ p_{t}^{'0}(s)} \rb \vo \;.
    \end{multline}
    Using Eq.~(\ref{eq:prod-exp}) we get
    \begin{multline}
      d_{D}^{(1)}(m,n;\text{quark}) = \uo  \sum_{i=1}^{m} \sum_{\rho = i}^{m}  \frac{1}{2m p_{t}^0(\rho)} \lb (-g)^{i} \bmat{q}\cdot \bmat{p}_t(i) \prod_{s=1}^{i} \frac{A^{0}_s}{ p_{t}^{0}(s)} \rb  \\ \lb  g^{\rho-i}  \prod_{s=i+1}^{\rho}\frac{A^{0}_s}{\sum_{\ell=i+1}^{s} p^0_{\rho+i+1-\ell}}  \rb  \lb (-g)^{m-\rho} \prod_{s=\rho+1}^{m} \frac{A^{0}_s}{\sum_{\ell=\rho+1}^{s} p^0_{\ell}}  \rb 
      \\ \Gamma^{(0)} (p_t(m),p'_t(n)) \lb g^{n} \prod_{s=1}^{n} \frac{A^{0}_{(n+1-s)'}}{ p_{t}^{'0}(s)} \rb \vo \;.
    \end{multline}
    Now summing over all permutations within each product, using the bosonic nature of gluons, and considering all values of $m$ and $n$ we have
    \begin{multline}
      d_{D}^{(1)}(\text{quark}) = \uo \lb \frac{1}{2m v\cdot \mathcal{P}}  \lb \bmat{q} \cdot \bmat{\mathcal{P}}S_{v}^{\dag}\rb S_v  \rb S_v^{\dag}\Gamma^{(0)} S_v \vo \\ = - \uo \lb \frac{1}{2m v\cdot \mathcal{P}}  S_{v}^{\dag} \lb \bmat{q} \cdot \bmat{\mathcal{P}} S_v\rb  \rb S_v^{\dag}\Gamma^{(0)} S_v \vo \;.
    \end{multline}
    Here we used the notation where the label momentum operator acts upon fields to the right which are enclosed within the same square brackets (including sub-brackets). A similar analysis for the antiquark yields
    \begin{equation}
      \label{eq:dDanti}
      d_{D}^{(1)}(\text{antiquark}) = - \uo  S_v^{\dag}\Gamma^{(0)} S_v \lb \frac{1}{2m v\cdot \mathcal{P}}  S_{v}^{\dag} \lb \bmat{q} \cdot \bmat{\mathcal{P}} S_v\rb  \rb \vo \;.
    \end{equation} 
    Therefore, the sum in Eq.~(\ref{eq:sumD}) gives an anti-commutator in the following form
    \begin{equation}
      d_{D}^{(1)} =  \uo \lbc  S_v^{\dag}\Gamma^{(0)} S_v, \lb \frac{-1}{2m v\cdot \mathcal{P}}  S_{v}^{\dag} \lb \bmat{q} \cdot \bmat{\mathcal{P}} S_v\rb  \rb \rbc \vo \;.
    \end{equation}
    Similarly from Eqs.~(\ref{eq:dg}) and (\ref{eq:dgb}) we have
    \begin{equation}
      d_{\gamma}^{(1)} =  \uo \lbc  S_v^{\dag}\Gamma^{(0)} S_v, \lb \frac{1}{2m v\cdot \mathcal{P}}  S_{v}^{\dag}  \bmat{q} \cdot \bmat{A} S_v  \rb - \bmat{\gamma}\cdot \bmat{q} \rbc \vo  \;.
    \end{equation}
    The total contribution to the P-wave $Q\bar{Q}$ states is given by
    \begin{align}
      \label{eq:d1G}
      d_{\Gamma}^{(1)} = d_{V}^{(1)} + d_{D}^{(1)}+d_{\gamma}^{(1)} = \frac{g}{2m} &\uo \lbc  S_v^{\dag}\Gamma^{(0)} S_v, \lb \frac{1}{ v\cdot \mathcal{P}} 
      \bmat{q} \cdot \bmat{B}_s \rb \rbc \vo \\\nn
      + &\uo S_v^{\dag}  \;\bmat{q} \cdot\lp \bmat{\Gamma}^{(1)}-\frac{1}{4m}\lbc\Gamma^{(0)},\bmat{\gamma}\rbc \rp S_v \vo \;,
    \end{align}
    where the building block, $B_s^{\mu}$, is defined as follows
    \begin{equation}
      B^{\mu}_s = - \frac{1}{g} S_v^{\dag} \lb (\mathcal{P}^{\mu}-g A^{\mu}) S_v \rb
    \end{equation}
    and was introduced and studied in the context of vNRQCD in Ref~\cite{Rothstein:2018dzq}. We demonstrate its importance for the gauge invariance of the EFT operators later in this section.  
    In the next subsection, we show how the form of the terms involving $\Gamma^{(0)}$ can be understood using reparametrization invariance (RPI).

    Eqs.~(\ref{eq:dG0}) and (\ref{eq:d1G}) are the main results of this section. Using these equations we can now easily identify the operators that do have nonvanishing matching with the full theory at any fixed order in the $\alpha_s$ expansion. Here we wish to make a few important observations. In Eq.~(\ref{eq:d1G}) there are three distinct contributions, each proportional to one of: $\bmat{B}_s$, $\bmat{\Gamma}^{(1)}$, and $\bmat{\gamma}$. We refer to those as Type I, Type II, and Type III contributions. For a given $\Gamma^{(0)}$
     Type I and Type III will never match onto the same operator due to the presence of $B_s$. They will also not contribute to $Q\bar{Q}$ with the same quantum numbers because the two operators have different spin angular momentum. If $\Gamma^{(0)}$ is a color-singlet then Type I contributions will be color-octets and Type II will be color singlets. Finally, as will see in a later subsection Type I and Type III are related to the S-wave operators
     responsible for Eq.~(\ref{eq:dG0}) so their matching will be the same to all orders in perturbation theory.  In Sect. 4 we will see that this has important implications for the IR finiteness of the EFT matching coefficients.
     
     It should be noted that in the current formulation of the SCETQ operators we cannot distinguish between the past and future going Wilson lines. Such a distinction is important when discussing universality of soft functions and properties of rapidity divergences. Although this analysis can be done in SCETQ we do not pursue this in this work. Such an analysis will be useful for discussing cancellation of rapidity divergences to all orders in perturbation theory, which we leave for future work. In addition, in our results no transverse gauge link is present, contrary to other soft functions such as those in Drell- Yan and electron-positron annihilation. This is a consequence of working in Feynman gauge (or any other covariant gauge). For full gauge invariance including axial gauges one should include the transverse gauge link.
    \subsection{A simple example}
    \label{simple}
    
\begin{figure}[!ht]
      \centerline{\includegraphics[width = 0.92 \textwidth]{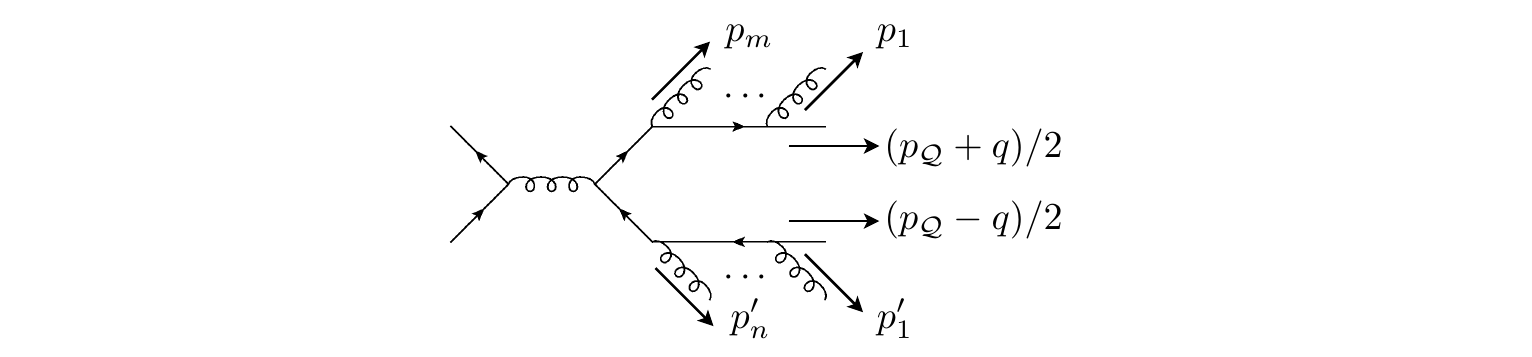}}
      \caption{QCD diagram for an arbitrary number of soft gluon emissions from the heavy quark-antiquark pair for the process $q\bar{q} \to Q\bar{Q}[n]$.}
      \label{fig:qq}
    \end{figure}

    Now we demonstrate how the analysis and results  of the previous section can be used in order to perform the matching for a particular process. As an example we study $Q \bar{Q}$ pair production(decay) from(to) a light quark-antiquark pair. The relevant QCD diagram is shown in Fig.~\ref{fig:qq}. The two on-shell light quarks have momenta with collinear scaling:
    \begin{align}
      p_n^{\mu} &\sim 2m(\lambda^{2},1,\lambda )  & p_{\bn}^{\mu} &\sim 2m(1,\lambda^{2},\lambda )  \;.
    \end{align}
    This indicates that the collinear approximation is appropriate, and many results from SCET can be recycled. The full theory diagram gives
    \begin{multline}
      d(q\bar{q} \to Q\bar{Q}) = \bar{v}(p_{\bn}) \lp  \gamma_{\rho} T^{c} \rp u(p_{n}) \frac{i g^2}{(P + p_{X})^{2}} \\
      \bar{u}(p_Q) \Vq{1} D(p_Q+p_1) \Vq{2} \cdots \Vq{m} D(p_Q+p_t(m))
      \; \gamma^{\rho} T^{c} \\ D(-p_{\bar{Q}}-p'_t(n)) \Vbq{n} \cdots \Vbq{2}   D(-p_{\bar{Q}}-p'_1) \Vbq{1} v(p_{\bar{Q}}) 
      \\\times A_{\nu_1}^{b_1}(p'_1) A_{\nu_2}^{b_2}(p'_2)\cdots A_{\mu_1}^{a_1}(p_1)  \;,
    \end{multline}
    from which we can directly read off the vertex $\Gamma = \gamma^{\rho} T^{c}  i g^2/s $, where $s=(P + p_{X})^{2}$ is the center of mass energy squared. We note that this expression is independent of $q^{\mu}$ and therefore $\Gamma^{(i >0)} =0 $. Expanding in $\lambda$ we find
    \begin{align}
      \label{eq:Gqq}
      \Gamma^{(0)} &= \gamma^{\rho} T^{c}  \frac{i g^2}{4 m^{2}}\;, & \bmat{\Gamma}^{(1)} &= 0 \;. 
    \end{align}
    Using this result along with Eq.~(\ref{eq:dG0}) we can easily obtain for the overlap with S-wave states:
    \begin{equation}
      d^{(0)}(q\bar{q} \to Q\bar{Q}) =i \alpha_s \frac{\pi}{m^{2}} \; \bar{v}_{\bn}(p_{\bn}) \lp  \gamma^{i} T^{c} \rp u_{n}(p_{n}) (\sqrt{2m} \xi^{\dag})\sigma^{i} S_{v}^{\dag} T^{c} S_{v} (\sqrt{2m} \eta^{\dag})\,,
    \end{equation}
    where we also expanded the collinear spinors in $\lambda$. This result is reproduced by the expectation value of a four fermion operator
    \begin{equation}
      d^{(0)}(q\bar{q} \to Q\bar{Q}) = i \alpha_s \frac{\pi}{m_{Q}^{2}} \; \Langle Q(p_Q) \bar{Q}(p_{\bar{Q}}) \Bvert \psi^{\dag}\sigma^{i} S_{v}^{\dag} T^{c} S_{v}\chi (0)\; \bar{\xi}_{\bn} \gamma^{i} T^{c} \xi_{n} (0)\; \Bvert q(p_n) \bar{q}(p_{\bn}) \Rangle\;\;\Bvert_{\text{tree-level}}\,.
    \end{equation}
    Until now we have not yet included soft or collinear gluon emissions from the light quark lines. This task was already tackled in SCET and here we simply use what is already well known. The collinear gluons along the $n$ direction can be organized in a  light-like Wilson line, $W^{\dag}_n$, acting on the collinear field $\xi_n$. The product is usually denoted with  
    \begin{equation}
      \chi_n(x) \equiv W_n^{\dag} \xi_n(x)\,.
    \end{equation}
    For more details on the collinear Wilson lines refer to Sect.~\ref{sec:efts} and references therein. The soft gluon attachments  can also be organized into a light-like Wilson line, $S_{n}$. The resulting operator is 
    \begin{equation}
       \mathcal{S}_{v}^{cd}\; \lp \psi^{\dag}\sigma^{i}  T^{d} \chi \rp \times \lp  \bar{\chi}_{\bn} \gamma^{i}  S^{\dag}_{\bn} T^{c} S_{n} \chi_{n} \rp \,,
    \end{equation}
    where we used
    \begin{equation}
      \mathcal{S}_{v}^{cd} T^{d} =  S_{v}^{\dag} T^{c} S_{v}\,.
    \end{equation}
    The tree-level matching coefficient for this operator is 
    \begin{equation}
      C_{q\bar{q}} ( \tensor*[^3]{S}{^{[8]}_1} ) = \alpha_s \frac{\pi}{m^{2}} \,.
    \end{equation}
    We note that the only term that contributes is the projection of  the heavy quark-antiquark pair onto a $\Q{3}{S}{1}{8}$ configuration. This is to  be expected from fixed order calculations in NRCQD, where the same mechanism is the only $\wave{S}$-wave contribution that has leading logarithmic growth in the small traverse momentum limit at NLO.
    
    We now repeat  the same procedure for the projection onto spin triplet $\wave{P}$-wave quarkonium states. Using Eqs.~(\ref{eq:d1G}) and (\ref{eq:Gqq}) we get
    \begin{multline}
      d^{(1)}(q\bar{q} \to Q\bar{Q}) =i  C_{q\bar{q}} ( \tensor*[^3]{S}{^{[8]}_1} ) \;\frac{g}{2 m} \; \mathcal{S}_{v}^{cd}\bar{v}_{\bn}(p_{\bn}) \lp  \gamma^{i} T^{c} \rp \\
      \times u_{n}(p_{n}) (\sqrt{2m} \xi^{\dag})\sigma^{i} \; \lbc T^{d}, T^{e}  \rbc \;(\sqrt{2m} \eta^{\dag})\lb \frac{ \bmat{q} \cdot \bmat{B}^{e}_s}{ v\cdot \mathcal{P}}   \rb \,,
    \end{multline}
    where we used $B^{\mu}_s = B_s^{e,\mu} T^{e}$~\cite{Rothstein:2018dzq}. The anti-commutator $\{ T^d, T^e \} = d^{def} T^f +\delta ^{de}/N_c $ will generate both color singlet and  octet projections of the heavy quark-antiquark pair. For this reason we break the contributions to this diagram into ones coming from two different operators: 
    \begin{equation}
       \sqrt{\frac{2}{N_c}} \times \mathcal{S}_{v}^{cd}\; \lb \frac{g \;B^{d,j}_s}{m\; v\cdot \mathcal{P}}   \rb \lb   \psi^{\dag}\frac{ \sigma^{i}\; \lrP^{j}}{2 \sqrt{2 N_c}} \; \chi \rb \times \lp  \bar{\chi}_{\bn} \gamma^{i}  S^{\dag}_{\bn} T^{c} S_{n}  \chi_{n} \rp
    \end{equation}
    and 
    \begin{equation}
       d^{def} \mathcal{S}_{v}^{cd}\; \lb \frac{g \; B^{e,j}_s}{m\; v\cdot \mathcal{P}}   \rb  \lb \psi^{\dag}\frac{ \sigma^{i}\; \lrP^{j} }{2 } \; T^{f} \chi \rb \times \lp  \bar{\chi}_{\bn} \gamma^{i}  S^{\dag}_{\bn} T^{c} S_{n}  \chi_{n} \rp \,,
    \end{equation}
    with
    \begin{equation}
      C_{q\bar{q}} (\tensor*[^3]{P}{^{[1]}_J} ) = C_{q\bar{q}} (\tensor*[^3]{P}{^{[8]}_J} ) = C_{q\bar{q}} ( \tensor*[^3]{S}{^{[8]}_1} ) \;. 
    \end{equation}
    The left-right label momentum operator is defined such that it acts on the left first, $[\psi^{\dag}\; \lrP^{\mu} \chi] =     [ \mathcal{P}^{\mu}\psi]^{\dag}  \chi-\psi^{\dag} [ \mathcal{P}^{\mu} \chi ]$,\footnote{Note this is in contrast with past definitions where the action of the derivative is  first on the right, see Ref.~\cite{Bodwin:1994jh}. With this definition we avoid having a minus sign in-front of the P-wave operator.} and as always the square brackets denote the range of action for the label momentum operator. Therefore, the leading term contributing to P-wave states can be written in terms of these two operators:
    \begin{eqnarray}
      \label{eq:tree_d1G}
      d^{(1)}(q\bar{q} \to Q\bar{Q})&=&   \\
     &&\hspace{-15ex} i  C_{q\bar{q}} ( \tensor*[^3]{S}{^{[8]}_1} ) \;\bigg( \Langle Q(p_Q) \bar{Q}(p_{\bar{Q}}) \Bvert 
      \sqrt{\frac{2}{N_c}} \times \mathcal{S}_{v}^{cd}\; \lb \frac{g \;B^{d,j}_s}{m\; v\cdot \mathcal{P}}   \rb \lb   \psi^{\dag}\frac{ \sigma^{i}\; \lrP^{j}}{2 \sqrt{2 N_c}} \; \chi \rb \times \lp  \bar{\chi}_{\bn} \gamma^{i}  S^{\dag}_{\bn} T^{c} S_{n}  \chi_{n} \rp
      \Bvert q(p_n) \bar{q}(p_{\bn}) \Rangle
      \nonumber \\
      && \hspace{-15ex} + 
     \Langle Q(p_Q) \bar{Q}(p_{\bar{Q}}) \Bvert  
      d^{def} \mathcal{S}_{v}^{cd}\; \lb \frac{g \; B^{e,j}_s}{m\; v\cdot \mathcal{P}}   \rb  \lb \psi^{\dag}\frac{ \sigma^{i}\; \lrP^{j} }{2 } \; T^{f} \chi \rb \times \lp  \bar{\chi}_{\bn} \gamma^{i}  S^{\dag}_{\bn} T^{c} S_{n}  \chi_{n} \rp 
      \Bvert q(p_n) \bar{q}(p_{\bn}) \Rangle\;\;\bigg)_{\text{tree-level}}\,.
      \nonumber
    \end{eqnarray}
    
    This gives the overlap with color singlet and octet $\tensor*[^3]{P}{_J}$ states which are the only ones which contribute in the light quark pair annihilation channel. Again, this is consistent with what is observed in fixed order NRQCD calculations. Note that Eq.~(\ref{eq:tree_d1G}) for the case of $\Q{3}{P}{J}{1}$ holds beyond leading order as discussed below Eq.~(\ref{eq:d1G}).
    \subsection{Reparametrization Invariance}
    
    In this subsection we explain how reparametrization invariance (RPI) \cite{Luke:1992cs,Manohar:2002fd} relates the NRQCD operators whose matrix elements  give
    the Type I and Type III terms in Eq.~(\ref{eq:d1G}) to the S-wave operators whose matrix elements give Eq.~(\ref{eq:dG0}). The operator which gives rise to Eq.~(\ref{eq:dG0}) is:
    \begin{equation}
      \label{eq:op_dG0}
      \psi^\dagger_{\bf p} S_{v}^{\dag} \;  \Gamma^{(0)} \; S_v \chi_{\bf p} \;.
    \end{equation}
    In order to match onto a diagram with the $Q\bar{Q}$ in the S-wave we took the four-velocity of the heavy quark and anti-quark to be $v^\mu=(1,\bf{0})$. In order to match onto $\wave{P}$-waves we need to give the heavy quark and anti-quark residual momentum $\pm q^\mu/2 = (0,\pm {\bf q}/2)$. The four velocities of the heavy quark and antiquark are 
    \begin{eqnarray}
      v_\pm^\mu &=& \left(\sqrt{1+\frac{{\bf q}^2}{4m^2}}, \pm \frac{{\bf q}}{2m} \right) \nn \\
      &=& v^\mu+\frac{q^\mu}{2m} + O\left(\frac{q^2}{m^2}\right) \, ,
    \end{eqnarray}
    so that $v_\pm^2=1$. The idea of this section is to match onto an operator like that in Eq.~(\ref{eq:op_dG0}), but instead of using $v^\mu=(1,\bf{0})$ for the Wilson lines and the heavy quark and antiquark fields label momentum, we use $v^\mu_+(v^\mu_-)$ for the heavy quark (antiquark) field and then expand the resulting operator in powers of ${\bf q}$. The NRQCD fields obey the constraint $\slashed{v}\psi_{\bf p} = \psi_{\bf p}$ (see Eq.~(\ref{eq:fieldredef})), this constraint is implicit in the notation. We introduce NRQCD fields that instead obey the constraints
    \begin{eqnarray}
        \label{eq:pm}
        \slashed{v}_\pm\psi_{\bf p,\pm} &=& \psi_{\bf p,\pm} \nn \\ \slashed{v}_\pm \chi_{\bf p,\pm} &=& -\chi_{\bf p,\pm} \, ,
    \end{eqnarray}
    as well as Wilson lines, $S_{v_\pm}$, along the $v^\mu_\pm$ directions. Matching in the $\wave{S}$-wave is a trivial generalization of the $\wave{S}$-wave matching performed earlier, the operator is 
    \begin{equation}
      \label{eq:op_dG0prime}
      \psi^\dagger_{{\bf p},+} S_{v_+}^{\dag} \;  \Gamma^{(0)} \; S_{v_-} \chi_{{\bf p},-} \;.
    \end{equation}
    Now insert $v^{\mu}_{\pm} = v^\mu \pm q^\mu/(2m)$ into the first equation in Eq.~(\ref{eq:pm}), write $\psi_{{\bf p},\pm} = \psi_{\bf p} + \delta \psi_{\bf p,\pm}$ then expanding to $O(q)$ and solving for $ \delta \psi_{\bf p,\pm}$,
    one obtains
    \begin{eqnarray}
    \label{eq:psichipm}
     \psi_{{\bf p},\pm} &=& \psi_{\bf p} \pm \frac{\slashed{q}}{4m} \psi_{\bf p} \nn\\
     \chi_{{\bf p},\pm} &=& \chi_{\bf p} \mp \frac{\slashed{q}}{4m} \chi_{\bf p} \, .
     \end{eqnarray}
     Similarly, writing $S_{v,+} = S_v +\delta S_v$ and using the equation of motion, we have
     \begin{eqnarray}
     \left(v + \frac{q}{2m}\right)\cdot ({\cal P} - g A) (S_v+\delta S_v) =0 \,.
     \end{eqnarray}
    Expanding to $O(q)$ and solving for $\delta S_v$ we find
    \begin{eqnarray}
    \label{eq:Svp}
    S_{v,+} &=& S_v+\frac{1}{2 m}\frac{1}{v\cdot({\cal P} - g A)} q\cdot({\cal P} - g A) S_v \nn \\
    &=& S_v+\frac{1}{2 m} S_v\frac{1}{v\cdot {\cal P}} S_v^\dagger q\cdot({\cal P} - g A) S_v \nn \\
    &=& S_v-\frac{g}{2 m} S_v\frac{1}{v\cdot {\cal P}} q\cdot B \,.
    \end{eqnarray}
    Inserting  Eqs.~(\ref{eq:psichipm}) and 
    Eqs.~(\ref{eq:Svp}) into Eq.~(\ref{eq:op_dG0prime}) we obtain Eq.~(\ref{eq:op_dG0}) plus operators whose matrix elements give the Type I and Type III terms in Eq.~(\ref{eq:d1G}).
    
    The RPI analysis explains the structure of the Type I and III terms in Eq.~(\ref{eq:d1G}), which otherwise seems obscure. Furthermore, since these terms are related to the operator in Eq.~(\ref{eq:op_dG0}) by RPI, we know now that the matching coefficients have to be the same to all orders in perturbation theory. That the Type I and III terms can be derived so easily using RPI does not mean that the rather complicated diagrammatic analysis performed earlier in this section was an unnecessary  effort. It could have been the case that expansion of the QCD diagrams led to other operators besides those that were related to the S-wave operator by RPI. The explicit matching calculation shows that, beside the Type II terms which come from the $O(q)$ terms in the vertex, the only other P-wave operators are related to the S-wave operators by RPI.

    \subsection{Gauge invariance}
    
    In this subsection, we investigate the gauge transformation properties of the three operators that appeared in the explicit matching calculation presented earlier in the example $q\bar{q} \to Q\bar{Q}$. Within the EFT there are three types of gauge transformation: collinear, soft, and ultra-soft. We demonstrate that all three operators are invariant under these transformations. Collinear fields transform only under collinear and ultra-soft gauge transformations and soft fields only under soft and ultra-soft. Potential (heavy quark/antiquark) and ultra-soft fields transform only under ultra-soft gauge transformations. The exact form of the  transformations for the relevant fields and building blocks are shown in Tab.~\ref{tb:gauge-tr}.
    
    We begin with the transformation of the collinear sector, which appears in all three of the relevant operators:
    \begin{equation}
       \bar{\chi}_{\bn} \gamma^{i}  S^{\dag}_{\bn} T^{a} S_{n}  \chi_{n}\;.
    \end{equation}
    This term is invariant under the collinear transformations, since the building block $\chi_n$ is the only operator that contains collinear fields and is invariant by construction. Under soft and ultra-soft gauge transformations:
    \begin{equation}
     \bar{\chi}_{\bn} \gamma^{i}  S^{\dag}_{\bn} T^{a}  S_{n}  \chi_{n}\to \bar{\chi}_{\bn} \gamma^{i}  S^{\dag}_{\bn} V_{s/us} T^{a} V^{\dag}_{s/us} S_{n}  \chi_{n} =
      \mathcal{V}_{s/us}^{ab} \bar{\chi}_{\bn} \gamma^{i}  S^{\dag}_{\bn} T^{b}S_{n}  \chi_{n}\,,
    \end{equation}
    where $\mathcal{V}_{s/us}^{ab}$ is the gauge transformation matrix in the adjoint representation. Since the generators of the adjoint representation, $t^c_{ab}$, are related to the structure constants of the Lie algebra, $t^{c}_{ab} =  i f^{acb}$, the following relation is satisfied 
    \begin{equation}
      \mathcal{V}_{ba} =  (\mathcal{V}^{-1})_{ab} =  (\mathcal{V}^{\dag})_{ab} \,.
    \end{equation}
    
    This gauge dependence needs to be cancelled by the heavy quark sector, which we now consider. Each of the operators contains heavy (potential) fields and soft fields, and the analysis is simpler if we consider them separately. Obviously since these terms do not contain any collinear fields, they do not transform under collinear gauge transformations.
    The elements that need to be considered are: the adjoint soft Wilson line $\mathcal{S}_v^{ad}$, the color-octet S-wave bilinear $ \psi^{\dag}\sigma^{i}  T^{a} \chi$, the soft gluon building block $B_{s}^{e,\mu}$, and the P-wave color-singlet and color-octet quark bilinears $\psi^{\dag} \sigma^{i} \; \lrP^{j}  \chi$ and $\psi^{\dag} \sigma^{i} \; \lrP^{j} T^{a} \chi$.
    \begin{itemize}
    \item  {\bf Soft Wilson line}. The ultra-soft transformation of the soft Wilson line is:
      \begin{equation}
        \mathcal{S}_v^{ad} T^d = S_v^{\dag} T^a S_v \to  V_{us} S_v^{\dag} V^{\dag}_{us} T^a V_{us} S_v V^{\dag}_{us} = \mathcal{V}^{ab}_{us}  \mathcal{S}_v^{bc} (\mathcal{V}^{\dag}_{us})^{cd} T^d \;.
      \end{equation}
     The soft transformation is:
      \begin{equation}
        \mathcal{S}_v^{ad} T^d = S_v^{\dag} T^a S_v \to  S_v^{\dag} V^{\dag}_{s} T^a V_{s} S_v = \mathcal{V}^{ab}_{s}  \mathcal{S}_v^{bd} T^d \;.
      \end{equation}
      \item {\bf Color-octet S-wave bilinear.}
      Since this involves only potential fields it does not transform under soft gauge transformations. For the ultra-soft transformation we have:
      \begin{equation}
         \psi^{\dag}\sigma^{i}  T^{a} \chi \to \psi^{\dag}\sigma^{i} V_{us}^{\dag} T^{a}  V_{us} \chi =  \mathcal{V}_{us}^{ab}\; \psi^{\dag}\sigma^{i}  T^{b} \chi\;.
      \end{equation}
      \item  {\bf The soft gluon building block.}
        The octet  $B_{s}^{e,\mu}$ is trivially soft gauge invariant since $B_s^{\mu} =B_{s}^{e,\mu} T^e$ is soft gauge invariant by construction. Under an ultra-soft gauge transformation
      \begin{equation}
        B_{s}^{e,\mu} T^e = B_s^{\mu} \to V_{us}  B_s^{\mu}  V_{us}^{\dag} = \mathcal{V}_{us}^{ef} B_{s}^{f,\mu} T^e \;.
      \end{equation}
      \item {\bf P-wave color-singlet bilinear.}
      The bilinear $\psi^{\dag} \sigma^{i} \; \lrP^{j}  \chi$ involves only potential fields which do not vary under soft gauge transformations, and is thus soft gauge invariant. It is also ultra-soft gauge invariant since the ultra-soft transformation commutes with the label momentum operator, $[\mathcal{P}^{\mu},V_{us}] = 0$. 
      \item  {\bf P-wave color-octet bilinear.}
    As for the color-singlet case this bilinear is invariant under soft gauge transformations. 
    Under ultra-soft transformations 
      \begin{equation}
        \psi^{\dag} \sigma^{i} \; \lrP^{j} T^{a} \chi \to  \mathcal{V}_{us}^{ab}  \psi^{\dag} \sigma^{i} \; \lrP^{j} T^{b} \chi \,.
      \end{equation}
      
    \end{itemize}
    We collect these results in Tab.~\ref{tb:gauge-tr}. Using them and the identity 
    \begin{equation}
      d^{def} \mathcal{V}_{us}^{dd'}\mathcal{V}_{us}^{ee'}\mathcal{V}_{us}^{ff'} = d^{d'e'f'}
    \end{equation}
    it is straightforward to show that the operators in Sect.~\ref{simple} are collinear, soft, and ultra-soft gauge invariant.  
    \begin{table}[!ht]
      \renewcommand{\arraystretch}{1.6}
      \begin{center}
        \begin{tabular}{|l|c|c|c|}
          \hline
          Operators  & collinear &  soft & ultra-soft \\ \hline \hline
          $ \mathcal{S}_v^{ad}                                $   & G.I.  & $ \mathcal{V}_{s}^{ab} \; \mathcal{S}_v^{bd}         $  & $\mathcal{V}_{us}^{ab}  \;\mathcal{S}_v^{bc} (\mathcal{V}_{s}^{\dag})^{cd} $  \\ \hline
          $B_s^{a,\mu} $                                           & G.I.  &  G.I.                                                &  $\mathcal{V}_{us}^{ab} \;B_s^{b,\mu}         $                                 \\ \hline
          $ \bar{\chi}_{\bn} \gamma^{i}  S^{\dag}_{\bn} T^{a} S_{n}  \chi_{n}$   &  G.I. & $ \mathcal{V}_{s}^{ab} \;  \bar{\chi}_{\bn} \gamma^{i}  S^{\dag}_{\bn} T^{a} S_{n}  \chi_{n}$ & $ \mathcal{V}_{us}^{ab}\;  \bar{\chi}_{\bn} \gamma^{i}  S^{\dag}_{\bn} T^{a} S_{n}  \chi_{n}$                        \\ \hline
          $\psi^{\dag}\sigma^{i}  T^{a} \chi $    & G.I.  &  G.I.                                                &  $\mathcal{V}_{us}^{ab} \;\psi^{\dag}\sigma^{i}  T^{a} \chi $ \\ \hline
          $\psi^{\dag} \sigma^{i} \; \lrP^{j}  \chi$      & G.I.  &  G.I.                                                &    G.I.                                                                   \\ \hline
          $ \psi^{\dag} \sigma^{i} \; \lrP^{j} T^{a} \chi$    & G.I.  &  G.I.                                                &  $\mathcal{V}_{us}^{ab} \; \psi^{\dag} \sigma^{i} \; \lrP^{j} T^{a} \chi$ \\ \hline
        \end{tabular}
        \caption{Gauge transformation of various operators and building blocks relevant for the light quark production/decay. Here G.I. stands for gauge invariant. }
        \label{tb:gauge-tr}
      \end{center}
    \end{table}
    
    With this we conclude our analysis for the light quark pair annihilation channel. We showed that at tree level and at leading order in $\lambda$ there is one relevant mechanism for S-wave states, $\tensor*[^3]{S}{^{[8]}_1}$,  and two mechanisms for P-wave states, $\tensor*[^3]{P}{^{[1]}_J}$ and $\tensor*[^3]{P}{^{[8]}_J}$.  We shoed that the corresponding operators, obtained through a diagrammatic analysis, are collinear, soft, and ultra-soft gauge invariant and satisfy the correct relative velocity scaling. We also showed how RPI relates different operators. As a last comment for this section notice that both from the diagrammatic analysis and from gauge invariance the P-wave bilinears involve the left-right label momentum operator instead of the covariant derivative usually seen in the NRQCD LDMEs. This is a direct consequence of power counting:
    \begin{equation}
      q^{\mu}= q^{\mu}_s + q^{\mu}_{us}  \simeq  q^{\mu}_s (1+\mathcal{O}(\lambda)) \to \mathcal{P}^{\mu} \,.
    \end{equation}
    The higher order corrections from $q_{us}$ can be included in the following way 
    \begin{equation}
      q^{\mu}= q^{\mu}_s + q^{\mu}_{us}  \to  \mathcal{P}^{\mu} +i \partial^{\mu} - g A_{us}^{\mu} \equiv \mathcal{D}_{us}^{\mu}\,,
    \end{equation}
    where we included the ultra-soft gluon field from gauge completion. 
    
    In the next section we use the results of this section to demonstrate how the relevant operators and the EFT Lagrangian can be used in order to obtain phenomenological quantities. Particularly how to obtain a factorization theorem, perform perturbative calculations, and resum large logarithmic enhancements that could potentially ruin the perturbative expansion. Simultaneously we perform some consistency checks of our approach and demonstrate with an explicit calculation that up to NLO our factorization holds.
    

    \section{P-wave decays to light quarks}
    \label{sec:NLL}
    In this section we show how our formalism can be 
    applied by considering a specific example: the semi-inclusive rate for $\chi_{J}$ quarkonium decay with two observed hadrons in the final state. As we will show, this process allows access to the TMD fragmentation function (TMDFF) of light quarks. Of course the decay rate includes contributions from gluons as well, however, since we are doing this analysis for illustrative purposes  these will not be included.  
    
    The events are chosen to contain two identified distinguishable hadrons, $H_1$ and $H_2$, and any number of additional particles. The hadrons are required to have large energy of order of the available center-of-mass energy $M_\chi$. We assign to $H_1$ the role of a trigger, where a measurement of the energy is carried out, and with the direction of the hadron fixing the $z$-axis. With this assignment a unique plane perpendicular to the $z$-axis can be chosen which splits the final state into hemispheres. Then $H_2$ is chosen to lie in the hemisphere not containing $H_1$, and a measurement of the energy and momentum transverse to the $z$-axis is carried out. 
    
    
    
    \subsection{Factorization}
    
    To begin we consider the kinematics of the proposed process, $\chi_J \to H_1 + H_2 +X$. Let $p_\chi^\mu$, $p_1^\mu$ and $p_2^\mu$ be the momenta of the $\chi_J$, $H_1$ and $H_2$ respectively. Define the dimensionless variables~\cite{Altarelli:1979kv}
    \begin{eqnarray}
    z_i &=& \frac{2 p_i \cdot p_\chi}{M_\chi^2}, \qquad i = 1,2 \nonumber \\
    u &=& \frac{2 p_1 \cdot p_2}{M_\chi^2 z_1} \,.
    \end{eqnarray}
    In the center-of-mass frame $z_i = 2 E_i/M\chi$ are the fraction of the hadron energy relative to the available energy for a back-to-back event. 
    Neglecting the mass of $H_1$
    \begin{equation}
        p_1^\mu = \frac{1}{2}M_\chi \, z_1 n^\mu\,.
    \end{equation}
    In this same frame
    \begin{equation}
        u = \frac{1}{2}z_2\big(1-\cos(\theta_{12})\big)\,,
    \end{equation}
    where $\theta_{12}$ is the angle between $\bmat{p}_1$ and $\bmat{p}_2$. For a back-to-back event $\theta_{12}= \pi$, and $u = z_2$. Treating $H_2$ as massless the magnitudes of the $z$ and transverse components of $H_2$ momenta are
    \begin{eqnarray}
    p_2^z &=& \frac{1}{2} M_\chi z_2 \cos\big(\pi-\theta_{12}\big)
    \nonumber \\
    p_2^\perp &=& \frac{1}{2} M_\chi z_2 \sin\big(\pi-\theta_{12}\big)\,.
    \end{eqnarray}
    To access the TMDFF we restrict ourselves to the kinematic region where $\theta_{12}\approx \pi$. To be specific we take $\theta_{12}- \pi \sim \lambda$, where $\lambda <<1$ is the SCET power counting parameter. Then $p_2^z \approx M_\chi z_2/2 + {\cal O} (\lambda^2)$, and $p_2^\perp \sim M_\chi z_2 \lambda$. In this regime we can express the $H_2$ momentum as a light-like vector in a direction approximately opposite to the direction of $H_1$:
    \begin{equation}
        p^\mu_2 = \frac{1}{2} M_\chi \, u \bn^\mu + p^\mu_\perp - \frac{p^2}{2 M_\chi u} n^\mu 
        = \frac{1}{2} M_\chi \, z_2 \bn^\mu + p^\mu_\perp +r^\mu \,,
    \end{equation}
    where the residual momentum $r^\mu$ is ${\cal O}(\lambda^2)$.
    The events picked out by our selection criteria will have 
    \begin{equation}
        p^\mu_X = p^\mu_\chi-p^\mu_1 -p^\mu_2 \approx 
        \frac{1}{2}M_\chi (1-z_1)n^\mu +\frac{1}{2}M_\chi (1-z_2)\bn^\mu-p^\mu_\perp \,,
    \end{equation}
    so $p^2_X \approx M^2_\chi (1-z_1)(1-z_2)+p^2_\perp$. For $z_1\,, z_2 \sim {\cal O}(1)$ (but not too close to one) the final state particles that are insensitive to the transverse momentum and off-shell on order of $M^2_\chi$ are integrated out. 
    
    Next we derive a factored form for the decay rate. To begin the $S$-matrix for the QCD decay amplitude $\langle H_1(p_1) H_2(p_2) + X \vert \chi_J \rangle$ needs to be matched onto an operator matrix element in the EFT. We did this in Sect.~\ref{simple} for the process in question. The result is:
    \begin{eqnarray}
    \langle H_1(p_1) H_2(p_2) + X \vert \chi_J \rangle & \to & \sum_{n={}^3S^{[8]}_1,{}^3P^{[1]}_J}\int d \omega_1 d\omega_2 \,\, C^{[n]}\big(\alpha_s(M_\chi),\omega_1,\omega_2\big) \\
    & & \hspace{-5ex} \times \langle H_{1\, n} X_n; H_{2\, \bn} X_{\bn}; X_{s}\vert J^{[n]}_{\omega_1,\omega_2}(0)\vert \chi_J\rangle \, \delta^{(4)}(p_{\chi} - \sum_{f_n} p_{f_n}-\sum_{f_{\bn}}p_{f_{\bn}}-\sum_{f_s}p_{f_s})\,, \nonumber
    \end{eqnarray}
    where there is an implicit sum over the directions of the collinear fields. The relevant SCET$_\text{Q}$ currents read
    \begin{eqnarray}
    J^{{}^3S^{[8]}_1}_{\omega_1,\omega_2} &=& \mathcal{S}_{v}^{cd}\; \lp \psi^{\dag}\sigma^{i}  T^{d} \chi \rp \times \lp  \bar{\chi}_{n \omega_1} \gamma^{i}  S^{\dag}_{\bn} T^{c} S_{n} \chi_{\bn \omega_2} \rp \,, \\
    J^{{}^3P^{[1]}_J}_{\omega_1,\omega_2} &=& \sqrt{\frac{2}{N_c}} \times \mathcal{S}_{v}^{cd}\; \lb \frac{g \;B^{d,j}_s}{m\; v\cdot \mathcal{P}}   \rb \lb   \psi^{\dag}\frac{ \sigma^{i}\; \lrP^{j}}{2 \sqrt{2 N_c}} \; \chi \rb \times \lp  \bar{\chi}_{n \omega_1} \gamma^{i}  S^{\dag}_{\bn} T^{c} S_{n}  \chi_{\bn \omega_2} \rp \,.\nonumber
    \end{eqnarray}
    The delta function can be decomposed into light-cone form:
    \begin{eqnarray}
      \delta^{(4)}(p_{\chi} - \sum_{f_n} p_{f_n}-\sum_{f_{\bn}}p_{f_{\bn}}-\sum_{f_s}p_{f_s}) &=& 2 \delta(M_\chi - \sum_{f_n} \bn\cdot p_{f_n})
      \delta(M_\chi - \sum_{f_{\bn}}n\cdot p_{f_{\bn}}) \\
    && \times  \delta^{(2)}(\sum_{f_n} \bmat{p}_{f_n\, \perp}+\sum_{f_{\bn}}\bmat{p}_{f_{\bn}\, \perp}+\sum_{f_s}\bmat{p}_{f_s \, \perp})\,. \nonumber
    \end{eqnarray}
    
    Squaring the amplitude to obtain the scattering probability, we derive the differential decay rate in the EFT:
    \begin{multline}
      \frac{d \Gamma}{dz_1 dz_2 d^2q_\perp} =
      {\cal N} \sum_{n={}^3S^{[8]}_1,{}^3P^{[1]}_J}\int d \omega_1 d\omega_2 \, \int d \omega'_1 d\omega'_2 C^{[n]}\big(\alpha_s(M_\chi),\omega_1,\omega_2\big)C^{\dagger[n]}\big(\alpha_s(M_\chi),\omega'_1,\omega'_2\big) \\
    \times
      \sum_{X_s} \sum_{X_n\neq H_1} \sum_{X_{\bn}\neq H_2} 
      \delta(M_\chi - \sum_{f_n} \bn\cdot p_{f_n})
      \delta(M_\chi - \sum_{f_{\bn}}n\cdot p_{f_{\bn}})\delta^{(2)}(\sum_{f_n} \bmat{p}_{f_n\, \perp}+\sum_{f_{\bn}}\bmat{p}_{f_{\bn}\, \perp}+\sum_{f_s}\bmat{p}_{f_s \, \perp})   \\
    \times  
    \Langle \chi_J \Bvert J^{\dagger[n]}_{\omega_1,\omega_2}(0)| H_{1\, n} X_n; H_{2\, \bn} X_{\bn}; X_{s}\rangle \langle H_{1\, n} X_n; H_{2\, \bn} X_{\bn}; X_{s}|J^{[n]}_{\omega_1,\omega_2}(0)\Bvert \chi_J\Rangle \,,
    \end{multline}
    where ${\cal N}$ is a normalization factor which we will fix after factorizing. In the absence of Glauber interactions the decay rate above can be split into three sectors: $n$-collinear, $\bn$-collinear and soft. The result is a factored form of the differential decay rate:
    \begin{eqnarray}
    \label{factdecay}
     \frac{d \Gamma}{dz_1 dz_2 d^2q_\perp} &=& 
     \Gamma_0 \sum_{n={}^3S^{[8]}_1,{}^3P^{[1]}_J} H_{[n]}(M_\chi,\mu)\int d^2 k_{\bar{n}\perp} \int d^2 k_{s\perp} \int d^2 k_{n\perp} \delta^{(2)}(\bmat{k}_{\bar{n}'\perp} + \bmat{k}_{n\perp} + \bmat{k}_{s\perp} - \bmat{q}_{\perp} )  \nn \\
     && \times S^\perp_{[n]}(\bmat{k}_{s\perp})D^{\perp}_{q/H_1}(z_1,\bmat{k}_{n\perp})  D^{\perp}_{\bar{q}/H_2}(z_2,\bmat{k}_{\bar{n}'\perp})\,,
    \end{eqnarray}
    where 
    \begin{equation}
        \bmat{q}_{\perp} \equiv-\frac{ p_{\perp}^{2}}{z_2}\Bvert_{\text{w.r.t}\; H_1}
    \end{equation}
    and $\Gamma_0$ is the LO tree level partonic decay rate. The direction $\bar{n}'$ is aligned with the motion of $H_2$ and appears after an RPI type I transformation. Note because of RPI the hard function $H$ is the same for ${}^3S^{[8]}_1$ and ${}^3P^{[1]}_J$. 
    
    The TMD functions (in $d$ dimensions) are defined next. The quark TMDFF is~\cite{Echevarria:2014rua,Bain:2016rrv}
    \begin{equation}
        D^{\perp}_{q/H}(z,\bmat{k}_{n\perp}) = \frac{z}{2N_c} \text{tr} \Langle 0 \Bvert \frac{\slashed{\bn}}{2} \chi_n(0)\delta(M_\chi - \bn\cdot\mathcal{P})\delta^{(d-2)}(\bmat{k}_{n\perp}- \bmat{\mathcal{P}}_\perp)a^{\dag}_H(z M_{\chi},\bmat{0}_{\perp})a_H(z M_{\chi},\bmat{0}_{\perp})\bar{\chi}_n(0)\Bvert 0\Rangle
    \end{equation}
    where $a^{\dag}_H(p^+,p^-,\bmat{p}_{\perp})$ is the creation operator for a hadron $H$ with label momentum $p^{\mu} = (p^+,p^-,\bmat{p}_{\perp})$. A similar definition exists for the antiquark TMD fragmentation function. The definition of the $\wave{S}$-wave color-octet quarkonium shape function,  which appears for the first time in this paper~\footnote{The corresponding color-singlet shape function first appeared in Ref.~\cite{Echevarria:2019ynx}.} is 
    \begin{equation}
    \label{shape83s1}
      S^{\perp}_{\chi_J \to  \Q{3}{S}{1}{8}}(\bmat{k}_\perp ) = \frac{d-2}{(d-1) t_F}  \text{tr} \Langle \chi_J  \Bvert \psi^{\dag}\sigma^{i}  T^{a} \chi \mathcal{S}^{ba}_v (S_{\bar{n}}^{\dag}T^{b}S_n) \delta^{(2)} (\bmat{k}_{\perp} - \bmat{\mathcal{P}}_{\perp} ) 
      \times (S_{n}^{\dag} T^c S_{\bar{n}}) \mathcal{S}_v^{dc}\chi^{\dag}  \sigma^{i}  T^{d} \psi\Bvert \chi_J \Rangle \,, 
    \end{equation}
    where the trace outside the matrix element refers to fundamental color indices. The NRQCD bilinears $\psi^{\dag} \sigma^i T^a \chi$, are individually traced  over color and Dirac indices. The $\vert \chi_J \rangle $ state is summed over all possible polarizations. For the $\wave{P}$-waves to project onto the individual $J=$0,1, and 2 we can use the projection operators discussed in Ref.~\cite{Petrelli:1997ge} or the helicity decomposition techniques in Ref.~\cite{Braaten:1996jt}. Note this projection is only possible since we sum over the polarization of the quarkonium state. For polarized quarkonium states interference terms of the same helicity but different values of $J$ are not excluded although they might be simplified using approximate heavy quark spin symmetry of NRQCD. Applying the projection operators at the matrix element level we obtain the $\wave{P}$-wave color-singlet quarkonium shape function 
    \begin{multline}
    \label{shape13pJ}
      S^{\perp}_{\chi_J \to {}^3P_J} (\bmat{k}_{\perp})  = (2J+1)
      \frac{g^2}{N^2_c
      \;t_F} \; \mathcal{A}_J^{ij}\; \text{tr} \Langle \chi_J  \Bvert 
      \psi^\dag \bmat{\sigma} \cdot \bmat{\lrP} \; \chi
      \lb \frac{B_s^{a,i}}{m \; v\cdot\mathcal{P}} \rb \mathcal{S}^{ba}_v (S_{\bar{n}}^{\dag}T^{b}S_n) \\
      \times \;\delta^{(2)} (\bmat{k}_{\perp} - \bmat{\mathcal{P}}_{\perp} ) (S_{n}^{\dag} T^c S_{\bar{n}}) \mathcal{S}_v^{dc} \lb \frac{B_s^{d,j}}{m \; v\cdot\mathcal{P}} \rb
      \chi^\dag \bmat{\sigma} \cdot \bmat{\lrP} \; \psi
      \Bvert \chi_J \Rangle\,,
    \end{multline}
    where  
    \begin{align}
      \mathcal{A}_0^{ij} =& \frac{1}{d-1} \delta_{\perp}^{ij} \nn\\
      \mathcal{A}_1^{ij} =& \frac{1}{d-1} \delta^{ij} - \frac{1}{(d-1)(d-2)} \delta_{\perp}^{ij} \nn\\
      \mathcal{A}_2^{ij} =& \frac{1}{d+1} \delta^{ij} - \frac{d-3}{(d+1)(d-1)(d-2)} \delta_{\perp}^{ij} \,,
    \end{align}
      where $\delta_{\perp}^{ij} = \delta^{ij}- \delta^{iz}$.  Similarly to the $\wave{S}$-wave case the NRQCD bilinears are individually traced  over color and Dirac indices. In the following sub-sections we give the ${\cal O}(\alpha_s)$ expressions for each of the various functions in Eq.~(\ref{factdecay}).
    
    
    \subsection{The hard function}
    
    The hard function for the quark-antiquark  fusion processes can be extracted from Eq.~(132) of Ref.~\cite{Petrelli:1997ge}:
    \begin{eqnarray}
      H(\mu) &=& 1- \frac{\alpha_s C_{F}}{2\pi} \lbc \ln^2 \lp \frac{\mu^2}{M^2}\rp + 3 \ln \lp \frac{\mu^2}{M^2}\rp +\frac{\pi^2}{6} - 2B (\Q{3}{S}{1}{8})\rbc - \frac{\alpha_s C_A}{2\pi} \ln \lp \frac{\mu^2}{M^2}\rp \,, \\
      B (\Q{3}{S}{1}{8}) &=&  C_{F} \bigg( -8 +\frac{2}{3} \pi^2 \bigg) + C_{A} \bigg( \frac{50}{9} + \frac{2}{3} \ln 2 -\frac{\pi^2}{4} \bigg) - \frac{10}{9} n_f t_F \,.
    \end{eqnarray}
    Note the Coulomb singularities in Eq.~(132) of Ref.~\cite{Petrelli:1997ge} need to be dropped as they are reproduced in the EFT, and the $1/\epsilon$ divergences are dropped as well, as they cancel against the combined counterterm of the EFT.
    The running of the hard function is the ``opposite'' of the combined running of the three TMD functions in Eq.~(\ref{factdecay}), and the combined counterterm can also be extracted from Eq.~(132) of Ref.~\cite{Petrelli:1997ge}:
    \begin{equation}
        Z = 1 - \frac{\alpha_s C_{F}}{\pi}\lbc \frac{1}{\epsilon^2}+\frac{1}{\epsilon}\ln\lp \frac{\mu^2}{M^2}\rp + \frac{3}{2 \epsilon}\rbc - \frac{\alpha_s C_{A}}{2 \pi} \frac{1}{\epsilon} \,.
    \end{equation}
    Then the anomalous dimension can be determined through standard methods, from which we infer
    \begin{align}
      \label{eq:gamma-H}
      \gamma_{\mu}^{H}(\mu) &= -2\frac{\alpha_s C_{F}}{\pi} \lbc \ln \lp \frac{\mu^2}{M^2}\rp + \frac{3}{2} \rbc -\frac{\alpha_s C_A}{\pi} \,.
    \end{align}
    The hard function then satisfies the RGE
    \begin{equation}
      \frac{d}{d\ln \mu} H(\mu) = \gamma_{\mu}^{H}(\mu) H(\mu)\,.
    \end{equation}
    It is important to note the additional divergence and associated logarithm that is proportional to $C_A$. Since the divergences from the TMDFFs at $\mathcal{O} (\alpha_s)$ are proportional to $C_F$  and since the total cross section needs to be independent of the renormalization scale, a corresponding term needs to be generated from the real soft emission diagrams. The only other element in the factorization theorem that does depend  on the production mechanism is the shape function, $S^{\perp}_{\chi \to [n]}$. Thus, it is expected that such a contribution will be present in the shape function.
    \subsection{The TMDFF}
    \label{tmdff}
    The TMD functions that appear in the factorization theorem are nonperturbative for $k_\perp \sim  \Lambda_{\text{QCD}}$, but for $k_{\perp} \gg \Lambda_{\text{QCD}}$ can be expanded in powers of $ \Lambda_{\text{QCD}}/k_{\perp}$. The expansion can be written as a convolution of short distance matching coefficients and the collinear fragmentation functions (FF). For quark fragmentation we have~\cite{Echevarria:2014rua,Bain:2016rrv}
    \begin{equation}
      \label{eq:beam_match}
      D^{\perp}_{q/h}(x,\bmat{k}_{\perp};\mu,\nu) = \int_{x}^1 \frac{dz}{z}\; \mathcal{J}_{q/j}(z, \bmat{k}_{\perp};\mu,\nu) \;D_{j/h}\lp \frac{x}{z};\mu  \rp \,.
    \end{equation}
    The  matching coefficient and FF need to be evaluated at the same factorization scale, $\mu$, before convolving. To this end we evaluate the matching coefficients at their characteristic scale and the FFs we evolve using the standard DGLAP renormalization group equations. The additional scale, $\nu$, that appears in the arguments of the  unsubtracted TMD fragmentation function and the matching coefficients in Eq.~(\ref{eq:beam_match}) comes from the  $\eta$-regulator (see Refs.~\cite{Chiu:2011qc, Chiu:2012ir}) which we use to regulate rapidity divergences. Rapidity divergences remain unregulated in pure dimensional regularization (dim-reg). We refer to this scale as the rapidity scale or simply rapidity.
    The renormalized matching coefficients at next-to-leading order (NLO) are given by (see Ref.~\cite{Echevarria:2014rua,Bain:2016rrv}):
    \begin{eqnarray}
      \tilde{\mathcal{J}}_{q/j}(z,\bmat{k}_\perp;\mu,\nu)&=& \delta_{qj}\delta(1-z)\delta^{(2)}(\bmat{k}_\perp) \\
    &&  + \frac{\alpha_s T_{qj}}{\pi} \lbc  \lb \delta_{ij}\delta(1-z)\ln\lp\frac{\omega^2}{\nu^2}\rp+\bar{P}_{jq}(z)\rb \frac{1}{2 \pi \mu^2}\bigg( \frac{\mu^2}{\bmat{k}_\perp^2}\bigg)_+
      +c_{qj}(z)\delta^{(2)}(\bmat{k}_\perp)
      \rbc, \nonumber
    \end{eqnarray}
   where $\omega$ is the minus component of the initiating parton with
    \begin{align}
      &\bar{P}_{qq}(z)=P_{qq}(z)-\bar{\gamma}_q\delta(1-z)=(1+z^2)\frac{1}{(1-z)_+}, \nn \\
      &\bar{P}_{gq}(z)=P_{gq}(z)=\frac{1+(1-z)^2}{z},
    \end{align}
    and
    \begin{equation}
      {c}_{qq}(z)=\frac{1-z}{2}, \;\;
      {c}_{qg}(z)=\frac{z}{2}\,.
    \end{equation}
    Here $T_{qq}=T_{qg}=C_F,$,  and $\bar{\gamma}_q =\bar{\gamma}_{\bar{q}} =3/2$.  The $\mu$-anomalous dimension for the TMDFF is:
    \begin{equation}
      \gamma_{\mu,q}^{D} (M,\nu) = \frac{\alpha_s C_F}{\pi} \lbc  \ln\lp \frac{\nu^2}{M^2}  \rp +\bar{\gamma}_q \rbc\;.
    \end{equation}
    The rapidity renormalization group (RRG) equation has a convolutional form
    \begin{equation}
    \label{tmdffrrge}
      \frac{d}{d \ln \nu} D^{\perp}_{i/h} (z,\bmat{k}_\perp;\mu,\nu) = \int \frac{d^2 k'_\perp}{(2\pi)^2} \, \gamma_{\nu,i}^{D} (\bmat{k}'_\perp,\mu) \; D^{\perp}_{i/h} (z,\bmat{k}_\perp - \bmat{k}'_\perp;\mu,\nu)
    \end{equation}
    where
    \begin{equation}
      \gamma_{\nu,k}^{D} (\bmat{k}_\perp,\mu) =  \frac{\alpha_s C_F}{\pi} \frac{1}{\pi \mu^2}\bigg( \frac{\mu^2}{\bmat{k}^2_\perp}\bigg)_+ \,.
    \end{equation}
    
    Eq.~(\ref{tmdffrrge}) can be put a multiplicative form by Fourier transform to transverse position space, $\bmat{b}$-space. Denoting $\bmat{b}$-space quantites with a tilde, Eq.~(\ref{tmdffrrge}) transforms to
    \begin{equation}
    \label{tmdffrrgebspace}
      \frac{d}{d \ln \nu} \tilde{D}^{\perp}_{i/h} (z,\bmat{b};\mu,\nu) = \tilde{\gamma}_{\nu,i}^{D} (\bmat{b},\mu) \; \tilde{D}^{\perp}_{i/h} (z,\bmat{b};\mu,\nu)\,
    \end{equation}
    where 
    \begin{equation}
      \tilde{\gamma}_{\nu,q}^{D} (\bmat{b},\mu) =  \frac{\alpha_s C_F}{\pi} \lbc \ln\bigg(\frac{b^2 \mu^2}{4 e^{-2 \gamma_E}}\bigg) + \bar{\gamma}_q \rbc\,.
    \end{equation}
    
    
    \subsection{The shape functions}
    The analysis of the shape function is the main topic of this paper. We calculate the TMD shape functions arising in the decay to NLO. The calculation is performed within the EFT approach and we use the EFT Lagrangian and Feynman rules for the calculations of the corresponding diagrams. At the end of this section our goal is to determine the RGEs  satisfied by the shape functions. As in the case of the TMDFF, in the regime of interest the shape functions are non-perturbative objects, but for $\vert \bmat{q}_{\perp} \vert \gg \Lambda_\textrm{QCD}$ the soft modes become perturbative and the TMD shape function can be expressed as a product of NRQCD LDMEs at the usoft scale and perturbative coefficients at the soft scale. Then the total shape function defined in Eqs.(\ref{shape83s1}) and (\ref{shape13pJ}) can be written as:
    \begin{equation}
      \label{eq:matching}
      {S}^{\perp}_{\chi_J \to  \Q{2s+1}{L}{J}{1,8}}(\bmat{k}_\perp;\mu_0,\nu_0) = \sum_n {C}_n (\bmat{k}_\perp;\mu_0,\nu_0, \mu) \times \langle  \mathcal{O}^{[n]} \rangle^\mu_{\chi_{J}} \;.
    \end{equation}
    This matching procedure it is essentially the standard NRQCD factorization applied to the shape function instead of the cross section. For the process we are considering and for each value of $J$ there are two LDMEs onto which we match:
    \begin{align}
      \label{eq:LDMEs}
      \langle \Q{3}{S}{1}{8} \rangle_{\chi_{J}} & = \langle \chi_J  \vert \psi^\dag \bmat{\sigma}T^a \chi \cdot \chi^\dag \bmat{\sigma}T^a \psi  \vert \chi_J \rangle \;,\nn \\
     \langle \Q{3}{P}{J}{1} \rangle_{\chi_{J}} & =  (2J+1)\frac{1}{2 N_c}  \langle  \chi_J  \vert \psi^\dag \bmat{\sigma} \cdot \bmat{\lrP} \; \chi   
      \chi^\dag \bmat{\sigma} \cdot \bmat{\lrP} \; \psi \vert \chi_J \rangle \;.
    \end{align}
    At NLO only these LDMEs appear in the matching but at higher orders there might be others as well. Note that the integrated shape functions reduce to 
    \begin{align}
      \int d^2q_{\perp}  S^{\perp}_{\chi_J \to  \Q{3}{S}{1}{8}}(\bmat{k}_\perp;\mu,\nu)  &= \frac{d-2}{d-1} \langle \Q{3}{S}{1}{8} \rangle_{\chi_{J}} \;, & \int d^2q_{\perp}  S^{\perp}_{\chi_J \to  \Q{3}{P}{J}{1}} (\bmat{k}_\perp;\mu,\nu) &=  \langle \Q{3}{P}{J}{1} \rangle^{1B}_{\chi_{J}} \,,
      \end{align}
    where
    \begin{equation}
         \langle \Q{3}{P}{J}{1} \rangle^{1B}_{\chi_{J}}  =  (2J+1)\frac{g^2}{N^2_c} \mathcal{A}_{J}^{ij} \Langle  \chi_J  \Bvert \psi^\dag \bmat{\sigma} \cdot \bmat{\lrP} \; \chi   \lb \frac{B_s^{a,i}}{m \; v\cdot\mathcal{P}} \rb \lb \frac{B_s^{a,j}} {m \; v\cdot\mathcal{P}} \rb
      \chi^\dag \bmat{\sigma} \cdot \bmat{\lrP} \; \psi \Bvert \chi_J \Rangle
    \end{equation}
    
    
    \subsubsection*{The color-octet S-wave shape function}
    The color-octet $\wave{S}$-wave shape function at LO is given by simply evaluating it at $g_s = 0$.   Therefore we have
    \begin{equation}
      S^{\perp, \text{LO}}_{\chi  \to \Q{3}{S}{1}{8} }(\bmat{k}_\perp;\mu,\nu) = \frac{d-2}{d-1} \delta^{(2)} (\bmat{k}_\perp) \, 4m^2 \eta^{\dag} \sigma^i  T^a  \xi \times \xi^{\dag}\sigma^i  T^a \eta \,,
    \end{equation}
    and immediately get
    \begin{equation}
      \label{eq:LO3s18}
      S^{\perp, \text{LO}}_{\chi  \to \Q{3}{S}{1}{8} }(\bmat{k}_\perp;\mu,\nu) = \frac{d-2}{d-1} \delta^{(2)} (\bmat{k}_\perp) \langle \Q{3}{S}{1}{8} \rangle _{\text{LO}} \,,
    \end{equation}
    where
    \begin{equation}
      \langle \Q{3}{S}{1}{8} \rangle _{\text{LO}} = 4m^2  \eta^{\dag} \sigma^i  T^a  \xi \times \xi^{\dag}\sigma^i  T^a \eta \,.
    \end{equation}
    
\begin{figure}[!ht]
      \centerline{\includegraphics[width = 0.92 \textwidth]{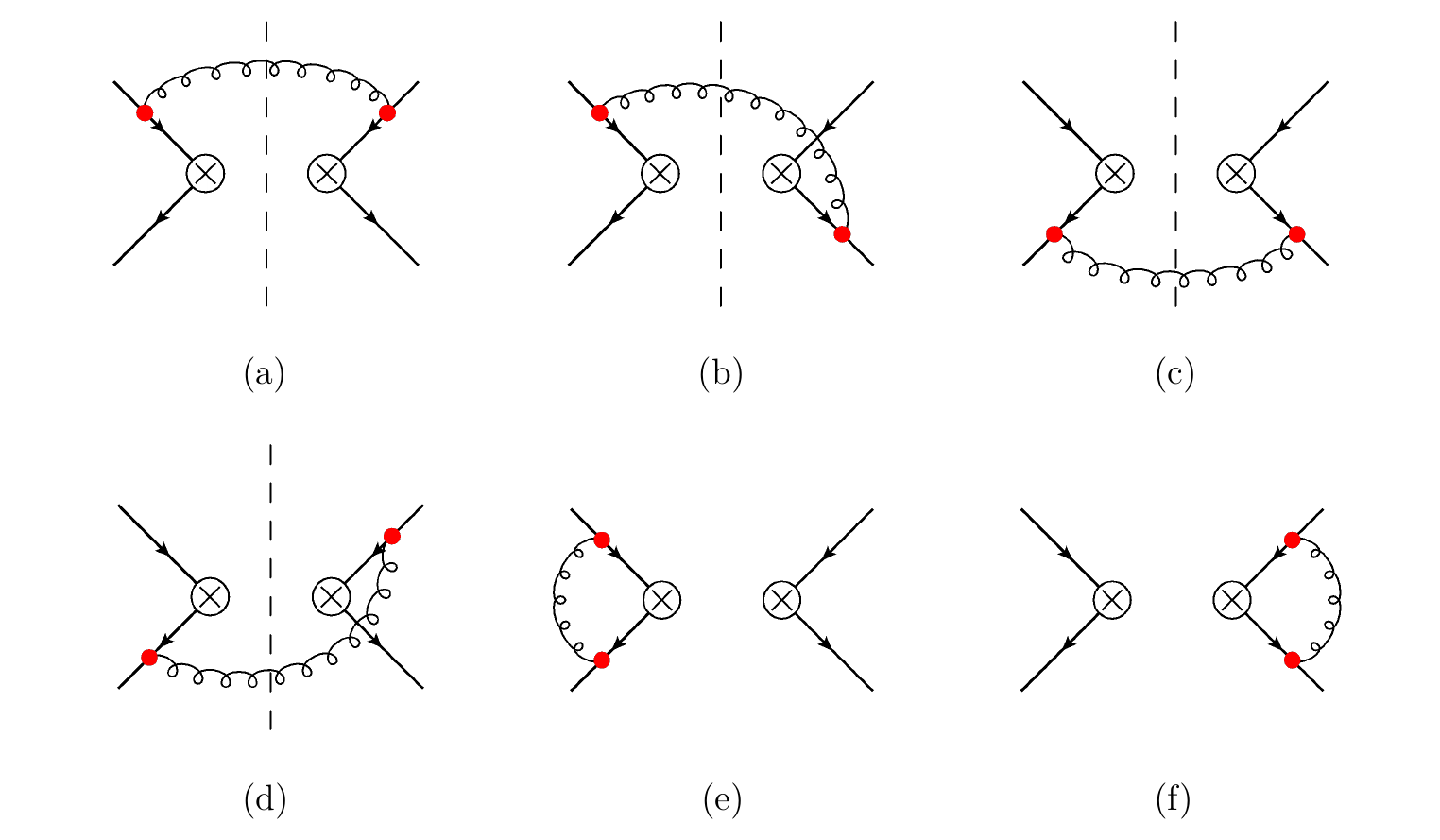}}
      \caption{NLO diagrams for heavy quark interaction through the chromo-electric dipole operator from the SCET$_{Q}$ Lagrangian.}
      \label{fig:heavy_heavy}
    \end{figure}

\begin{figure}[!ht]
      \centerline{\includegraphics[width = 0.92 \textwidth]{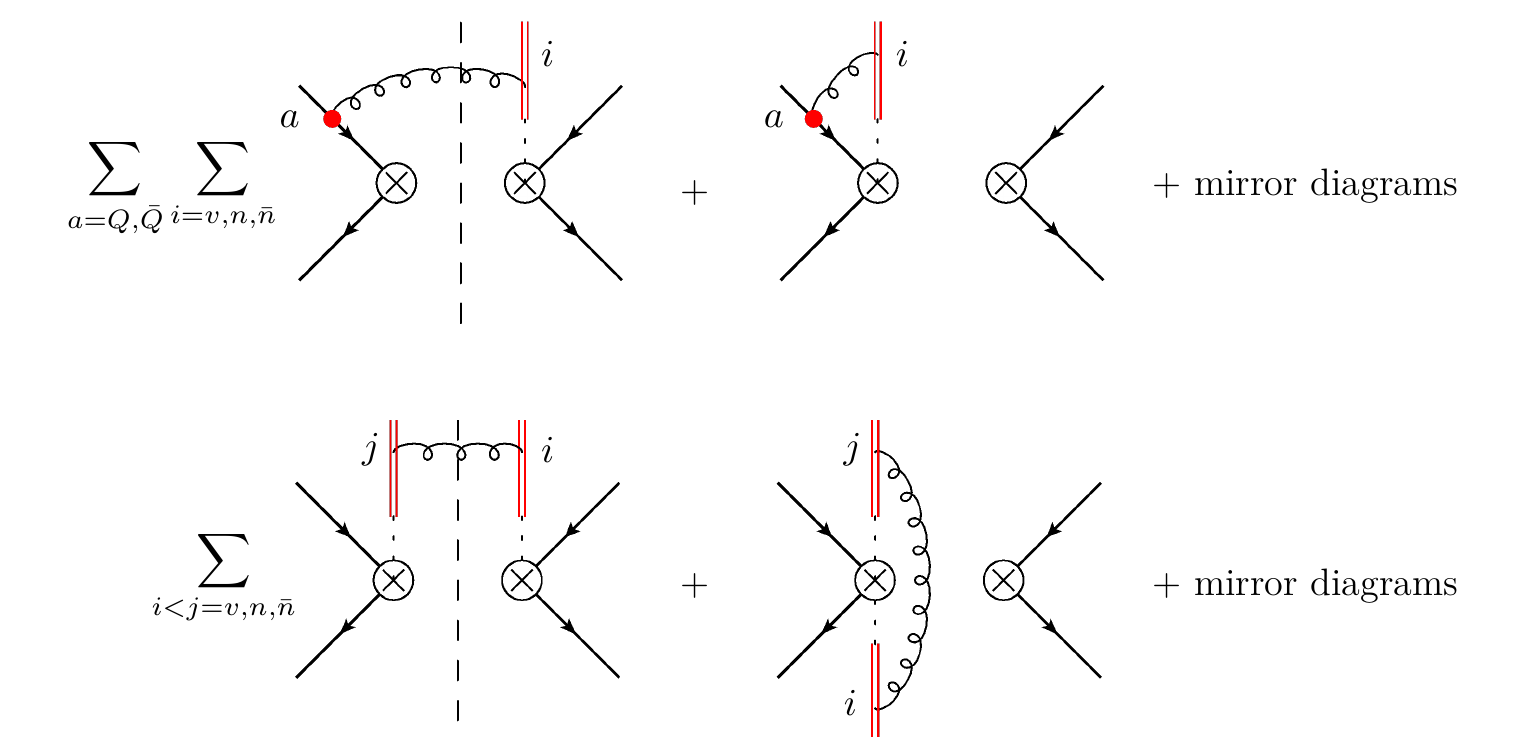}}
      \caption{NLO diagrams that involve ultra-soft Wilson lines. The first row are the real and virtual contributions from ultra-soft Wilson lines contracted with the heavy quark/antiquark through an insertion of the chromo-electric dipole operator. In the second row are the diagrams that involve only contractions of ultra-soft Wilson lines. }
      \label{fig:us-us}
    \end{figure}

\begin{figure}[!ht]
      \centerline{\includegraphics[width = 0.92 \textwidth]{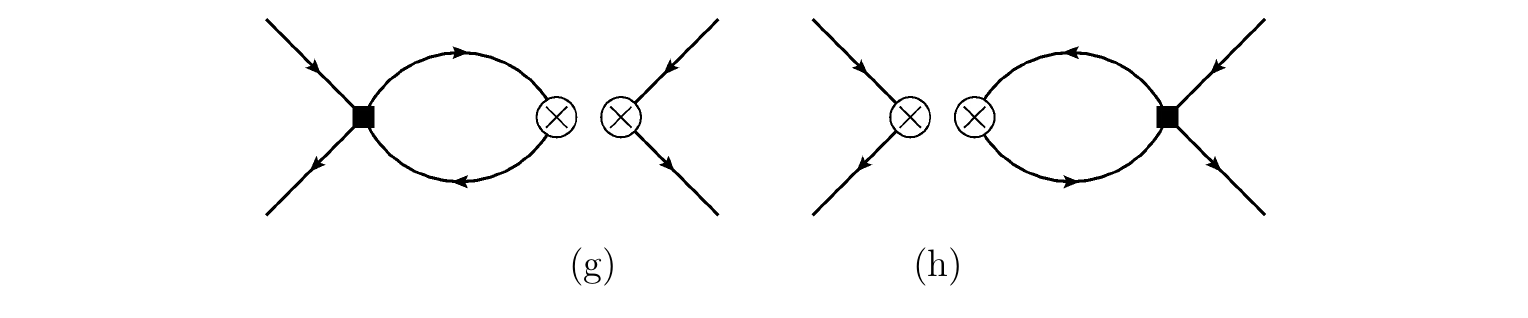}}
      \caption{Coulomb interactions at NLO.}
      \label{fig:coulomb}
    \end{figure}

\begin{figure}[!ht]
      \centerline{\includegraphics[width = 0.92 \textwidth]{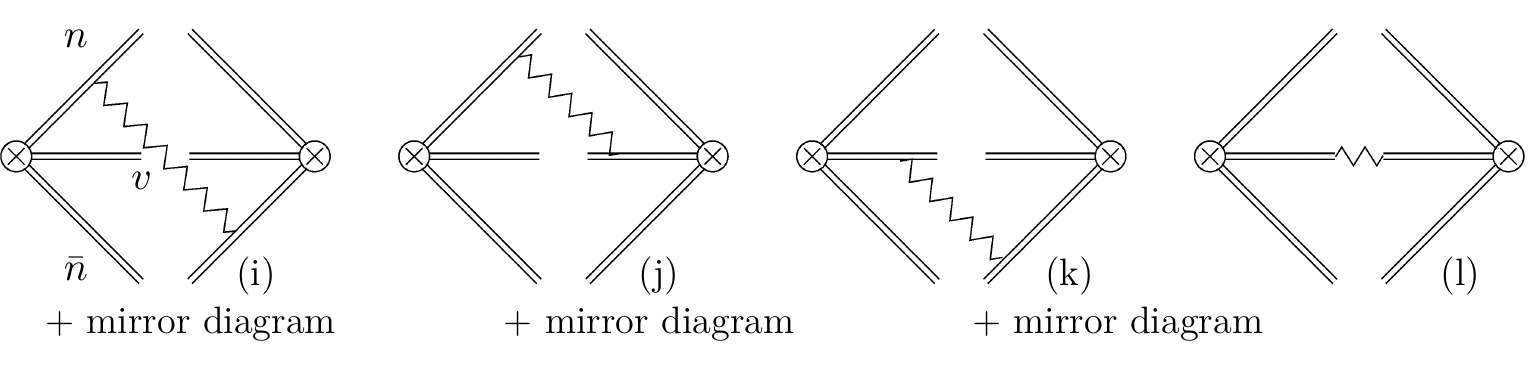}}
      \caption{The NLO real soft gluon shape function diagrams. At this order real gluon emissions can only come from soft Wilson lines. We use zigzag lines to indicate soft gluons.}
      \label{fig:soft-soft}
    \end{figure}

    At NLO where only diagrams with one soft or ultra-soft gluon are needed, the contributions can be categorized as follows (note we do not show the heavy-quark self energy contributions as they are well known): 
    \begin{enumerate}
    \item Ultra-soft gluon exchanges between heavy quarks and/or antiquarks. After the BPS field redefinition the interactions of the heavy quarks through ultra-soft gluons are pushed to the soft sector of the vNRQCD Lagrangian as well as to subleading terms in the Lagrangian. This induces usoft Wilson lines along the directions $v$, $n$ and $\bn$ to appear in operators. We will suppress these Wilson lines in the operators since, with the exception of the interaction we consider here, usoft interactions only produce uninteresting scaleless integrals that we drop in our scheme.  However, we will consider one subleading operator, the so-called chromo-electric dipole transition, as it leads to mixing between the $\Q{3}{P}{1}{J}$ and $\Q{3}{P}{8}{J}$ channels. After the BPS field redefinition this operator has the form: 
      \begin{equation}
        \label{eq:B-us}
        \sum_{\bmat{p}}- g \psi^{\dag}_{\bmat{p}} \frac{\bmat{B}_{us} \cdot \bmat{\mathcal{P}}}{m} \psi_{\bmat{p}} + (\psi \to \chi, T \to \bar{T}) \,.
      \end{equation}
    The diagrams associated with this contribution are shown in Fig.~\ref{fig:heavy_heavy}. Note that the virtual contributions, diagrams (e) and (f), will mix operators that contribute to $\chi_{J}$ decay only at higher orders in the relative velocity expansion. Thus we only consider real contributions, diagrams (a)-(d).
    \item Ultra-soft gluon exchanges between heavy quarks and ultra-soft Wilson lines. These contributions involve a single iteration of the interaction term in Eq.(\ref{eq:B-us}) and a single insertion of an ultra-soft gluon from any of the corresponding Wilson lines. Adding the corresponding diagrams shown in the first line of Fig.~\ref{fig:us-us} gives a contribution that vanishes.  
    \item Ultra-soft gluon exchanges between ultra-soft Wilson lines. The corresponding diagrams are shown in the second line of Fig.~\ref{fig:us-us}. They result in scaleless integrals which are zero in our scheme. In other schemes these diagrams cancel against the zero-bin subtraction of the corresponding soft diagrams. Thus, independent of scheme these diagrams may be set to zero if the corresponding zero-bin subtractions are set to zero as well. 
    \item Coulomb interactions between heavy quarks and antiquarks. This contribution is generated through the coulomb interaction term in the vNRQCD Lagrangian~\cite{Luke:1999kz}, 
      \begin{equation}
        \sum_{\bmat{p}, \bmat{q}} \frac{4 \pi \alpha_s }{(\bmat{p} -\bmat{q})^2}  \psi_{\bmat{p}} T^{a}  \psi^{\dag}_{\bmat{q}} \chi^{\dag}_{-\bmat{p}} \bar{T}^{a}  \chi_{-\bmat{q}}\,.
      \end{equation}
     The associated diagrams are shown in Fig.~\ref{fig:coulomb}.
    \item Soft gluon exchanges between soft Wilson lines. The real contributions are the only pieces with non-trivial dependence on transverse momentum. The virtual gluons contributions only give scaleless integrals that we set to zero in our scheme. The relevant (real emission) diagrams contributing at this order are shown in Fig.~\ref{fig:soft-soft}.
    \end{enumerate}
    
    Next we calculate each of the contributions we listed. For the regularization of ultraviolet and infrared divergences we use dim-reg in $d=4-2\epsilon$ dimensions and for the rapidity divergences we use the so called $\eta$-regulator.
    First we consider ultra-soft gluon exchanges between heavy quarks and/or antiquarks. 
    We calculate the diagram in Fig.~\ref{fig:heavy_heavy}(a), by first expanding the ultra-soft gluon building block $B_{us}^{\mu}$:
    \begin{equation}
      B_{us}^{\mu} = A_{us}^{\mu} - \frac{k^{\mu} A_{us}^{0}}{k^0} + \mathcal{O}(g)\,,
    \end{equation}
    and obtain the NLO amplitude
    \begin{equation}
      d_{\text{a}} = \frac{d-2}{d-1}\frac{\alpha_s}{6 \pi^2} \delta^{(2)} (\bmat{k}_{\perp})  \;  \eta^{\dag} \sigma^i q^{k} T^a T^b \xi \times \xi^{\dag}\sigma^i q^{k} T^b T^a \eta  \int \frac{d^{d-1} \bmat{k}}{\bmatn{k}^3}  \,.
    \end{equation}
    In Fig.~\ref{fig:heavy_heavy} the contributions of diagrams (b), (c), and (d) are
    \begin{align}
       d_{\text{b}} &= d_{\text{a}}(T^a T^b \otimes T^b T^a \to T^a T^b \otimes T^a T^b )\;, &  d_{\text{c}} &= d_{\text{a}}\;, & d_{\text{d}} &= d_{\text{b}} \,.
    \end{align}
    Adding these results we find that the amplitude for real ultra-soft gluon emission from heavy quarks and antiquarks, is proportional to the P-wave LO matrix element. However, the final phase space integral is scaleless. In our scheme we throw this contribution out, but it is important to appreciate its importance. Thus for now we use the $1/\epsilon_{\UV} -1/\epsilon _{\IR}$ prescription and obtain for the sum of amplitudes:
    \begin{equation}
      d_{\text{a+b+c+d}} = \frac{8\alpha_s}{9 \pi m^2} \delta^{(2)} (\bmat{k}_{\perp}) \lbc C_F \sum_J \langle \Q{3}{P}{J}{1}  \rangle_{\text{LO}} +  B_F \sum_J \langle \Q{3}{P}{J}{8}  \rangle_{\text{LO}}  \rbc \; \lp \frac{1}{\epsilon_{\UV}} -  \frac{1}{\epsilon_{\IR}}\rp \,,
    \end{equation}
    where
    \begin{equation}
    B_F = \frac{N_c^2 -4}{4N_c}\;.
    \end{equation}
    We later show how the IR divergence that appears here cancels an IR divergence from the $\wave{P}$-wave shape function, leaving only a UV divergence in this contribution. Clearly this is equivalent to setting this contribution to zero and interpreting the $1/\epsilon$ term in a real emission term of the 
    $\wave{P}$-wave shape function as UV. However, keeping this term momentarily non-zero makes it clear that there is a mixing between the color-octet $\wave{S}$-wave operator and the $\wave{P}$-wave operators. 
    
    We next consider the coulomb diagrams which involve the LO insertion of the coulomb interaction term from the vNRQCD Lagrangian, diagrams (g) and (h) in Fig.~\ref{fig:coulomb}. For diagram (g) we obtain:
    \begin{equation}
      d_{\text{g}} = \frac{2}{3} (C_F - \frac{1}{2} C_A)  \; I_g(q_0,\bmat{q})\; \langle \Q{3}{S}{1}{8} \rangle _{\text{LO}}   \; \delta^{(2)} (\bmat{k}_{\perp})
    \end{equation}
    where 
    \begin{equation}
      I_{\text{g}} = -i g^2 \int \frac{d^4 p}{(2\pi)^4} \frac{1}{(\bmat{p} - \bmat{q}/2)^2} \frac{1}{q_0 + p_0 - \bmat{p} ^2/(2m)} \frac{1}{q_0 - p_0 - \bmat{p}^2/(2m)}\,.
    \end{equation}
    This contribution and the corresponding one from diagram (h), are evaluated in Refs.~\cite{Bodwin:1994jh,Petrelli:1997ge} and the total contribution is given by:
    \begin{equation}
      d_{\text{g+h}} = \frac{2}{3} (C_F - \frac{1}{2} C_A)\; \frac{\pi \alpha_s}{\ve}\; \langle \Q{3}{S}{1}{8} \rangle _{\text{LO}}  \; \delta^{(2)} (\bmat{k}_{\perp})\,.
    \end{equation}
    This result which scales as $\ve^{-1}$ relative to the LO order shape function reproduces the coulomb singularities that arise in QCD. 
    
     Next we discuss the soft gluon contributions. The corresponding diagrams are shown in Fig.~\ref{fig:soft-soft}. These diagrams will contribute to the transverse momentum measurement since they involve a soft real gluon in the final state. The corresponding virtual contributions yield scaleless integrals and as usual are set to zero.
    We denote the contribution of each diagram by
    \begin{equation}
      d_r = \frac{2}{3} \langle \Q{3}{S}{1}{8} \rangle _{\text{LO}}\;C_r\; I_{r}(k_{\perp})\; \,,
    \end{equation}
    where $C_r $ are the color coefficients and $I_r$ the phase space integrals. For the coefficients $C_r$ and the coefficients of the mirror diagrams $C_{\bar{r}}$ we find
    \begin{align}
      C_{\text{i}} = C_{\bar{\text{i}}} &= -\frac{1}{2 N_c}\;,& C_{\text{j}} = C_{\bar{\text{j}}}= C_{\text{k}} = C_{\bar{\text{k}}} &= \frac{N_c}{2} \;, & C_{\text{l}} = N_c \,.
    \end{align}
    The integrals $I_r$ are defined and evaluated in Appendix~\ref{app:defs}, see Eqs.(\ref{eq:Ii}), (\ref{eq:Ij/k}), and  (\ref{eq:Il}). Summing all the contributions yields 
    \begin{equation}
    \label{eq:softdia}
      d_{(\text{i} + \bar{\text{i}})+ (\text{j} + \bar{\text{j}}) +(\text{k}+ \bar{\text{k}}) +\text{l}}  = \frac{2}{3} \langle \Q{3}{S}{1}{8} \rangle _{\text{LO}}\lp S^{\perp}_{\text{DY}}(\bmat{k}_\perp) + \frac{\alpha_s C_A}{2\pi} \lbc \frac{1}{\epsilon} \delta^{(2)}(\bmat{k}_\perp) - 2 \mathcal{L}_0 (\bmat{k}_\perp^2,\mu^2)  \rbc \rp \,,
    \end{equation}
    where
    \begin{multline}
      S^{\perp}_{\text{DY}}(\bmat{k}_\perp) = \frac{\alpha_s C_{F}}{2\pi} \lbc \frac{4}{\eta} \lb 2\mathcal{L}_0(\bmat{k}_\perp^2,\mu^2) -\frac{1}{\epsilon} \delta^{(2)}(\bmat{k}_\perp) \rb + \frac{2}{\epsilon} \lb \frac{1}{\epsilon} - \ln \lp \frac{\nu^2}{\mu^2} \rp  \rb \delta^{(2)}(\bmat{k}_\perp) -\frac{\pi^2}{6} \delta(\bmat{p}_\perp) \\
      -4 \mathcal{L}_1 (\bmat{k}_\perp^2, \mu^2) +4 \mathcal{L}_0 (\bmat{k}_\perp^2, \mu^2) \ln \lp \frac{\nu^2}{\mu^2} \rp  \rbc
    \end{multline}
    is the unsubtracted Drell-Yan TMD soft function. 
    The definition of the distributions is~\cite{Chiu:2012ir}
    \begin{equation}
        {\cal L}_n(\mu,\bmat{k}^2_\perp;\lambda) = \frac{1}{2 \pi \mu^2}\bigg[\frac{\mu^2}{\bmat{k}_{\perp}^2}\ln^{n}\bigg(\frac{\mu^2}{\bmat{k}_{\perp}^2}\bigg)\bigg]^\lambda_+ \,.
    \end{equation}
    If the domain $\lambda = \mu$ we write ${\cal L}_n(\mu,\bmat{k}_\perp^2;\lambda=\mu) = {\cal L}_n(\bmat{k}^2_\perp,\mu)$.
    
    Adding all the contributions together we get
    \begin{eqnarray}
      \label{eq:NLO3s18}
      S^{\perp, \text{NLO}}_{\chi  \to \Q{3}{S}{1}{8} }(\bmat{k}_\perp;\mu,\nu) &=& \frac{d-2}{d-1}\Bigg\{ \bigg[ S^{\perp}_{\text{DY}}(\bmat{k}_\perp) + \frac{\alpha_s C_A}{2\pi} \bigg( \frac{1}{\epsilon} \delta^{(2)}(\bmat{k}_\perp) - 2 \mathcal{L}_0 (\bmat{k}_\perp^2,\mu^2)\bigg) \bigg] \langle \Q{3}{S}{1}{8} \rangle _{\text{LO}} \\
    && \hspace{-20ex} +\delta^{(2)}(\bmat{k}_\perp) \bigg[ (C_F - \frac{1}{2} C_A)\; \frac{\pi \alpha_s}{\ve}\langle \Q{3}{S}{1}{8} \rangle _{\text{LO}} \nonumber \\
    &+& 
    \frac{4\alpha_s}{3 \pi m^2}\bigg( C_F \sum_J \langle \Q{3}{P}{J}{1}  \rangle_{\text{LO}} +  B_F \sum_J \langle \Q{3}{P}{J}{8}  \rangle_{\text{LO}}  \bigg) \; \bigg( \frac{1}{\epsilon_{\UV}} -  \frac{1}{\epsilon_{\IR}} \bigg)
    \bigg] \Bigg\} \nonumber \,.
    \end{eqnarray}
    
    We note that compared to the usual Drell-Yan or lepton fusion to di-hadron processes the shape function contains an additional divergence and thus an additional logarithmic scale dependence. The origin of this term is diagram (l) in Fig.~\ref{fig:soft-soft} which encodes the gluon self-exchanges of the $Q\bar{Q}$ state. We discuss this divergence later and how the logarithmic scale dependence cancels in the fixed order cross section against the virtual contributions from the hard process.
    
    
    \subsubsection*{The $\wave{P}$-wave shape function}
    
\begin{figure}[!ht]
      \centerline{\includegraphics[width = 0.92 \textwidth]{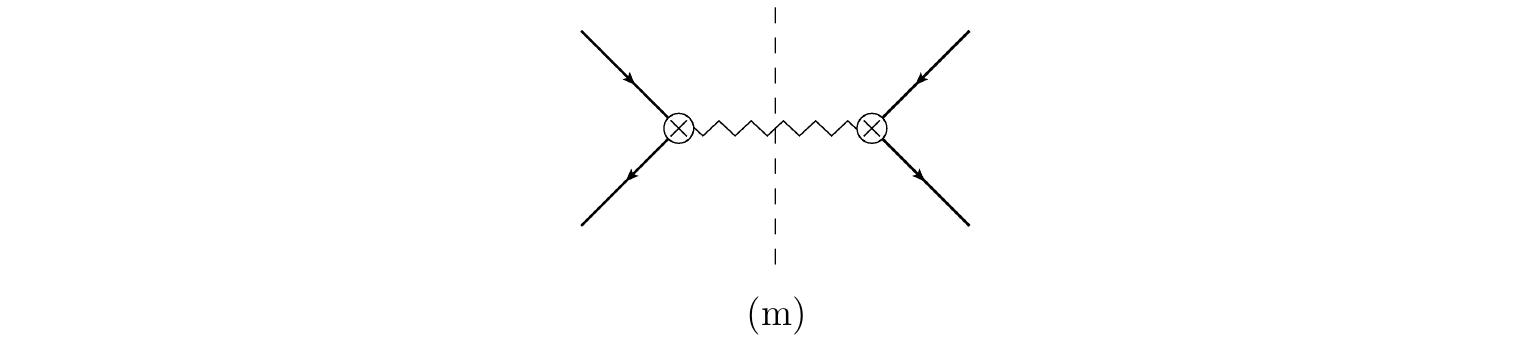}}
      \caption{The only $\mathcal{O}(\alpha_s)$ (real) soft gluon diagram contributing to the $\Q{3}{P}{J}{1}$ shape function.}
      \label{fig:P-shape}
    \end{figure}

    The contribution of the $\wave{P}$-wave starts at order $\alpha_s$ because of the presence of the soft gluon field operator, $B_s^{a,k}$ in the shape function. Thus, as we define LO as ${\cal O}(\alpha_s^0)$ this shape function starts at NLO. The diagram corresponding to this contribution is given in Fig.~\ref{fig:P-shape}, and we have:
    \begin{equation}
      \label{eq:LO3PJ}
      S^{\perp,\;\text{LO}}_{\chi \to \Q{3}{P}{J}{1/8}} = 0
    \end{equation}
    and 
    \begin{equation}
      S^{\perp,\;\text{NLO}}_{\chi \to \Q{3}{P}{J}{1}}(\bmat{k}_{\perp}) =  \frac{4 C_F}{m^2} \; \mathcal{A}_{J}^{ij} \; I^{ij}(\bmat{k}_{\perp}) \Langle \Q{3}{P}{J}{1} \Rangle_{\text{LO}}\,,
    \end{equation}
    where
    \begin{equation}
      I^{ij}(\bmat{k}_{\perp}) = g^{2} \lp\frac{e^{\gamma_E} \mu^2}{4\pi} \rp^{\epsilon} \int \frac{d^{d}\ell}{(2\pi)^3} \frac{1}{\ell_0^4} \delta(\ell^2)\delta^{(2)}(\bmat{\ell}_{\perp} - \bmat{k}_{\perp}) (\delta^{ij}\ell_0^2 - \ell^{i} \ell^{j}) \,.
    \end{equation}
    After contracting with $\mathcal{A}^{ij}$ the integral can be expressed in terms of $I_{\text{l}}$ and $I_{\text{m}}$ in Eqs.~(\ref{eq:Il}) and (\ref{eq:Im}). Expanding in $\epsilon$ and keeping the non vanishing terms  we get
    \begin{equation}
      \label{eq:NLO3PJ1}
      S^{\perp, \text{NLO}}_{\chi \to \Q{3}{P}{J}{1}} (\bmat{k}_{\perp})=  \frac{8 \alpha_s C_F}{9 \pi m^2}  \sum_J \langle \Q{3}{P}{J}{1} \rangle_{\text{LO}} \lp \frac{1}{\epsilon} \delta^{(2)}(\bmat{k}_\perp)-2 \mathcal{L}_0 (\bmat{k}_{\perp}^2,\mu^2)  + c_J \rp\,,
    \end{equation}
    where
    \begin{align}
      c_{0} &= - \frac{5}{6}\;,& c_{1} &= - \frac{7}{12}\;,& c_{2} &= - \frac{41}{60}\;.
    \end{align}
    Up to  color factors the calculation of the color octet shape function is the same. Reevaluating the color factors using
    \begin{equation}
      d^{abc}d^{abd} = 4B_F \delta^{cd}
    \end{equation}
    we get
    \begin{equation}
      \label{eq:NLO3PJ8}
      S^{\perp,\;\text{NLO}}_{\chi \to \Q{3}{P}{J}{8}}(\bmat{k}_{\perp}) =  \frac{8 \alpha_s B_F}{9 \pi m^2}  \sum_J \langle \Q{3}{P}{J}{8} \rangle_{\text{LO}} \lp \frac{1}{\epsilon} \delta^{(2)}(\bmat{k}_\perp)-2 \mathcal{L}_0 (\bmat{k}_{\perp}^2,\mu^2)  + c_J \rp \,.
    \end{equation}
    
    
    \subsubsection*{Matching in the perturbative region at NLO}
    
    As discussed before, in the region where $\bmat{k}_\perp \ll \Lambda_\textrm{QCD}$ we can match the shape function onto usoft vNRQCD operators. The matching is performed by expanding the left and right hand side of Eq.~(\ref{eq:matching}) to the same order. This results in recursive relations
    \begin{align}
      C^{i}_{\text{LO}} \langle \mathcal{O}^{[i]} \rangle_{\text{LO}}& = S^{\perp,\;\text{LO}} \nn\\
      C^{i}_{\text{NLO}} \langle \mathcal{O}^{[i]} \rangle_{\text{LO}}& = S^{\perp,\;\text{NLO}} - C^{i}_{\text{LO}} \langle \mathcal{O}^{[i]} \rangle_{\text{NLO}} \nn\\
      & \vdots
    \end{align}
    From a direct comparison with Eqs.(\ref{eq:LO3s18}) and (\ref{eq:LO3PJ}) we find:
    \begin{align}
      C^{\Q{3}{S}{1}{8}}_{\text{LO}} &= \frac{2}{3} \delta^{(2)} (\bmat{k}_\perp)\;,&  C^{\Q{3}{P}{J}{1/8}}_{\text{LO}} &=0 \;.
    \end{align}
    Next to obtain the NLO matching coefficients we need the NLO $\wave{S}$-wave matrix element. These diagrams have already been calculated; namely diagrams (a)-(h) in Figs.~\ref{fig:heavy_heavy}, \ref{fig:us-us} and \ref{fig:coulomb}. We find
    \begin{multline}
      \label{eq:NLO3S18LDME}
      \langle \Q{3}{S}{1}{8} \rangle_{\text{NLO}} =   \frac{4\alpha_s}{3 \pi m^2}  \lbc C_F \sum_J \langle \Q{3}{P}{J}{1}  \rangle_{\text{LO}} +  B_F \sum_J \langle \Q{3}{P}{J}{8}  \rangle_{\text{LO}}  \rbc \; \lp \frac{1}{\epsilon_{\UV}} -  \frac{1}{\epsilon_{\IR}}\rp \\ +  (C_F - \frac{1}{2} C_A)\; \frac{\pi \alpha_s}{\ve}\; \langle \Q{3}{S}{1}{8} \rangle _{\text{LO}}\,.
    \end{multline}
    Note that this matrix element introduces a non-trivial dependence on the $\wave{P}$-wave matrix elements at leading order. We discuss the significance of this in the following sections. Using this result we obtain the matching coefficients at NLO
    \begin{align}
      C^{\Q{3}{S}{1}{8}}_{\text{NLO}}(\bmat{k}_\perp;\mu,\nu) &= \frac{2}{3} \lp S^{\perp}_{\text{DY}}(\bmat{k}_\perp;\mu,\nu) - \frac{\alpha_s C_A}{\pi}  \mathcal{L}_0 (\bmat{k}_\perp^2,\mu^2) \rp \;, \nn \\
      C^{\Q{3}{P}{J}{1}}_{\text{NLO}}(\bmat{k}_\perp;\mu) &=  \frac{8 \alpha_s C_F}{9 \pi m^2}  \lp -2 \mathcal{L}_0 (\bmat{k}_\perp^2,\mu^2)  + c_J \rp \;, \nn \\
      C^{\Q{3}{P}{J}{8}}_{\text{NLO}}(\bmat{k}_\perp;\mu) &=  \frac{8 \alpha_s B_F}{9 \pi m^2}  \lp -2 \mathcal{L}_0 (\bmat{k}_\perp^2,\mu^2)  + c_J \rp \;.
    \end{align}
    Here we introduce the renormalized Drell-Yan function:
    \begin{equation}
    S^{\perp}_{\text{DY}}(\bmat{k}_\perp;\mu,\nu)= 
      \frac{\alpha_s C_{F}}{2\pi} \lbc 4 \mathcal{L}_0 (\bmat{k}_\perp^2, \mu^2) \ln \lp \frac{\nu^2}{\mu^2} \rp-4 \mathcal{L}_1 (\bmat{k}_\perp^2, \mu^2)  -\frac{\pi^2}{6} \delta(\bmat{k}_\perp) 
        \rbc  \,.
    \end{equation}
    
    Next we discuss the form of various divergences and how those cancel or renormalized in order to obtain finite and meaningful predictions. At the end of this section we describe the RG evolution properties of the shape function and give the prescription to resum logarithms in the TMD spectrum of $\chi_J$ decay.
    
    \subsubsection*{Treatment of divergences}
    Before proceeding with renormalization and RG evolution we wish to point out some salient points regarding divergences in our calculation. First we discuss the importance of the operator mixing terms arising from diagrams (a)-(d) in Fig.~\ref{fig:heavy_heavy}. This is followed by a discussion of the soft-soft interaction in diagram (l) of Fig.~\ref{fig:soft-soft}.
    \\

    \noindent\emph{Channel mixing:}
    
    As mentioned earlier, diagrams (a)-(d) give part of the NLO contribution of the $\Q{3}{S}{1}{8}$ LDME and the result has the spinor structure of the $\wave{P}$-wave LDME. Although they give a scaleless contribution and should be set to zero in our scheme, doing so would lead one to miss the significant role these diagrams play. Specifically, they lead to operator mixing under renormalization. If we had set this contribution to zero and taken the $1/\epsilon$ in Eq.~(\ref{eq:NLO3PJ8}) as a UV divergence needing to be absorbed into the $\wave{P}$-wave shape function we would have missed this mixing. This is why we choose not to discard this particular contribution.
    \\

    \noindent\emph{Non-singlet divergences from diagram (l):}
    
    As pointed out earlier diagram (l) which describes soft radiation from the color octet $\wave{S}$-wave $Q\bar{Q}$ state results in a divergence in Eq.~(\ref{eq:softdia}) proportional to $C_A$. This divergence is accompanied by $\ln( q_\perp / \mu)$ that needs to be resummed through RG evolution. The presence of this divergence (and the associated logarithm) serves also as a non-trivial check of factorization at NLO. The scale independence of the cross section requires a corresponding logarithm proportional to $C_A$ to be present in another of the functions in the factorized decay rate. Since the (un-subtracted) TMDFFs are universal and channel independent this term needs to appear in the corresponding hard function. Indeed, in the virtual cross section calculation (see Ref.~\cite{Petrelli:1997ge}) we find such a term. 

    \subsubsection*{Renormalization, evolution, and resummation }
    
    The shape function we are considering here is related to the transverse momentum measurement of the quarkonium state. As expected the perturbative calculation suffers from rapidity divergences along with the usual IR and UV divergences. The rapidity scale dependence introduced by the necessary regulator is described by the rapidity renormalization group evolution (RRGE). In this section we address both the virtuality and rapidity evolution equations one at a time. 
    
    As discussed in Sect.~\ref{tmdff} running which involves a convolution in $\bmat{k}_\perp$, can be made multiplicative under a Fourier transform to $\bmat{b}$-space. In this section, for simplicity we will work with quantities in $\bmat{b}$-space, which are denoted by the tilde. Note some quantities like the LDMEs do not depend on transverse momentum and are unaffected by transforming. 
    
    First we consider the virtuality RGE which has non-trivial matrix structure. The RG equations satisfied by the matching coefficients $C_n(\bmat{k}_\perp;\mu,\nu)$ in Eq.~(\ref{eq:matching}) and the LDMEs they multiply are:
    \begin{align}
      \label{eq:RGEs}
      \frac{d}{d \ln\mu} \tilde{C}_{n} (b;\mu,\nu) & = \sum_m \lp \tilde{\gamma}_{\mu}^{S} \delta^{nm} + \gamma^{nm}_C \rp \tilde{C}_{m}(b;\mu,\nu)\;,& \frac{d}{d \ln\mu} \langle \mathcal{O}^{[n]}_{\chi_{J}} \rangle^{(\mu)} &=  \sum_m \gamma^{nm}_{\mathcal{O}} \langle \mathcal{O}^{[m]} \rangle^{(\mu)}_{\chi_{J}} \;,
    \end{align}
    where by consistency $\tilde{\gamma}^{S}_{\mu} = -\tilde{\gamma}^{H+2D}_{\mu}$, and $\tilde{\gamma}^{H+2D}_{\mu}$ is the sum of the hard and TMDFF anomalous dimensions. From consistency of the factorization theorem we also have
    \begin{equation}
      \label{eq:matr_ad}
      \gamma^{T}_{C} = -  \gamma_{\mathcal{O}}\;.
    \end{equation}
    In this section we calculate the anomalous dimensions in Eq.~(\ref{eq:RGEs}) and solve the corresponding RG equations.  We begin our analysis with the LDMEs and and then perform the analogous analysis for the matching coefficients. At the order we are working we need only to consider two channels: $\Q{3}{S}{1}{8}$ and $\Q{3}{P}{0}{1}$. These two channels are the leading channels in the  velocity expansion based on the NRQCD power-counting.
    
    We define the renormalized LDMEs, $\langle \mathcal{O}^{[n]} \rangle ^{(\mu)}_{\chi_{J}}$ as follows,
    \begin{equation}
      \langle \mathcal{O}^{[n]} \rangle_{\chi_{J}}  = Z_{\mathcal{O}}^{nm} \langle \mathcal{O}^{[m]} \rangle ^{(\mu)}_{\chi_{J}}\,,
    \end{equation}
    and the anomalous dimension is given in terms of the renormalization matrix $Z$:
    \begin{equation}
    \gamma_{\mathcal{O}}^{nk}  \equiv -(Z_{\mathcal{O}}^{-1})^{nm} \frac{d}{d\ln\mu} Z_{\mathcal{O}}^{mk} \,.
    \end{equation}
    Since at this order only $\langle \mathcal{O}^{[1]} \rangle_{\chi_{J}}  \equiv \langle \Q{3}{S}{1}{8}\rangle_{\chi_{J}} $ needs to be renormalized we have
    \begin{align}
      Z_{\Q{3}{S}{1}{8}} &=
      \begin{pmatrix}
        1 & 0\\
        0 & 1
      \end{pmatrix}
      +\frac{4 \alpha_s(\mu) C_F}{3 \pi m^2} \frac{1}{\epsilon}
       \begin{pmatrix}
        0 & 1\\
        0 & 0
      \end{pmatrix} \,,
    \end{align}
    which gives the LDMEs anomalous dimension matrix:
    \begin{align}
      \gamma_{\Q{3}{S}{1}{8}} &=
       \frac{8 \alpha_s(\mu) C_F}{3 \pi m^2}
      \begin{pmatrix}
       0 & 1 \\
       0 & 0
      \end{pmatrix}\,.
    \end{align}
    We can now solve the RG equation to obtain the evolved LDMEs. At next-leading-logarithmic accuracy (NLL or $\textrm{NLL}'$) we have
    \begin{equation}
       \langle \Q{3}{S}{1}{8} \rangle^{(\mu)}_{\chi_J}  = \mathcal{U}_{\Q{3}{S}{1}{8}}^{1 m}(\mu,\mu_f) \langle \mathcal{O}^{[m]} \rangle ^{(\mu_f)}_{\chi_J}\,.
    \end{equation}
    where in practice the scale $\mu_f$ is the scale at which the LDMEs are extracted and $\mu$ the scale at which they are evaluated. Most recent extraction use $\mu_f = 2 m_{Q} \simeq M$. The evolution kernel $\mathcal{U}_{\Q{3}{S}{1}{8}} (\mu,\mu_f)$ is
    \begin{align}
      \mathcal{U}_{\Q{3}{S}{1}{8}}(\mu,\mu_f) &=
      \begin{pmatrix}
        1 & \omega_{\Q{3}{S}{1}{8}}(\mu,\mu_f)\\
        0 & 1
      \end{pmatrix}
      \;, &  \omega_{\Q{3}{S}{1}{8}}(\mu,\mu_f) &= - \frac{8 C_F}{3 m^2 \beta_0} \ln \lp \frac{\alpha_s(\mu)}{\alpha_s(\mu_f)}\rp \,.
    \end{align}
    As mentioned already, this is not a new results since the scale dependence of the NRQCD long distance matrix elements is a well known fact, see for example Refs.~\cite{Bodwin:1994jh,Petrelli:1997ge}. Nevertheless it is instructive to reproduce these results here.
    
    To perform the analogous analysis for the matching coefficients $C_n(\bmat{k}_\perp;\mu,\nu)$ in Eq.~(\ref{eq:matching}) we work in Fourier transform space where the RG equations simplify from convolutions to simple multiplications. We use the formulas collected in Appendix~\ref{app:evolution}. The renormalized matching coefficients are defined as follows
    \begin{equation}
      \tilde{C}_{n} (b)  = Z_{C}^{nm}(b;\mu,\nu) \tilde{C}_{m} (b;\mu,\nu) \,.
    \end{equation}
    Since $C_2$ starts at order $\mathcal{O}(\alpha_s^2)$ the renormalization matrix cannot be determined  uniquely at this accuracy. From the perturbative calculations of the matching coefficients we have 
    \begin{align}
      Z_C(b;\mu,\nu) = Z_{S}(b;\mu,\nu)
      \begin{pmatrix}
       1 & 0 \\ 
       0 & 1
      \end{pmatrix}
      -\frac{4 \alpha_s(\mu) C_F}{3 \pi m^2} \frac{1}{\epsilon}
      \begin{pmatrix}
       0 & 0 \\ 
       1 & 0
      \end{pmatrix} \,,
    \end{align}
    where
    \begin{equation}
    Z_{S}(b;\mu,\nu) =  1+ \frac{\alpha_s(\mu) C_{F}}{2\pi} \lbc - \frac{4}{\eta} \lb L_{b} +\frac{1}{\epsilon} \rb + \frac{2}{\epsilon} \lb \frac{1}{\epsilon} - \ln \lp \frac{\nu^2}{\mu^2} \rp  \rb \rbc + \frac{\alpha_s(\mu) C_A}{2\pi}  \frac{1}{\epsilon}\,,
    \end{equation}
    with
    \begin{equation}
      L_b = \ln \lp \frac{b^2 \mu^2}{4 e^{-2\gamma_E}} \rp\;.
      \end{equation}
    The corresponding anomalous dimension is given in terms of the renormalization matrix
    \begin{equation}
     \tilde{\gamma}_{\mu}^{S} \delta^{nk} + \tilde{\gamma}^{nk}_C  \equiv -(Z_{C}^{nm}(b,\mu))^{-1} \frac{d}{d\ln\mu} Z_{C}^{mk}(b,\mu)  \;.
    \end{equation}
    The $\mathcal{O}(\alpha_s)$ contribution to the $Z_S$ renormalization kernel multiplying the leading $^3P_0^{[1]}$ shape function will result in $\mathcal{O}(\alpha_s^2)$ terms which we have not calculated yet. Therfore the renormalization matrix cannot uniquely be determined at this order and it is easier to determine the anomalous dimension by simply taking the derivative of the renormalized matching coefficients. Note that this means that the order $\mathcal{O}(\alpha_s)$ contribution to the corresponding  diagonal term (in this case the one associated with $\Q{3}{P}{0}{1}$) cannot be determined. We complete the missing element by consistency of factorization. To show that this is the correct completion through explicit calculation one needs to perform the NNLO calculation of the associated shape function. This gives
    \begin{align}
      \gamma_{\mu}^{S} &= - (\gamma_{\mu}^{H} + 2\gamma_{\mu,q}^{D}) = -2 \frac{\alpha_s C_F}{\pi}   \ln\lp \frac{\nu^2}{\mu^2}  \rp + \frac{\alpha_s C_A}{\pi}\;, &
      \gamma_C &= 
      - \frac{8 \alpha_s(\mu) C_F}{3 \pi m^2}
      \begin{pmatrix}
       0 & 0 \\
       1 & 0
      \end{pmatrix}\,.
    \end{align}
    The solution of the RG equation for the matching coefficients is then given by
    \begin{equation}
     \tilde{ C}^{n}(b;\mu,\nu) = \mathcal{U}_S(\mu,\mu_0,\nu) \times \mathcal{U}_C^{nm}(\mu,\mu_0)  \tilde{C}^{m}(b;\mu_0,\nu)\;,
    \end{equation}
    where
    \begin{equation}
      \mathcal{U}_C^{T}(\mu,\mu_0)  = \mathcal{U}_{\mathcal{O}}^{-1}(\mu,\mu_0) = \mathcal{U}_{\mathcal{O}}(\mu_0 ,\mu) =
      \begin{pmatrix}
       1 &  \omega_{\mathcal{O}}(\mu_0,\mu)\\
       0 & 1
      \end{pmatrix}\,,
    \end{equation}
    and $ \mathcal{U}_S$ is a function we give in the Appendix~\ref{app:evolution}. We now have all the ingredients to write the virtuality evolution of the total shape function
    \begin{equation}
      \tilde{S}^{\perp}(b;\mu,\nu) = \mathcal{U}_{S}(\mu,\mu_0) \lb \tilde{C} (b;\mu_0,\nu) \mathcal{U}_{\mathcal{O}} (\mu_0,\mu_f) \langle \mathcal{O} \rangle^{(\mu_f)} \rb\,.
    \end{equation}
    
    This is our  result for the virtuality evolved shape function. We next consider the rapidity-RG evolution which is the same as in traditional TMD observables. The rapidity evolution of the shape function should match the evolution of the  unsubtracted TMDFFs. Furthermore the rapidity scale dependence is isolated in the matching coefficients and thus we can obtain the corresponding anomalous dimension considering only the diagonal element of the renormalization matrix $Z_S$. The rapidity scale dependence of the shape function is described by:
    \begin{align}
      \frac{d}{d\nu} \tilde{S}^{\perp}(b;\mu,\nu) &=  \gamma_{\nu}^{S} (\bmat{b},\mu,\nu) \times \tilde{S}^{\perp}(b;\mu,\nu)\;, &  \gamma_{\nu}^{S} (\bmat{b},\mu) & = Z_S^{-1} \frac{d}{d\nu}  Z_S\,,
    \end{align}
    where
      \begin{equation}
      \gamma_{\nu}^{S} (\bmat{b},\mu) = - 2 \gamma_{\nu,q}^{D} (\bmat{b},\mu) = -2 \frac{\alpha_s C_F}{\pi} L_b \,.
      \end{equation}
    The fully evolved and resummed shape function is then given in terms of the rapidity and virtuality evolution kernels,
    \begin{equation}
      \label{eq:final}
      \tilde{S}^{\perp}(b;\mu,\nu) = \mathcal{U}_{S}(\mu,\mu_0,\nu) \mathcal{V}_{S}(\nu,\nu_0,\mu_0) \lb \tilde{C} (b;\mu_0,\nu_0) \mathcal{U}_{\mathcal{O}} (\mu_0,\mu_f) \langle \mathcal{O} \rangle^{(\mu_f)} \rb\,,
    \end{equation}
    where
    \begin{equation}
       \mathcal{V}_{S}(\nu,\nu_0,\mu) = \exp \lb  \gamma_{\nu}^{S} (\bmat{b},\mu) \ln \lp \frac{\nu}{\nu_0}\rp \rb\,.
      \end{equation}
    Note that the rapidity and virtuality evolution are coupled and the order or path we wish to choose for the two-dimensional evolution changes the form of the evolution kernels. See Ref.~\cite{Scimemi:2018xaf} on a recent treatment of double scale evolution. The path of evolution we choose here is $(\mu_0,\nu_0) \to (\mu_0,\nu)  \to (\mu,\nu) $. In the  RG  and RRG evolution the initiating scales $\mu_0$ and $\nu_0$ are usually chosen to minimize large logarithms in the Fourier space.
    
    In the small transverse momentum limit when $q_{\perp} \sim \Lambda_{\text{QCD}}$ the expression in the square brackets of Eq.(\ref{eq:final}) is sensitive to non perturbative effects and the matching coefficients $C_{n}$ should be convolved with a model function, $f_n$
    \begin{equation}
      \tilde{C}_n (b;\mu_0,\nu_0) \to \tilde{C}_n (b;\mu_0,\nu_0) \times f_n(b)\;.
    \end{equation}
     Since the two shape functions in this process mix under RG evolution we assign the same model function to the matching coefficient for both shape functions. Thus at the order we are working (NLL) only one model function is introduced. 
    \subsection{Resummed cross section and numerics}
\begin{figure}[!ht]
      \centerline{\includegraphics[width =  \textwidth]{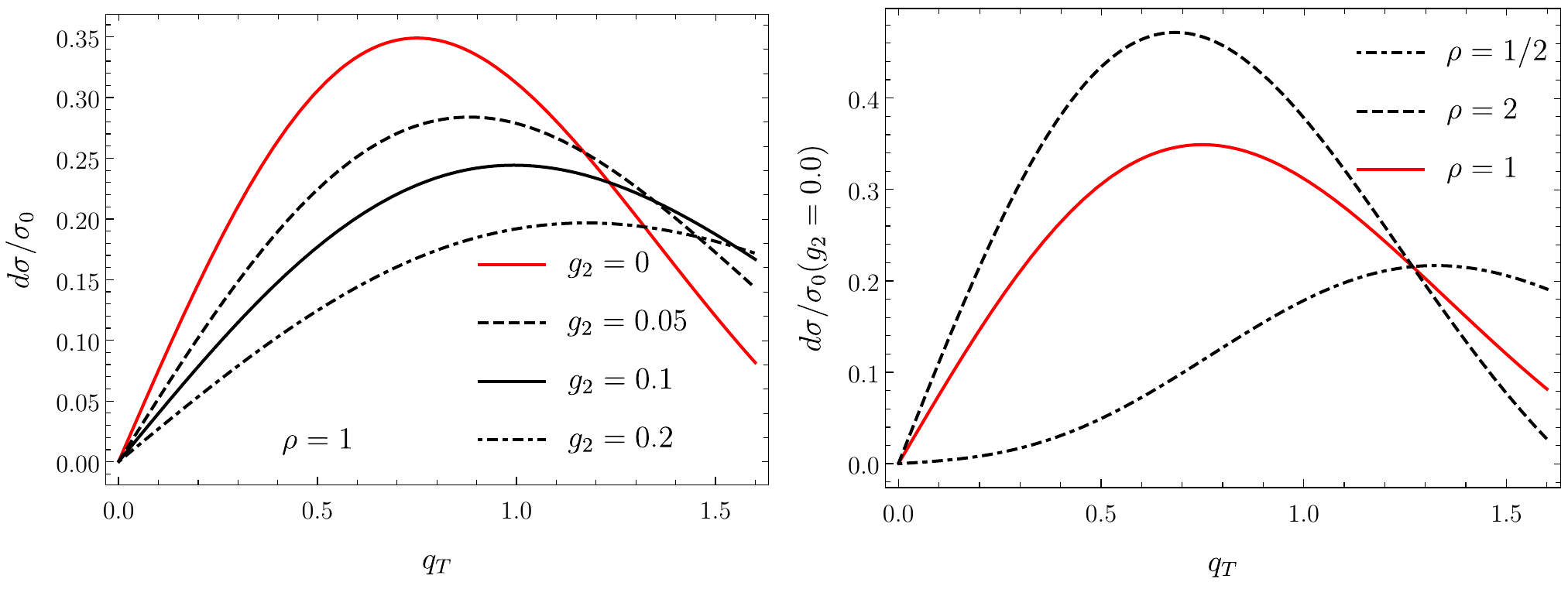}}
      \caption{Differential cross section for the process $\chi_{b0} \to \pi^{+}+\pi^{-} +X$ at NLL. We demonstrate the effect of varying the various non-perturbative parameters such as the LDMEs and the rapidity anomalous dimension.}
      \label{fig:variations}
    \end{figure}

    Here we combine the results of the past section to obtain the NLL resummed cross section for the annihilation of $\chi_{J}$ to light quarks which further decay into hadrons. 
    As was already discussed in the past section, we consider both rapidity and virtuality renormalization  group equations. In order to resum all logarithms up to NLL, the various elements of factorization need to be evaluated at the same virtuality and rapidity scales. We choose the final scales of evolution to be $\mu = \mu_b = 2 \, \textrm{Exp}(-\gamma_E)/b$ and $\nu = M_{\chi}$, in impact parameter space. Starting from the factorization theorem and using the results of Appendix~\ref{app:evolution} we have
    \begin{multline}
      \label{eq:factorization_final}
      \frac{1}{\Gamma_0} \frac{d\Gamma^{\chi_{J}}}{d^2q_\perp dz_1 dz_2} = \int_{0}^{\infty} bdb \;J_0(b q_\perp) \;\mathcal{U}_H ( \mu_b, M_{\chi})  \mathcal{V}_{S}(M_{\chi},\mu_b,\mu_b) D_{q/H_1} (z_1;\mu_b, M_\chi) D_{\bar{q}/H_2} (z_2;\mu_b,M_{\chi}) \\ f_{\Q{3}{S}{1}{8}}(b) \times \lb 1+ \frac{\langle \Q{3}{P}{J}{1} \rangle}{\langle \Q{3}{S}{1}{8} \rangle} \omega_{^3S_1^{[8]}}(\mu_b,M_{\chi}) \rb\,.
    \end{multline}
    For the discussion that follows the model function $f_{\Q{3}{S}{1}{8}}$ is not implemented. 
    The perturbative evolution kernels $\mathcal{U}_H$ and $\mathcal{V}_S$ are constructed as described in Appendix~\ref{app:evolution}. The non-perturbative part of the rapidity evolution is usually implemented through a model in the rapidity anomalous dimension:
    \begin{equation}
      \gamma_{\nu}^{S} (b,\mu) \to \gamma_{\nu}^{S} (b,\mu) + g_{K}(b)\,.
    \end{equation}
    The model function $g_K(b)$ is chosen such that it vanishes in the small $b$ limit so that we recover the perturbative prediction. Also from the operator product expansion we can show that at leading order $g_K(b)$ is quadratic in $b$. The model we use here was introduced in Ref.~\cite{COLLINS1985199} and subsequently used in various phenomenological studies,
    \begin{equation}
      g_K(b) = - g_2 b^2 \,.
    \end{equation}
    The parameter $g_2$ is a non-perturbative parameter that needs to be extracted from the experimental data but is universal among various processes. The effect of this implementation is to widen and shift the differential spectrum to larger values, see the plot on the left-hand-side of Fig.~\ref{fig:variations}.
    
    The rapidity anomalous dimension is not the only non-perturbative parameter needed to obtain a  prediction. The LDMEs also need to be extracted and are also considered universal parameters. It is important to notice that for the shape of the spectrum only the ratio 
    \begin{equation}
    \rho = \langle \Q{3}{S}{1}{8} \rangle m^2 / \langle \Q{3}{P}{J}{1} \rangle 
    \end{equation}
    is needed. The value of this ratio also modifies the shape of the distribution. The smaller the ratio we use the wider the distribution becomes, see the right-hand-side of Fig.~\ref{fig:variations}.

    To demonstrate the numerical impact of the non-perturbative parameters we show the differential cross section  for various values of these parameters in Fig.~\ref{fig:variations}. For these plot we choose $H_{1,2} = \pi^{+/-}$ with $z_1=z_2 =0.6$. The collinear fragmentation functions at NLO are taken from Ref.~\cite{Hirai:2007cx}. It is also important to note that the scale $\mu_b$ becomes non-perturbative in the region of integration and eventually the integrand diverges as we reach the Landau pole. For this reason we implement an upper cutoff in the $b$ integral in Eq.(\ref{eq:factorization_final}) at $b_{\text{max}} = 5\;\text{GeV}^{-1}$. This value is determined by searching where the integral in Eq.~(\ref{eq:factorization_final}) converges. What we note from these results is that both the rapidity anomalous dimension and the ratio of LDMEs have significant impact on the distribution. This suggests that for well constrained LDMEs one can use quarkonium production or decays to access non-perturbative aspects of TMD distributions.

    
    \section{Summary and outlook}
    \label{sec:conclusions}
    
    The goal of this paper is to study factorization and resummation for transverse momentum dependent observables involving quarkonium production or decay process. In the first part of the paper we give a diagrammatic analysis involving an  arbitrary number of soft gluon emissions from a heavy quark pair. We derived the general form of the operators involving soft Wilson Lines for both S- and P-wave production and decay (see Eqs.~\ref{eq:dG0} and \ref{eq:d1G}). We showed how RPI constrains the form of these operators and explains their structure. This symmetry constrains the matching coefficients of these operators to be the same to all orders in perturbation theory.  Using this analysis we obtained the operators and the associated  matching coefficients contributing at NLL for the process $q\bar{q} \to Q\bar{Q}[n]$. Finally we showed that the operators are invariant under collinear, soft, and ultra-soft gauge transformation.
    
    As an application of our methods, we derived the factorization and resummation for the process $\chi_J \to q \bar{q}$, followed by fragmentation to identified hadrons $H_1$ and $H_2$. The cross section studied is differential in the relative  transverse momentum of the two hadrons. This particular process demonstrates some new features one encounters in the studies of quarkonium with observables that are transverse momentum dependent. We showed that a new TMD element appears in the place of soft functions. The new functions which we refer to as quarkonium shape functions encode all the information about soft radiation from light (collinear) partons but as well as from the heavy quark pair (see also Ref.~\cite{Echevarria:2019ynx}). The contribution from the heavy quarks can be non-trivial even in the case of color singlet channels. Mixing between different production mechanisms can be generated by RG evolution. We derive the mixed renormalization group equations satisfied by the quarkonium shape functions for this particular example and solve those up to NLL accuracy.
    
    At the end of this paper we construct the NLL resummed cross section. An important result of our analysis is that we find that the differential cross section is sensitive to both the values of the LDMEs and the non-perturbative model of the rapidity anomalous dimension. This demonstrates that the LDMEs need to be well-constrained in order to extract parameters related to the TMDs. This has important implications for phenomenological studies of transverse momentum dependent observables involving quarkonium. Our analysis does not include gluon decay channel but it  would be straight forward to include them and this does not change this qualitative observation.  
    
    Extending our analysis to spin-triplet S-wave quarkonia such as $\Upsilon(1S)$ and $J/\psi$ involves one more non-trivial step. For example the process $\Upsilon(1\text{S}) \;\text{or}\; J/\psi \to gg+X$ gets significant contribution from the color singlet mechanism $\Q{3}{S}{1}{1}$. Although this channel does not contribute at leading power in $\lambda$ it is enhanced by $\ve^{-4}$ compared to the color octet and $\lambda$-leading mechanisms:  $\Q{1}{S}{0}{8}$ and  $\Q{3}{P}{0,2}{8}$. Therefore, in order to obtain reasonable and consistent results resummation of sub-leading logarithms from the color singlet channel is necessary. 
    
    Expanding this analysis to the deep inelastic scattering  process which is relevant for for the proposed electron-ion collider (EIC) is of major importance \cite{Bacchetta:2018ivt}. The partonic hard scattering process  $g+\gamma^{*} \to Q\bar{Q}[n]+ X$ which is the dominant mechanism for quarkonium production at $-q^2 \gtrsim M_{\mathcal{Q}}^2$ can give access to gluon TMDs. As we have shown in this paper quarkonium shape functions are important for this process and need to be included in future analyses. Resummation of large logarithms similar to what we have performed in this paper will be important as well. We will study this issue in future work.

    \begin{acknowledgments}
       YM was supported by the U.S. Department of Energy through the Office of Science, Office of Nuclear Physics under Contract DE-AC52-06NA25396 and by an Early Career Research Award (PI: Christopher Lee), through the LANL/LDRD Program.  YM is financially supported by the European Union's Horizon 2020 research and innovation programme under the Marie Sk\l{}odowska-Curie grant agreement No. 754496 - FELLINI. TM is supported by Director, Office of Science, Office of Nuclear Physics, of the U.S. Department of Energy under grant number DE-FG02-05ER41368. The work of SF was supported in part by the Director, Office of Science, Office of Nuclear Physics, of the U.S. Department of Energy under grant number DE-FG02-04ER41338. SF would also like to acknowledge the Mainz Institute for Theoretical Physics (MITP) of the Cluster of Excellence PRISMA+ (Project ID 39083149) for enabling us to complete this work.

    \end{acknowledgments}
    
    \appendix
    
    \section{Definitions, notation, and formulae}
    \label{app:defs}
    
    We will work in the chiral representation of Dirac matrices:
    \begin{align}
      \gamma^{\mu} &=
      \begin{pmatrix}
        0 & \sigma^{\mu}\\
        \bar{\sigma}^{\mu} & 0
      \end{pmatrix}
      \,,&
      \text{where}\;\;\; \sigma^{\mu} &= (1,\bmat{\sigma})\;, \;\;\bar{\sigma}^{\mu} = (1,-\bmat{\sigma}) \,.
    \end{align}
    The Dirac spinors in this representation then take the form:
    \begin{align}
      u(p) &=
      \begin{pmatrix}
        \sqrt{p\cdot \sigma} \;\xi\\
        \sqrt{p\cdot \bar{\sigma}}\; \xi
      \end{pmatrix}
      \,,&
      v(p) &=
      \begin{pmatrix}
        \sqrt{p\cdot \sigma} \;\eta \\
        -\sqrt{p\cdot \bar{\sigma}}\; \eta
      \end{pmatrix} \,.
    \end{align}
    Using $p_{Q/\bar{Q}} = (P +/- q)/2$, the $q$ and $\lambda$ expansion of the spinors is given by
    \begin{align}
      \label{eq:spinors-BST}
      u(p_{Q}) &=  \lp 1 - \frac{\bmat{q} \cdot \bmat{\gamma}}{4 m} +\cdots \rp u^{(0)} (1+\mathcal{O} (\lambda))\;, &  v(p_{\bar{Q}})= \lp 1 - \frac{\bmat{q} \cdot \bmat{\gamma}}{4 m}  + \cdots \rp v^{(0)} (1+\mathcal{O} (\lambda))
      \,,
    \end{align}
    where the ellipsis denotes terms of order $\mathcal{O}(q^2)$ or higher. The normalized rest frame spinors $u^{(0)}$ and $v^{(0)}$ are given by
    \begin{align}
      \label{eq:spinors-RF}
      u^{(0)} &=  \sqrt{m}
      \begin{pmatrix}
        \xi\\
        \xi
      \end{pmatrix}
      \,,&
      v^{(0)} &=  \sqrt{m}
      \begin{pmatrix}
        \eta \\
        -\eta
      \end{pmatrix}
    \end{align}
    and satisfy the equation of motion
    \begin{align}
      \label{eq:EoM-RF}
      (1-\slashed{v}) u^{(0)} &= 0\;,& (1+\slashed{v}) v^{(0)} &= 0 \,,
    \end{align}
    with $v^{\mu} = (1, \bmat{0})$.
    
    The expansion of the  heavy quark and antiquark propagator when soft momenta is inserted to the quark line is given  by,
    \begin{equation}
      \label{eq:prob}
      D(p_Q + p_{soft}) = i \frac{\slashed{p}_Q+\slashed{p}_{soft} +m}{(p_Q+p_{soft})^2 - m^2} =  \lb  i\frac{1+\slashed{v}}{2p_{soft}^{0}} + \frac{i}{4m p_{soft}^{0}} \lp \frac{1+\slashed{v}}{p_{soft}^{0}} \bmat{p}_{soft} - \bmat{\gamma}  \rp \cdot \bmat{q}  + \cdots \rb (1+ \mathcal{O}(\lambda)) \,.
    \end{equation}
    Similarly for the antiquark we have 
    \begin{equation}
      \label{eq:anti-prob}
      D(- p_{\bar{Q}} - p_{soft}) = i \frac{-\slashed{p}_{\bar{Q}}-\slashed{p}_{soft} +m}{(p_Q+p_{soft})^2 - m^2} =  \lb  i\frac{1-\slashed{v}}{2p_{soft}^{0}} - \frac{i}{4m p_{soft}^{0}} \lp \frac{1-\slashed{v}}{p_{soft}^{0}} \bmat{p}_{soft} + \bmat{\gamma}  \rp \cdot \bmat{q}  + \cdots \rb (1+ \mathcal{O}(\lambda)) \,.
    \end{equation}
    A useful formula that we used extensively is
    \begin{multline}
      \label{eq:prod-exp}
      \frac{1}{p_{t}^{0}(i)} \lb (-g)^{m-i} \prod_{s=i+1}^{m} \frac{A^{0}(p_s)}{p_t^0(s)} \rb = \sum_{\rho = i}^{m} \frac{1}{p_t^{0} (\rho)}  \lb g^{\rho-i}\prod_{s=i+1}^{\rho} \frac{A^{0}(p_s)}{\sum_{\ell=i+1}^{s} p^0_{\rho+i+1-\ell}}  \rb \\
      \times \lb (-g)^{m-\rho} \prod_{s=\rho+1}^{m} \frac{A^{0}(p_s)}{\sum_{\ell=\rho+1}^{s} p^0_{\ell}}  \rb \,.
    \end{multline}

    \subsection{Useful integrals}
    Here we give all the integrals necessary for the calculation of the real soft emissions of the shape function. This corresponds to diagrams (i)-(m):
    \begin{multline}
      \label{eq:Ii}
    I_{\text{i}} = -g^2 \lp \frac{e^{\gamma_E} \mu^2}{4\pi} \rp^{\epsilon} \int \frac{d^d k}{(2\pi)^{d-1}} \frac{\bar{n} \cdot n}{(\bar{n} \cdot k) (n\cdot k)} \frac{w^2 \nu^\eta}{\vert 2k_3 \vert^{\eta}} \delta^{(2)} (\bmat{q}_{\perp} - \bmat{k}_{T}(k^{\mu})) \delta (k^2) \\
     = - \frac{\alpha_s w^2}{2\pi}  \frac{2 e^{\epsilon \gamma_E}}{2^{\eta}\sqrt{\pi}} \frac{\Gamma(1/2-\eta/2) \Gamma(\eta/2) \Gamma(1+\eta/2 +\epsilon)}{\Gamma(1+\eta/2)} \lp \frac{\nu}{\mu} \rp^{\eta}  \frac{1}{2\pi \mu^2} \lp \frac{\mu^2}{\bmat{p}_T^2}  \rp^{1+\epsilon+\eta/2}
    \end{multline}
      
    \begin{multline}
        \label{eq:Ij/k}
      I_{\text{j/k}} = -g^2 \lp \frac{e^{\gamma_E} \mu^2}{4\pi} \rp^{\epsilon} \int \frac{d^d k}{(2\pi)^{d-1}} \frac{v \cdot n}{(v\cdot k) (n\cdot k)} \frac{w^2 \nu^\eta}{\vert 2k_3 \vert^{\eta}} \delta^{(2)} (\bmat{q}_{\perp} - \bmat{k}_{T}(k^{\mu})) \delta (k^2) \\
      =-  \frac{\alpha_sw^2}{2\pi} \frac{e^{\epsilon \gamma_E}\Gamma (1+\eta/2+\epsilon)}{ \Gamma (1+\eta/2)} \frac{2}{\eta} \lp \frac{\nu}{\mu} \rp^{\eta}  \frac{1}{2\pi \mu^2} \lp \frac{\mu^2 }{\bmat{q}_{\perp}^{2} } \rp ^{1+\epsilon+\eta/2} + \mathcal{O}(\eta)
    \end{multline}
    where we use the following relation for $c>0$
      \begin{multline}
        \int_{-\infty}^{+\infty} \frac{ dx }{(x^2 + c) (\sqrt{x^2 + c}-x)\;\vert x \vert^{\eta}} =  \int_{-\infty}^{+\infty} \frac{ dx }{(x^2 + c)^{1+\eta/2} (\sqrt{x^2 + c}-x)}  + \mathcal{O}(\eta) \\ = \frac{1}{c^{1+\eta/2}} \lp \frac{2}{\eta}+2\ln(2) +\mathcal{O}(\eta) \rp
      \end{multline}
    
      \begin{equation}
          \label{eq:Il}
      I_{\text{l}} = - g^2 \lp \frac{e^{\gamma_E } \mu^2}{4\pi} \rp^{\epsilon} \int \frac{d^d k}{(2\pi)^{d-1}} \frac{v^2}{(v\cdot k)^2} \delta^{(2)} (\bmat{q}_{\perp} - \bmat{k}_{T}(k^{\mu})) \delta (k^2) 
      =  \alpha_s \;  \frac{e^{\epsilon \gamma_E}  \csc(\pi \epsilon)}{ \Gamma (-\epsilon)} \frac{1}{2\pi \mu^2} \lp \frac{\mu^2}{\bmat{q}_{\perp}^2} \rp^{1+\epsilon}
    \end{equation}
    
      \begin{equation}
          \label{eq:Im}
      I_{\text{m}} = - g^2 \lp \frac{e^{\gamma_E } \mu^2}{4\pi} \rp^{\epsilon} \int \frac{d^d k}{(2\pi)^{d-1}} \frac{v^2 \bmat{k}_{\perp}^2}{(v\cdot k)^4} \delta^{(2)} (\bmat{q}_{\perp} - \bmat{k}_{T}(k^{\mu})) \delta (k^2) 
      = \frac{2}{3} I_{\text{l}}\,.
      \end{equation}
    
    \section{Evolution and resummation}
    \label{app:evolution}
    In this appendix we discuss both virtuality and rapidity renormalization group equations and the solutions of those equations. All factorization elements (hard, soft, soft-collinear, collinear-soft, and jet) satisfy renormalization group equations, but only transverse momentum dependent quantities have rapidity RGEs.
    
    
    \subsection{Renormalization group evolution }
    \label{ap:C1}
    The RGEs we consider have the following form
    \begin{equation}
      \label{eq:unmeasRG}
      \frac{d}{d \ln\mu}F(\mu)= \gamma^{F}_{\mu} (\mu, \alpha_s) F(\mu)= \lb \Gamma^{F}_{\mu}[\alpha_S] \ln \lp \frac{\mu^2}{m^2_F} \rp + \Delta \gamma^F_{\mu}[\alpha_S]\rb F(\mu),
    \end{equation}
    where $\gamma_{\mu}^{F}$ is the virtuality anomalous dimension. We refer to the first term in the square brackets as the cusp part since $\Gamma^{F}_{\mu}[\alpha_s]$ is proportional to the cusp anomalous dimension, $\Gamma_{\text{cusp}}$, and the second term, $\Delta \gamma^F_{\mu}[\alpha_S]$, as the non-cusp part. Both the cusp and the non-cusp terms have an expansion in the strong coupling. For the cusp term we have
    \begin{equation}
      \label{eq:G}
      \Gamma^{F}_{\mu}[\alpha_s] =  (\Gamma_F^0/\Gamma_{\text{cusp}}^0) \Gamma_{\text{cusp}} = (\Gamma_F^0/\Gamma_{\text{cusp}}^0) \sum_{n=0}^{\infty} \left(\frac{\alpha_s}{4 \pi} \right)^{1+n} \Gamma_{\text{cusp}}^n,
    \end{equation}
    and similarly the non-cusp part is given by
    \begin{equation}
      \label{eq:g}
      \Delta \gamma^F_{\mu}[\alpha_S] =  \sum_{n=0}^{\infty} \left(\frac{\alpha_s}{4 \pi} \right)^{1+n} \gamma_{F}^n\,.
    \end{equation}
    The solution to the RGE in Eq.(\ref{eq:unmeasRG}) is
    \begin{align}
      \label{eq:U}
      F(\mu)&= \mathcal{U}_F(\mu,\mu_0) F(\mu_0) \, , &\mathcal{U}_F(\mu,\mu_0)=\exp \left( K_F (\mu, \mu_0) \right) \left( \frac{\mu_0}{m_F} \right) ^{2 \;\omega_F(\mu, \mu_0)},
    \end{align}
    with
    \begin{align}
      \label{eq:Kt}
      K_F(\mu, \mu_0) &= 2 \int_{\alpha (\mu_0)}^{\alpha(\mu)} \frac{d \alpha}{\beta[\alpha]} \Gamma^{F}_{\mu}[\alpha] \int_{\alpha(\mu_0)}^{\alpha}
      \frac{d \alpha'}{\beta[\alpha']} +\int_{\alpha (\mu_0)}^{\alpha(\mu)} \frac{d \alpha}{\beta[\alpha]} \Delta \gamma^F_{\mu} [\alpha] ,\\
      \label{eq:wt}
      \omega_F(\mu, \mu_0) &=  \int_{\alpha (\mu_0)}^{\alpha(\mu)} \frac{d \alpha}{\beta[\alpha]} \Gamma^{F}_{\mu} [\alpha].
    \end{align}
    Since in this work we are interested only in the NLL and $\textrm{NLL}'$ result we keep only the first two terms in the perturbative expansion of the cusp part (i.e., $\Gamma_0^F$, $\Gamma^{0}_{\text{cusp}}$, and $\Gamma^{1}_{\text{cusp}}$) and only the first term form the non-cusp part ($\gamma_{F}^{0}$). Performing this expansion we get
    \begin{align}
      \label{eq:K}
      K_F(\mu, \mu_0) &=-\frac{\gamma_F^0}{2 \beta_0} \ln r -\frac{2 \pi \Gamma_F^0}{(\beta_0)^2} \Big{\lbrack} \frac{r-1-r\ln r}{\alpha_s(\mu)}
      + \left( \frac{\Gamma^1_{\text{cusp}}}{\Gamma^0_{\text{cusp}}}-\frac{\beta_1}{\beta_0} \right) \frac{1-r+\ln r}{4 \pi}+\frac{\beta_1}{8 \pi \beta_0}
      \ln^2 r  \Big{\rbrack}, \\
      \label{eq:w}
      \omega_F(\mu, \mu_0) &= - \frac{\Gamma_F^0}{ 2 \beta_0} \Big{\lbrack} \ln r + \left( \frac{\Gamma^1_{\text{cusp}}}{\Gamma^0_{\text{cusp}}} -
      \frac{\beta_1}{\beta_0}  \right) \frac{\alpha_s (\mu_0)}{4 \pi}(r-1)\Big{\rbrack},
    \end{align}
    where $r=\alpha(\mu)/\alpha(\mu_0)$ and $\beta_n$ are the coefficients of the QCD $\beta$-function
    \begin{equation}
      \beta[\alpha_s] = \mu \frac{d \alpha_s}{d \mu}= -2 \alpha_s \sum_{n=0}^{\infty} \left( \frac{\alpha_s}{4 \pi} \right)^{1+n} \beta_n \; .
    \end{equation}
    The expressions for all ingredients necessary to perform the evolution of any function that appears in the factorization theorems we considered in this paper are given in Tab.~\ref{tb:evolution}.
    \begin{table}[t!]
      \renewcommand{\arraystretch}{1.5}
      \begin{center}
        \begin{tabular}{|c|c|c|c|}
          \hline
          Function                   & $H(q\bar{q} \to \Q{3}{S}{1}{8})$  & $S^{\perp}$              &  $S^{\perp} (\otimes D^{\perp})^2$ \\ \hline \hline
          $\Gamma_{F}^{0}$            & $-8 C_F$                       & $ 8C_F$                    &  $8C_F$                               \\\hline
          $\gamma_{F}^{0}$            & $-12C_F - 4C_A $               & $4C_A$                     &  $12C_F + 4C_A$                  \\\hline
          $m_{F}$                    & $Q$                            & $\nu_{s}$                  &   $Q$                             \\\hline
          $\xi_{F}$                  & n.a.                           & 2                          & n.a.                             \\\hline
          $\Delta\gamma_{\nu}^{F}$    & n.a.                           & $ \mathcal{O}(\alpha_s^2)$ & n.a.                               \\\hline
        \end{tabular}
        \caption{Anomalous dimensions coefficients up to NLL accuracy.}
        \label{tb:evolution}
      \end{center}
    \end{table}
    The coefficients for the expansion of the cusp anomalous dimension are 
    \begin{align}
      \Gamma_{\text{cusp}}^{0} &= 4 C_F \nn \\
      \Gamma_{\text{cusp}}^{1} &= 4 C_F \lb \lp \frac{67}{9}-\frac{\pi ^2}{3}\rp C_A -\frac{20 } {9} n_f T_R\rb \nn \\
      \Gamma_{\text{cusp}}^{2} &= 4 C_F \lb \lp \frac{245}{6} - \frac{134}{27} \pi^2 + \frac{11}{45} \pi^4 + \frac{22}{3} \zeta_3 \rp C_A^2 + \lp -\frac{209}{108} + \frac{5}{27} \pi^2 - \frac{7}{3} \zeta_3  \rp 8 C_A n_f T_R  \nn \\
      &+ \lp 16 \zeta_3 -\frac{15}{3} C_F n_f T_R - \frac{64}{27} T_R^2 n_f^2 \rp  \rb \,.
    \end{align}
    
    
    \subsection{Rapidity renormalization group evolution }
    \label{ap:C2}
    In this section we summarize the solution for the rapidity renormalization group equations for the soft TMD functions.  The RRG equation for transverse momentum measurements (in Fourier space) of the function $ S^{\perp}$ takes the following form,
    \begin{equation}
      \frac{d}{d\ln\nu} \tilde{S}^{\perp}(\mu,\nu) = \tilde{\gamma}^{S}_{\nu} (\mu) \times  \tilde{S}^{\perp}(\mu,\nu),
    \end{equation}
    where
    \begin{equation}
      \label{eq:rg}
      \tilde{\gamma}_{S}^{\nu} (\mu) = 2\Gamma_{\nu}^S [\alpha_s] \ln \lp \frac{\mu_E}{\mu} \rp + \Delta\gamma_{\nu}^S [\alpha_s].
    \end{equation}
    The solution of this equations is
    \begin{equation}
      S^{\perp}(\mu,\nu) = \mathcal{V}_S(\mu,\nu,\nu_0) S^{\perp}(\mu,\nu_0),
    \end{equation}
    where
    \begin{equation}
      \label{eq:RRGEsoln}
      \mathcal{V}_S(\mu,\nu;\nu_0) =  \exp(\kappa_S(\mu,\nu,\nu_0)) \times \lp \frac{\mu_E}{\mu} \rp^{ \eta_S(\mu,\nu,\nu_0)}
    \end{equation}
    and 
    \begin{align}
      \eta_S(\mu,\nu,\nu_0) &= 2 \Gamma_{\nu}^S[\alpha] \ln \left( \frac{\nu}{\nu_0}  \right),  &  \kappa_S(\mu,\nu,\nu_0) &=  \Delta\gamma_{\nu}^s[\alpha] \ln \left( \frac{\nu}{\nu_0} \right),
    \end{align}
    with $\nu_0$ the characteristic scale from which we start the evolution. This scale is chosen such that rapidity logarithms are minimized.  The first term in the rapidity anomalous dimension in Eq.(\ref{eq:rg}) is proportional to the cusp anomalous dimension and the proportionality constant we denote with $\xi_S$,  ({\it i.e.} $\Gamma_{\nu}^{S} = \xi_S \Gamma_{\text{cusp}}$). The two loop non-cusp part of the soft rapidity anomalous dimensions which we need for the NNLL resummation is
    \begin{align}
      \Delta\gamma_{\nu}^s = - \lp \frac{\alpha_s(\mu)}{4\pi}\rp^2  C_i \lb \lp  \frac{128}{9} - 56 \zeta_3 \rp C_A + \frac{112}{9} \beta_0  \rb + \mathcal{O}(\alpha_s^3) \,.
    \end{align}
    
    
    \section{Details for the matching calculation}
    \label{app:details}
    
    In this section we give some of the details of the calculations for the results presented in the main sections. We begin with the calculation of $d_{D}^{(1)}(\text{antiquak})$ which is  presented in Eq.~(\ref{eq:dDanti}). We begin by expanding one of the propagators in the antiquary line at leading order in $\bmat{q}$ and summing over all possible expansions. After some simplifications we have
    \begin{multline}
      d_{D}^{(1)}(m,n;\text{antiquark}) = -\uo   \lb (-g)^{m}  \prod_{s=1}^{m} \frac{A^{0}_s}{ p_{t}^{0}(s)} \rb \Gamma^{(0)} (p_t(m),p'_t(n)) \\ \times  \sum_{i=1}^{n} \lbc \lb g^{i} \prod_{s=i+1}^{n} \frac{A^{0}_{(n+i+1-s)'}}{ p_{t}^{'0}(s)} \rb 
      \frac{\bmat{q}\cdot \bmat{p}'_t(i)}{2 m p_{t}^{'0}(i)} \lb g^{n-i} \prod_{s=1}^{i} \frac{A^{0}_{(i+1-s)'}}{p_t^{'0}(s)} \rb \rbc \vo \,. 
    \end{multline}
    Using Eq.~(\ref{eq:prod-exp}) we have
    \begin{multline}
      d_{D}^{(1)}(m,n;\text{antiquark}) = -\uo   \lb (-g)^{m}  \prod_{s=1}^{m} \frac{A^{0}_s}{ p_{t}^{0}(s)} \rb \Gamma^{(0)} (p_t(m),p'_t(n)) \sum_{i=1}^{n} \sum_{\rho=i}^{n} \lbc \frac{1}{2mp_t^{'0}(\rho)}\\
      \times \lb g^{n-\rho} \prod_{s=\rho+1}^{n} \frac{A^{0}_{(n+\rho+1-s)'}}{ \sum_{\ell = \rho+1}^{s} p_{\ell}^{'0}} \rb 
      \lb (-g)^{n-\rho-i} \prod_{s=i+1}^{\rho} \frac{A^{0}_{(\rho+i+1-s)'}}{ \sum_{\ell = i+1}^{s} p_{\rho+i+1-\ell}^{'0}} \rb \bmat{q}\cdot \bmat{p}'_t(i)
      \lb g^{i} \prod_{s=1}^{i} \frac{A^{0}_{(i+1-s)'}}{p_t^{'0}(s)} \rb \rbc \vo \,,
    \end{multline}
    which after summing over permutations and normalizing correspondingly for each product and considering all possible values of $n$ and $m$ we have
    \begin{equation}  
        d_{D}^{(1)}(\text{antiquark}) = - \uo  S_v^{\dag}\Gamma^{(0)} S_v \lb \frac{1}{2m v\cdot \mathcal{P}}  S_{v}^{\dag} \lb \bmat{q} \cdot \bmat{\mathcal{P}} S_v\rb  \rb \vo\,.
    \end{equation}
    This is the result quoted in Eq.~(\ref{eq:dDanti}). We now proceed with the evaluation of $d_{\gamma}^{(1)}$ which as mentioned earlier is the contribution from the terms proportional to $\bmat{q} \cdot \bmat{\gamma}$ from the expansion of the propagators and spinors. Similar to the case of $d_{D}^{(1)}$ we break down the calculation into expansions along the quark and antiquark lines: 
    \begin{equation}
      d_{\gamma}^{(1)} = d_{\gamma}^{(1)}(\text{quark}) + d_{\gamma}^{(1)}(\text{antiquark})\,.
    \end{equation}
    Starting with $d_{\gamma}^{(1)}(\text{quark})$ we performed the following decomposition 
    \begin{equation}
      \label{eq:deco-g}
      d_{\gamma}^{(1)}(m,n;\text{quark}) = d_{\gamma}^{(1)}(m,n;\text{quark})\Bvert_{u} + \sum_{i=1}^{m} d_{\gamma}^{(1)}(m,n;\text{quark})\Bvert_{i} 
    \end{equation}
    corresponding to the expansion of the spinor $\bar{u}(p_{Q})$ and the $i$th propagator. We then immediately have for the general case $i\neq m$
    \begin{multline}
      d_{\gamma}^{(1)}(m,n;\text{quark})\Bvert_{i\neq m} = -(-g)^{m} \uo \lb  \prod_{s=1}^{i-1} \frac{A^{0}_s}{ p_{t}^{0}(s)} \rb \frac{\slashed{A}_i \bmat{q} \cdot \bmat{\gamma} \slashed{A}_{i+1}(1+\slashed{v})}{8 m p_{t}^{0}(i) p_{t}^{0}(i+1)} \lb  \prod_{s=i+2}^{m} \frac{A^{0}_s}{p_t^0(s)} \rb  \\
     \times  \Gamma^{(0)} (p_t(m),p'_t(n)) \lb g^{n} \prod_{s=1}^{n} \frac{A^{0}_{(n+1-s)'}}{ p_{t}^{'0}(s)} \rb \vo \,.
    \end{multline}
    Then using
    \begin{equation}
      \label{eq:gamma-exp}
      \gamma^{\mu_i}\;\bmat{\gamma}\;\gamma^{\mu_{i+1}} (1+\slashed{v}) = (1-\slashed{v})  \gamma^{\mu_i}\;\bmat{\gamma} \; \gamma^{\mu_{i+1}} + 2 v^{\mu_i} \bmat{\gamma} \; \gamma^{\mu_{i+1}} +2 v^{\mu_{i+1}}  \gamma^{\mu_i}\;\bmat{\gamma}
    \end{equation}
    along with the Eq.~(\ref{eq:EoM-RF}) we have
    \begin{multline}
      d_{\gamma}^{(1)}(m,n;\text{quark})\Bvert_{i\neq m} = -(-g)^{m} \uo \lb  \prod_{s=1}^{i-1} \frac{A^{0}_s}{ p_{t}^{0}(s)} \rb \frac{A^{0}_i \bmat{q} \cdot \bmat{\gamma} \slashed{A}_{i+1} + \slashed{A}_i \bmat{q} \cdot \bmat{\gamma} A^{0}_{i+1}}{4 m p_{t}^{0}(i) p_{t}^{0}(i+1)} \\
      \times \lb  \prod_{s=i+2}^{m} \frac{A^{0}_s}{p_t^0(s)} \rb  \Gamma^{(0)} (p_t(m),p'_t(n)) \lb g^{n} \prod_{s=1}^{n} \frac{A^{0}_{(n+1-s)'}}{ p_{t}^{'0}(s)} \rb \vo \,.
    \end{multline}
    For the special case $i=m$ we get 
    \begin{multline}
      d_{\gamma}^{(1)}(m,n;\text{quark})\Bvert_{i= m} = -(-g)^{m} \uo \lb  \prod_{s=1}^{m-1} \frac{A^{0}_s}{ p_{t}^{0}(s)} \rb \frac{ \slashed{A}_m \bmat{q} \cdot \bmat{\gamma} }{4 m p_{t}^{0}(m) } \\
     \times  \Gamma^{(0)} (p_t(m),p'_t(n)) \lb g^{n} \prod_{s=1}^{n} \frac{A^{0}_{(n+1-s)'}}{ p_{t}^{'0}(s)} \rb \vo \,.
    \end{multline}
    Expanding the spinor we find
    \begin{multline}
      d_{\gamma}^{(1)}(m,n;\text{quark})\Bvert_{u} = -(-g)^{m} \uo \lb   \frac{\bmat{q} \cdot \bmat{\gamma} \slashed{A}_1 (1+\slashed{v})  }{8 m p_1^0 } \prod_{s=2}^{m} \frac{A^{0}_s}{ p_{t}^{0}(s)} \rb \\
     \times  \Gamma^{(0)} (p_t(m),p'_t(n)) \lb g^{n} \prod_{s=1}^{n} \frac{A^{0}_{(n+1-s)'}}{ p_{t}^{'0}(s)} \rb \vo 
    \end{multline}
    and using
    \begin{align}
      \bmat{\gamma}\;\gamma^{\mu} (1+\slashed{v}) &= (1+\slashed{v})  \;\bmat{\gamma} \; \gamma^{\mu} + 2 v^{\mu} \bmat{\gamma}\;, & \uo  (1+\slashed{v}) = 2 \uo
    \end{align}
    we have
    \begin{multline}
      d_{\gamma}^{(1)}(m,n;\text{quark})\Bvert_{u} = -(-g)^{m} \uo \lb   \frac{\bmat{q} \cdot \bmat{\gamma} \slashed{A}_1 + \bmat{q} \cdot \bmat{\gamma} A^{0}_1 }{4 m p_1^0 } \prod_{s=2}^{m} \frac{A^{0}_s}{ p_{t}^{0}(s)} \rb\\
     \times  \Gamma^{(0)} (p_t(m),p'_t(n)) \lb g^{n} \prod_{s=1}^{n} \frac{A^{0}_{(n+1-s)'}}{ p_{t}^{'0}(s)} \rb \vo \,.
    \end{multline}
    Therefore from Eq.~(\ref{eq:deco-g})we have
    \begin{multline}
      d_{\gamma}^{(1)}(m,n;\text{quark}) = -(-g)^{m} \uo \sum_{i=1}^{m} \lbc \lb  \prod_{s=1}^{i-1} \frac{A^{0}_s}{ p_{t}^{0}(s)} \rb \frac{ \bmat{q} \cdot \bmat{A}_i}{2 m p_{t}^{0}(i) } \lb  \prod_{s=i+1}^{m} \frac{A^{0}_s}{p_t^0(s)} \rb \rbc \\
     \times  \Gamma^{(0)} (p_t(m),p'_t(n)) \lb g^{n} \prod_{s=1}^{n} \frac{A^{0}_{(n+1-s)'}}{ p_{t}^{'0}(s)} \rb \vo  \\
      -(-g)^{m} \uo  \lb  \prod_{s=1}^{m} \frac{A^{0}_s}{ p_{t}^{0}(s)} \rb \frac{ \bmat{q} \cdot \bmat{\gamma}}{4 m  }  \Gamma^{(0)} (p_t(m),p'_t(n)) \lb g^{n} \prod_{s=1}^{n} \frac{A^{0}_{(n+1-s)'}}{ p_{t}^{'0}(s)} \rb \vo \,.
    \end{multline}
    Using Eq.~(\ref{eq:prod-exp}):
    \begin{multline}
      d_{\gamma}^{(1)}(m,n;\text{quark}) = \uo  \sum_{i=1}^{m} \sum_{\rho = i}^{m}\lbc \frac{g}{2m p_{t}^0(\rho)} \lb (-g)^{i-1}  \prod_{s=1}^{i-1} \frac{A^{0}_s}{ p_{t}^{0}(s)} \rb  \bmat{q} \cdot \bmat{A}_i  \\
      \times \lb  g^{\rho-i}  \prod_{s=i+1}^{\rho}\frac{A^{0}_s}{\sum_{\ell=i+1}^{s} p^0_{\rho+i+1-\ell}}  \rb  \lb (-g)^{m-\rho} \prod_{s=\rho+1}^{m} \frac{A^{0}_s}{\sum_{\ell=\rho+1}^{s} p^0_{\ell}}  \rb \rbc
      \Gamma^{(0)} (p_t(m),p'_t(n)) \\
      \times \lb g^{n} \prod_{s=1}^{n} \frac{A^{0}_{(n+1-s)'}}{ p_{t}^{'0}(s)} \rb \vo \\
      - \uo  \lb (-g)^{m}  \prod_{s=1}^{m} \frac{A^{0}_s}{ p_{t}^{0}(s)} \rb \frac{ \bmat{q} \cdot \bmat{\gamma}}{4 m  }  \Gamma^{(0)} (p_t(m),p'_t(n)) \lb g^{n} \prod_{s=1}^{n} \frac{A^{0}_{(n+1-s)'}}{ p_{t}^{'0}(s)} \rb \vo \,.
    \end{multline}
    After summing over permutations within each product, normalizing with the corresponding number of permutations and considering all values of $n$ and $m$ we find
    \begin{equation}
      \label{eq:dg}
        d_{\gamma}^{(1)}(\text{quark}) = + \uo \lb \frac{g}{2m v\cdot \mathcal{P}}  S_{v}^{\dag} \,\bmat{q} \cdot \bmat{A} \,S_v  \rb S_v^{\dag}\Gamma^{(0)} S_v \vo -\uo  S_v^{\dag}\,\frac{\bmat{q} \cdot \bmat{\gamma}}{4m}\,\Gamma^{(0)} S_v \vo\,.
    \end{equation}
    The last element we calculate in this section is  $d_{\gamma}^{(1)}(\text{antiquark})$ which corresponds to the $\bmat{q} \cdot \bmat{\gamma}$ proportional terms from the properer and spinor expanding along the antiquark line. Following similar decomposition as in Eq.~(\ref{eq:deco-g}) we have
    \begin{equation}
      \label{eq:deco-g-2}
      d_{\gamma}^{(1)}(m,n;\text{antiquark}) = d_{\gamma}^{(1)}(m,n;\text{antiquark})\Bvert_{u} + \sum_{i=1}^{m} d_{\gamma}^{(1)}(m,n;\text{antiquark})\Bvert_{i} 
    \end{equation}
    where
    \begin{multline}
      d_{\gamma}^{(1)}(m,n;\text{antiquark}) \Bvert_{i} = - g^{n}\uo   \lb (-g)^{m}  \prod_{s=1}^{m} \frac{A^{0}_s}{ p_{t}^{0}(s)} \rb \\
      \times \Gamma^{(0)} (p_t(m),p'_t(n)) \sum_{i=1}^{n} \lbc \lb  \prod_{s=i+2}^{n} \frac{A^{0}_{(n+i+2-s)'}}{ p_{t}^{'0}(s)} \rb 
      \frac{(1-\slashed{v})\slashed{A}_{(i+1)'}\bmat{q}\cdot \bmat{\gamma} \slashed{A}_{i'}}{8 m p_{t}^{'0}(i) p_{t}^{'0}(i+1)} \lb  \prod_{s=1}^{i-1} \frac{A^{0}_{(i-s)'}}{p_t^{'0}(s)} \rb \rbc \vo \,.
    \end{multline}
    Then using Eqs.~(\ref{eq:EoM-RF}) and (\ref{eq:gamma-exp}) we  rewrite the above as
    \begin{multline}
      d_{\gamma}^{(1)}(m,n;\text{antiquark}) \Bvert_{i\neq n} = g^{n}\uo   \lb (-g)^{m}  \prod_{s=1}^{m} \frac{A^{0}_s}{ p_{t}^{0}(s)} \rb \\
      \times \Gamma^{(0)} (p_t(m),p'_t(n))   \lb  \prod_{s=i+2}^{n} \frac{A^{0}_{(n+i+2-s)'}}{ p_{t}^{'0}(s)} \rb 
      \frac{\slashed{A}_{(i+1)'}\bmat{q}\cdot \bmat{\gamma} A^{0}_{i'} + A^{0}_{(i+1)'}  \bmat{q}\cdot \bmat{\gamma}\slashed{A}_{i'} }{4 m p_{t}^{'0}(i) p_{t}^{'0}(i+1)} \lb  \prod_{s=1}^{i-1} \frac{A^{0}_{(i-s)'}}{p_t^{'0}(s)} \rb  \vo \,.
    \end{multline}
    For the special case $i=n$ we have
    \begin{multline}
      d_{\gamma}^{(1)}(m,n;\text{antiquark}) \Bvert_{i\neq n} = g^{n}\uo   \lb (-g)^{m}  \prod_{s=1}^{m} \frac{A^{0}_s}{ p_{t}^{0}(s)} \rb \Gamma^{(0)} (p_t(m),p'_t(n))  \\
      \times \frac{\bmat{q}\cdot \bmat{\gamma}\slashed{A}_{n} }{4 m p_{t}^{'0}(n) } \lb  \prod_{s=1}^{n-1} \frac{A^{0}_{(n-s)'}}{p_t^{'0}(s)} \rb  \vo \,.
    \end{multline}
    Expanding the spinor we get
    \begin{multline}
      d_{\gamma}^{(1)}(m,n;\text{antiquark}) \Bvert_{u} = g^{n}\uo   \lb (-g)^{m}  \prod_{s=1}^{m} \frac{A^{0}_s}{ p_{t}^{0}(s)} \rb \Gamma^{(0)} (p_t(m),p'_t(n))  \\
     \times \lb  \prod_{s=2}^{n} \frac{A^{0}_{(n+2-s)'}}{p_t^{'0}(s)} \rb  \frac{(1-\slashed{v})\slashed{A}_{1'})\bmat{q}\cdot \bmat{\gamma} }{4 m p_{1}^{'0} }  \vo 
    \end{multline}
    and using
    \begin{align}
      (1-\slashed{v})\gamma^{\mu} \; \bmat{\gamma} & = \gamma^{\mu} \; \bmat{\gamma}\; (1-\slashed{v}) -2 v^{\nu}\;\bmat{\gamma} \,, & (1-\slashed{v}) \vo &= 2 \vo
    \end{align}
    we arrive at
    \begin{multline}
      d_{\gamma}^{(1)}(m,n;\text{antiquark}) \Bvert_{u} = g^{n}\uo   \lb (-g)^{m}  \prod_{s=1}^{m} \frac{A^{0}_s}{ p_{t}^{0}(s)} \rb \Gamma^{(0)} (p_t(m),p'_t(n))  
      \lb  \prod_{s=2}^{n} \frac{A^{0}_{(n+2-s)'}}{p_t^{'0}(s)} \rb\\
      \times \frac{\slashed{A}_{1'})\bmat{q}\cdot \bmat{\gamma} - A^{0}_{1'}\bmat{q}\cdot \bmat{\gamma})  }{4 m p_{1}^{'0} }  \vo \,.
    \end{multline}
    Adding all contribution from Eq.~(\ref{eq:deco-g-2})
    \begin{multline}
      d_{\gamma}^{(1)}(m,n;\text{antiquark}) =  g^{n} \uo  \lb (-g)^{m} \prod_{s=1}^{m} \frac{A^{0}_s}{ p_{t}^{0}(s)} \rb \\
    \times \Gamma^{(0)} (p_t(m),p'_t(n)) \sum_{i=1}^{n} \lbc \lb  \prod_{s=i+1}^{n} \frac{A^{0}_{(n+i+1-s)'}}{ p_{t}^{'0}(s)} \rb 
      \frac{ \bmat{q} \cdot \bmat{A}_{i'}}{2 m p_{t}^{'0}(i) } \lb  \prod_{s=1}^{i-1} \frac{A^{0}_{(i-s)'}}{p_t^{'0}(s)} \rb \rbc\vo  \\
      - \uo  \lb (-g)^{m}  \prod_{s=1}^{m} \frac{A^{0}_s}{ p_{t}^{0}(s)} \rb  \Gamma^{(0)} (p_t(m),p'_t(n)) \frac{ \bmat{q} \cdot \bmat{\gamma}}{4 m  }  \lb g^{n} \prod_{s=1}^{n} \frac{A^{0}_{(n+1-s)'}}{ p_{t}^{'0}(s)} \rb \vo \,.
    \end{multline}
    Which can be rewritten using Eq.~(\ref{eq:prod-exp}) 
    \begin{multline}
      d_{D}^{(1)}(m,n;\text{antiquark}) = \uo   \lb (-g)^{m}  \prod_{s=1}^{m} \frac{A^{0}_s}{ p_{t}^{0}(s)} \rb \Gamma^{(0)} (p_t(m),p'_t(n)) \sum_{i=1}^{n} \sum_{\rho=i}^{n} \lbc \frac{g}{2mp_t^{'0}(\rho)}\\
      \lb g^{n-\rho} \prod_{s=\rho+1}^{n} \frac{A^{0}_{(n+\rho+1-s)'}}{ \sum_{\ell = \rho+1}^{s} p_{\ell}^{'0}} \rb 
      \lb (-g)^{n-\rho-i} \prod_{s=i+1}^{\rho} \frac{A^{0}_{(\rho+i+1-s)'}}{ \sum_{\ell = i+1}^{s} p_{\rho+i+1-\ell}^{'0}} \rb \bmat{q}\cdot \bmat{A}_{i'}
      \lb g^{i-1} \prod_{s=1}^{i-1} \frac{A^{0}_{(i-s)'}}{p_t^{'0}(s)} \rb \rbc \vo  \\
      - \uo  \lb (-g)^{m}  \prod_{s=1}^{m} \frac{A^{0}_s}{ p_{t}^{0}(s)} \rb  \Gamma^{(0)} (p_t(m),p'_t(n)) \frac{ \bmat{q} \cdot \bmat{\gamma}}{4 m  }  \lb g^{n} \prod_{s=1}^{n} \frac{A^{0}_{(n+1-s)'}}{ p_{t}^{'0}(s)} \rb \vo \,.
    \end{multline}
    After summing over permutations within each product, normalizing with the corresponding number of permutations and considering all values of $n$ and $m$ we obtain
    \begin{equation}
      \label{eq:dgb}
        d_{\gamma}^{(1)}(\text{antiquark}) = + \uo  S_v^{\dag}\Gamma^{(0)} S_v \lb \frac{g}{2m v\cdot \mathcal{P}}  S_{v}^{\dag} \,\bmat{q} \cdot \bmat{A} \,S_v  \rb \vo -\uo  S_v^{\dag}\Gamma^{(0)} \,\frac{\bmat{q} \cdot \bmat{\gamma}}{4m}\, S_v \vo \,.
    \end{equation}
    From Eqs.~(\ref{eq:dg}) and (\ref{eq:dgb}) we finally obtain
    \begin{equation}
      d_{\gamma}^{(1)} =  \uo \lbc  S_v^{\dag}\Gamma^{(0)} S_v, \lb \frac{1}{2m v\cdot \mathcal{P}}  S_{v}^{\dag} \lb \bmat{q} \cdot \bmat{A} S_v\rb  \rb - \bmat{\gamma} \rbc \vo\,.
    \end{equation}
    
    \bibliography{./scetq}
    \end{document}